\documentclass[aps, prx, showpacs,preprintnumbers,amsmath,amssymb,twocolumn,superscriptaddress,floatfix, nofootinbib, 10pt]{revtex4-2}

\usepackage{tikz}
\usetikzlibrary{arrows.meta,positioning,fit,calc}
\usepackage{float}
\usepackage{comment}
\usepackage{amsfonts}
\usepackage{amsmath}
\usepackage{amssymb}
\usepackage{graphicx}
\usepackage{dcolumn}
\usepackage{bm}
\usepackage{xkcdcolors}
\usepackage{xcolor}
\usepackage{mathrsfs}
\usepackage{mathtools}

\usepackage{verbatim}
\usepackage{hyperref}
\usepackage{cleveref}
\usepackage{array}
\usepackage{algorithm}
\usepackage{algpseudocode}
\usepackage{etoolbox}

\newtheorem{theorem}{Theorem}
\newtheorem{definition}{Definition}

\graphicspath{{Images/}}

\begin{document}

\title{Testing maximum entropy models with e-values}

\author{Francesca Giuffrida}
\email{francesca.giuffrida@imtlucca.it}
\affiliation{IMT School for Advanced Studies, Lucca (Italy)}
\affiliation{Lorentz Institute for Theoretical Physics (LION), Leiden University, Leiden (The Netherlands)}
\affiliation{Dipartimento di Fisica e Chimica, University of Palermo, Palermo (Italy)}
\author{Diego Garlaschelli}
\affiliation{IMT School for Advanced Studies, Lucca (Italy)}
\affiliation{Lorentz Institute for Theoretical Physics (LION), Leiden University, Leiden (The Netherlands)}
\author{Peter Gr\"{u}nwald}
\affiliation{Centrum Wiskunde \& Informatica, Amsterdam (The Netherlands)}
\affiliation{Mathematical Institute, Leiden University, Leiden (The Netherlands)}

\begin{abstract}
E-values have recently emerged as a robust and flexible alternative to p-values for hypothesis testing, especially under optional continuation, i.e., when additional data from further experiments are collected. In this work, we define optimal e-values for testing between maximum entropy models, both in the microcanonical (hard constraints) and canonical (soft constraints) settings. 
We show that, when testing between two hypotheses that are both microcanonical, the so-called growth-rate optimal e-variable admits an exact analytical expression, which also serves as a valid e-variable in the canonical case. 
For canonical tests, where exact solutions are typically unavailable, we introduce a microcanonical approximation and verify its excellent performance via both theoretical arguments and numerical simulations. 
We then consider constrained binary models, focusing on $2 \times k$ contingency tables --- an essential framework in statistics and a natural representation for various models of complex systems. Our microcanonical optimal e-variable performs well in both settings, constituting a new tool that remains effective even in the challenging case when the number $k$ of groups grows with the sample size, as in models with growing features used for the analysis of real-world heterogeneous networks and time-series.
\end{abstract}

\maketitle
\section{Introduction}

In recent years,  scientific interest in complex data modeling has surged, due to the increasing availability of both global-scale structured data and computational power. At the same time, rising concerns about the misuse of p-values and significance testing~\cite{ioannidis05, benjamin17, mcshane19} underscore the need for reliable statistical methods to extract knowledge from data. As a robust and flexible alternative to p-values for hypothesis testing, \textit{e-values}~\cite{ramdas2023savi,RamdasW25} have recently gained considerable attention. Having been independently (re)discovered several times in different contexts (including by physicists ~\cite{zhang2011asymptotically} --- see~\cite{ramdas2023savi} for early history) over the past decades, interest suddenly exploded in 2019 when the first versions of several breakthrough papers~\cite{safe_testing, wasserman20, vovk21, shafer21} appeared on arXiv.  

An \textit{e-variable} is simply a nonnegative random variable whose expected value under the null hypothesis is at most one. 
The value it takes on the given sample is called the e-value.
This simple definition yields several desirable properties: e-values provide rigorous control of the Type I error, retain it under optional continuation (i.e., when data from additional experiments become available), and can be interpreted as a measure of evidence against the null hypothesis.
However, not all e-variables are equally useful as test statistics. To address this, a notion of \textit{optimality} is introduced. An \textit{optimal} e-variable is one that grows quickly under the alternative hypothesis, accumulating strong evidence against the null when the latter is false. In this paper, we focus specifically on \textit{growth-rate optimal} (GRO) e-variables~\cite{safe_testing}. The results in~\cite{safe_testing}, later extended in 
\cite{LardyHG24,larsson2024numeraire}, provide a general theoretical framework for constructing GRO e-variables in broad testing scenarios.

The aim of this work is to develop optimal e-variables for hypothesis testing between maximum entropy models (MEMs). These models are derived by considering each possible realization $\mathbf{x}\in\mathcal{X}$ of the data (where $\mathcal{X}$ is the set of allowed realizations) and looking for the probability distribution $P(\mathbf{x})$ that maximizes Shannon entropy 
\begin{equation}\label{ent_func_intro} 
\mathcal{S}[P]=-\sum_{\mathbf{x}\in\mathcal{X}}P(\mathbf{x})\log P(\mathbf{x}) 
\end{equation} 
under a set of constraints, typically defined through a vector of observables $\mathbf{c}(\mathbf{x})$ over the data. This approach, due to Gibbs~\cite{gibbs} and Jaynes~\cite{jaynes}, outputs ensembles of data reproducing the constrained quantities and randomizing everything else maximally.

Two main formulations of MEMs exist, depending on how the constraints are enforced. If the constraints are imposed as exact values, i.e., $\mathbf{c}(\mathbf{x}) = \mathbf{c}^*$ on each realizable $\mathbf{x}$, one obtains a \textit{microcanonical model}, where only configurations satisfying the constraints are assigned nonzero probability. If, instead, the constraints are satisfied only on average, i.e., $\mathbb{E}_P[\mathbf{c}(\mathbf{x})] = \mathbf{c}^*$, one obtains a \textit{canonical model}, where fluctuations are allowed and the probability distribution has exponential form. In statistical terminology, by varying $\mathbf{c}^*$ one obtains an {\em exponential family with discrete outcome space and uniform carrier\/}~\cite{Brown86}. 
In both cases, the probability of $\mathbf{x}$ is entirely determined by the value of $\mathbf{c}(\mathbf{x})$, which plays the role of sufficient statistic. 

MEMs are commonly used to model complex systems that give rise to structured data, e.g., in network science~\cite{squartini2017, squartini2018, cimini2019} and time-series analysis~\cite{marcaccioli2020, marcaccioli2020maximum}, where they capture structural properties such as (heterogeneous) node degrees in networks or empirical trends in (non-stationary) temporal data, respectively. 
However, while statistical tests for exponential family models are well established, testing procedures specifically tailored to maximum entropy models remain far less developed, especially when applied in the microcanonical setting. 
In fact, e-variables for testing between general MEMs have so far not been developed at all: the only related works we are aware of are~\cite{TurnerG22,TurnerLG24} and~\cite{hao2024evaluesexponentialfamiliesgeneral, GrunwaldLHBJ24}. The former concentrates on the very specific sub-case of $2 \times 2$ tables (to re-appear as Example A--C in our paper later on), but uses e-variables which are designed for purely sequential purposes, and are therefore not optimal in the sense we define below, neither in the canonical nor in the microcanonical setting. 
The latter works,~\cite{hao2024evaluesexponentialfamiliesgeneral, GrunwaldLHBJ24}, studied e-variables for testing between two exponential families with the same sufficient statistic but different carriers.
By contrast, testing between MEMs amounts to testing exponential families with different sufficient statistics but the same (uniform) carrier. This is exactly the aim of this work.

This paper is organized as follows. 
In section \ref{sec:intro_evariables}, we introduce e-variables and growth-rate optimality. 
In section \ref{sec:application_to_MEMS}, we address the problem of finding optimal e-variables for testing between two maximum entropy models, either microcanonical (\ref{subsec:micro_test}) or canonical (\ref{subsec:cano_test}), that differ in their sufficient statistics. 
We introduce a method to construct optimal e-variables in both Bayesian and non-Bayesian settings. We show that the microcanonical GRO e-variable is also a valid canonical e-variable, and that in some cases, it asymptotically coincides with the optimal one. 
This is particularly relevant: while canonical models are way more commonly used in the literature, calculating the optimal canonical e-variable is usually analytically impossible and computationally infeasible. 
Here, we provide a method to explicitly compute the optimal microcanonical e-variable and to further verify how well it approximates the optimal canonical e-variable. 
In section \ref{sec:contingency_tables}, we explicitly apply these results to contingency tables, underlying connections with important problems in network science. 
We first analyze the case of $2\times2$ contingency tables (\ref{subsec:2x2}) and then generalize to $2 \times k$ (\ref{subsec:2xk}). 
We show that, in these cases and for both Bayesian and non-Bayesian examples, the GRO microcanonical e-variable is not only a valid canonical e-variable but also an excellent approximation of the optimal canonical one. 

\section{Introduction to E-variables}
\label{sec:intro_evariables}
Consider the typical hypothesis testing scenario, where the goal is to test a \textit{null hypothesis} $\mathcal{M}_0$ against an \textit{alternative hypothesis} $\mathcal{M}_1$. Both $\mathcal{M}_0$ and $\mathcal{M}_1$ are assumed to be parametric statistical models, i.e., families of distributions sharing the same functional form: 
\begin{equation}
\mathcal{M}_j=\{P_{j}(\mathbf{x};\bm{\theta})\}_{\bm{\theta}\in\bm{\Theta}_j},\quad j \in \{0,1\},
\end{equation}
where 
$\bm{\theta}$ represents the vector of model parameters and
$\bm{\Theta}_j$ the corresponding parameter space for model $\mathcal{M}_j$.

An \textit{e-variable} $E$ is a non-negative random variable that satisfies the following condition under all distributions in the null hypothesis:
\begin{equation}\label{e_definition}
\mathbb{E}_0 [\,E\,] \leq 1 \quad \forall P_0 \in \mathcal{M}_0.
\end{equation}
The realized value of $E$ evaluated on data $\mathbf{x}$ is called an \textit{e-value}. Unlike $p$-values, {\em larger\/} e-values indicate stronger evidence against the null. This follows directly from their defining property that, under the null, their expectation is bounded by one. This simple yet powerful definition has several important implications~\cite{ramdas2023savi,RamdasW25}:
\begin{itemize}
\item \textbf{Type I error control}: The condition $\mathbb{E}_0[\,E\,] \leq 1$ ensures that a test based on e-values controls the Type I error, that is, the probability of rejecting the null hypothesis when it is actually true. Given a significance level $0\leq \alpha \leq 1$, by Markov's inequality, we have
\begin{equation}
P_0(E \geq 1/\alpha) \leq \alpha,
\end{equation}
for all $P_0 \in \mathcal{M}_0$. This guarantees that the probability of wrongly rejecting the null hypothesis does not exceed the significance level $\alpha$, regardless of the true parameter value within the null model.
\item \textbf{Post-hoc error control:} 
E-values allow a variation of valid Type-I error control even when the significance level is chosen \textit{after} observing the data~\cite{grunwald24beyond}. Specifically, if $e$ is the observed e-value, then rejecting the null hypothesis at level $1/e$ preserves a Type I risk bound despite this level being data-dependent. This contrasts with traditional p-values, which only guarantee valid inference when the significance level is fixed in advance.  
\item \textbf{Optional continuation:}
E-values support valid testing under \emph{optional continuation}, making them well-suited for sequential analyses and meta-analyses across independent studies. If $ e_{(1)}, e_{(2)}, \ldots $ are e-values computed on independent data batches (e.g., studies), their product remains a valid e-value --- even if the decision to analyze further batches, to perform tests on them, or to incorporate specific prior knowledge into the e-values is guided by the outcomes of earlier batches. In this way, Type I error control is preserved, enabling flexible and robust hypothesis testing across repeated or cumulative experimental settings.
\end{itemize}

While all random variables satisfying condition~\eqref{e_definition} qualify as e-variables, not all of them are informative. For instance, the constant random variable $ E(\mathbf{x}) \equiv 1 $ satisfies the definition but provides no information. To address this, a notion of \textit{optimal} e-variables was introduced in~\cite{safe_testing}. In particular, the authors define the \textit{Growth Rate Optimal} (GRO) e-variable as the unique solution to a specific optimization problem based on a growth criterion, which we present below.

As a first step toward understanding GRO e-variables, we introduce the concept of \textit{Bayesian evidence} (also known as the \textit{Bayesian marginal likelihood}) of model $j$ with prior density $w_j$, defined as:
\begin{equation}
P_j^{w_j}(\mathbf{x}) = \int_{\bm{\Theta}_j} P_j(\mathbf{x}; \bm{\theta}) w_j(\bm{\theta}) d\bm{\theta}.
\end{equation}
This quantity reflects the overall support the data provide for model $j$, by averaging the likelihood over the prior; it is widely used in Bayesian model selection, where models with higher evidence are preferred.
We shall mostly work with prior densities $w_j$ defined on convex parameter spaces $\bm{\Theta}_j \subset \mathbb{R}^d$ ($d > 0$), assuming they are continuous and strictly positive for all $\bm{\theta} \in \bm{\Theta}_j$. We refer to such priors as \textit{regular priors}.

Given a fixed prior $w_1$ (regular or not) on the alternative hypothesis, the GRO e-variable $S^{\text{GRO}}$ is the unique solution to the following optimization problem:
\begin{equation}\label{GRO_problem}
S^{\text{GRO}} = \arg \max_{E \in \mathcal{E}_0} \mathbb{E}_{P_1^{w_1}} [\,\log E\,],
\end{equation}
where  $\mathcal{E}_0$ denotes the set of all e-variables relative to the null model $\mathcal{M}_0$, i.e., the set of all random variables satisfying \eqref{e_definition}.

This optimization can be interpreted as a growth criterion: while the expected value of any e-variable is bounded under the null, a well-designed e-variable should grow rapidly assuming the alternative is true, when the prior $ w_1 $ is correctly specified. The use of the logarithmic growth in this criterion is motivated and discussed in more detail in~\cite{safe_testing}. The quantity $ \mathbb{E}_{P_1^{w_1}} [\log E] $, known as the \textit{e-power}~\cite{zhang24, wang24, vovk24} of $E$, has become a standard measure for evaluating the performance of an e-variable.

In the most common case, a GRO e-variable solving the aforementioned optimization problem takes the form of a \textit{Bayes factor}~\cite{morey2016philosophy}, i.e., the ratio between two Bayesian evidences:
\begin{equation}\label{bayes_GRO_form}
S(\mathbf{x}) = \frac{P_1^{w_1}(\mathbf{x})}{P_0^{w_0}(\mathbf{x})}.
\end{equation}
Equation \eqref{bayes_GRO_form} represents the Bayes factor comparing models $\mathcal{M}_1$ and $\mathcal{M}_0$. It measures the relative support that the data provide for one model over the other. However, not all Bayes factors qualify as e-variables; to ensure the e-variable property
(\ref{e_definition}), while $w_1$ may be chosen freely, a specific prior $w_0^*$, depending on $w_1$, must then be chosen for the null hypothesis. Specifically, $w_0^*$ is the solution to the following optimization problem:
\begin{equation}\label{DKL_problem}
w_0^* = \arg \min_{w\in \mathcal{W}_{\bm{\theta}_0}}D_{\text{KL}}(P_1^{w_1} \Vert P_0^{w_0})
\end{equation}
where $\mathcal{W}_{\bm{\theta}_0}$ is the space of all priors on $\bm{\theta}_0$, and {$D_{\text{KL}}(P_1^{w_1} \Vert P_0^{w_0})$ denotes the \textit{Kullback-Leibler divergence}:
\begin{align}
\nonumber D_{\text{KL}}(P_1^{w_1} \Vert P_0^{w_0}) &= \sum_{\mathbf{x}\in \mathcal{X}} P_1^{w_1}(\mathbf{x}) \log \frac{P_1^{w_1}(\mathbf{x})} {P_0^{w_0}(\mathbf{x})}\\ 
&= \mathbb{E}_{P_1^{w_1}} [\,\log S(\mathbf{x})\,].
\end{align}
}
Theorem 1 in~\cite{safe_testing} proves that given $w_1$, among all Bayes factors of the form \eqref{bayes_GRO_form}, the random variable
\begin{equation}\label{eq:sgro}
S^{\text{GRO}}(\mathbf{x}) = \frac{P_1^{w_1}(\mathbf{x})}{P_0^{w_0^*}(\mathbf{x})},
\end{equation}
is the \textit{only} e-variable,
assuming that a $w^*_0$ achieving the minimum in (\ref{DKL_problem}) exists \footnote{As shown in~\cite{safe_testing}, multiple distinct minimizers $w^*_0$ may exist, but they yield the same $P_0^{w^*_0}$. Even when a minimizer does not exist, $P_0^{w^*_0}$ can be defined as a limit along a minimizing sequence $w_j$, ensuring that (\ref{eq:sgro}) remains a valid e-variable. }.

\begin{figure}[H]
\center
\includegraphics[width=0.4\textwidth]{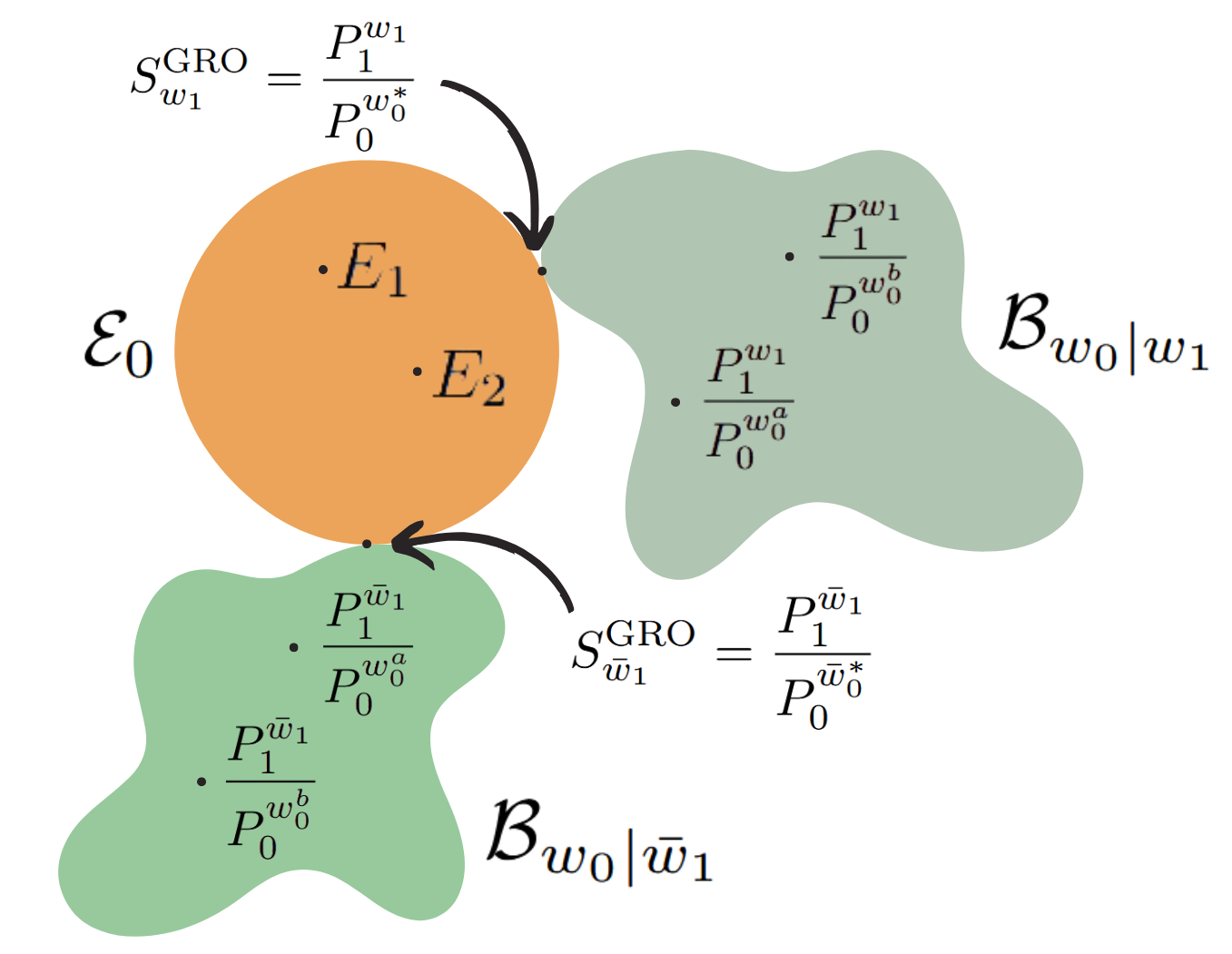}
\caption{The GRO e-variable $S^{\text{GRO}}$ is the unique intersection between the set $\mathcal{B}_{w_0 \mid w_1}$ of all Bayes factors for a given $P_1^{w_1}$ and varying $P_0^{w_0}$, and the set $\mathcal{E}_0$ of all e-variables relative to model $\mathcal{M}_0$. At the same time, it is the unique e-variable maximizing the e-power relative to $P_1^{w_1}$.  This schematic representation considers two possible alternative priors, $w_1$ and $\bar{w}_1$.}
\label{fig_intersection}
\end{figure}

To sum up, the GRO e-variable is the unique solution of two different optimization problems, defined on two different sets: for a given $P_{1}^{w_1}$ and null model $\mathcal{M}_0$, $S^{\text{GRO}}$ is the only e-variable among Bayes factors of the form \ref{bayes_GRO_form}; at the same time it is the only e-variable maximizing the e-power (see Figure~\ref{fig_intersection}).

The reader may have noticed a seeming asymmetry: while e-values are defined in a frequentist sense --- requiring Type I error control for \emph{all} $P_0 \in \mathcal{M}_0$ --- our optimality criterion for GRO relies on a prior $w_1$ over the alternative model $\mathcal{M}_1$, and thus relies on a Bayesian formulation.

It would be conceptually appealing to define an optimality criterion that, like the e-variable condition, provides performance guarantees over \textit{all $P_1 \in \mathcal{M}_1$} rather than \textit{on average according to a prior $w_1$}. As it turns out, this is indeed possible by drawing on ideas from information theory.

To move in that direction, we first note that the result of~\cite{safe_testing} is not limited to Bayes factors. It applies to more general e-variables of the form
\begin{equation}\label{GRO_form}
S(\mathbf{x}) = \frac{\bar{P}_1(\mathbf{x})}{P_0^{w_0^*}(\mathbf{x})},
\end{equation}
where $\bar{P}_1$ is any probability distribution over the data space. For such constructions to be useful, however, the choice of $\bar{P}_1$ must be guided by an appropriate extension of the GRO criterion.

This leads us to the concept of \emph{regret}, also referred to as \emph{relative growth} (\emph{regrow}) in~\cite{safe_testing}, which we adopt here using more common terminology. Regret quantifies the power loss incurred when using a candidate e-variable instead of the ideal one, which is designed for the true data-generating distribution.

To define it precisely, suppose that the data are generated according to a fixed but unknown distribution $P_1(\mathbf{x}; \bm{\theta}_1) \in \mathcal{M}_1$. 
If we knew $\bm{\theta}_1$, we could construct the GRO e-variable $S^{\text{GRO}(\bm{\theta}_1)}$ optimal for that specific alternative:
\begin{equation}
\label{eq:gro_a_posteriori}
S^{\text{GRO}(\bm{\theta}_1)} = \frac{P_1(\mathbf{x}; {\bm{\theta}_1})}{P_0^{\tilde{w}^{
*}_0}(\mathbf{x})}
\end{equation}
where 
\begin{equation}
\tilde{w}^{
*}_0 = \arg \min_{w_0 \in \mathcal{W}_{\bm{\theta}_0}} \mathbb{E}_{\bm{\theta}_1}\left[ \log \frac{P_1(\mathbf{x}; \bm{\theta}_1)}{P_0^{w_0}}\right]
\end{equation}
and $\mathbb{E}_{\bm{\theta}_j}$ denotes the expected value under $P_j(\,\cdot\,, \bm\theta_j)$.  Here, the alternative hypothesis reduces to a \textit{singleton} --- a statistical model containing only one distribution --- and in some cases, such as the $2\times k$ contingency tables considered later in this paper, $S^{\text{GRO}(\bm{\theta}_1)}$ can be computed exactly. The regret of a candidate e-variable $S_{\text{cand}}$ is then given by:
\begin{equation}
\label{regret_with_theta1}
\text{REG}_1(\bm{\theta}_1; S_{\text{cand}}) := \mathbb{E}_{\bm{\theta}_1} \left[ \log S^{\text{GRO}(\bm{\theta}_1)} - \log S_{\text{cand}} \right],
\end{equation}
which quantifies the expected loss in log-growth due to not knowing the true parameter.

Since $\bm{\theta}_1$ is unknown, a natural robustness criterion is to consider the \emph{worst-case regret} across the entire alternative:
\begin{equation}
\label{regret}
\text{REG}_1(\bm{\Theta}_1; S_{\text{cand}}) := \max_{\bm{\theta}_1 \in \bm{\Theta}_1} \text{REG}_1(\bm{\theta}_1; S_{\text{cand}}).
\end{equation}

This leads to a new optimality principle: among all e-variables, one would seek the \emph{minimax optimal} e-variable --- i.e., the e-variable that minimizes $\text{REG}_1(\bm{\Theta}_1 ; S_{\text{cand}})$ over all valid choices of $S_{\text{cand}}$.
However, computing this minimax-optimal e-variable is generally infeasible in practice, as we currently lack efficient algorithms for solving the corresponding optimization problem.
Nevertheless, when the models $\mathcal{M}_0$ and $\mathcal{M}_1$ exhibit sufficient regularity --- as is the case for Maximum Entropy models, discussed in the next section --- GRO e-variables constructed from \eqref{GRO_form} with appropriately chosen $\bar{P}_1$ can closely approximate the minimax optimal solution.
In particular, one can consider e-variables of the form~\eqref{GRO_form}, where the numerator $\bar{P}_1$ is set to a \emph{universal distribution} relative to the alternative model $\mathcal{M}_1$. Universal distributions, which include Bayesian mixtures $P_1^{w_1}$ as special cases, arise naturally in the theory of the \emph{Minimum Description Length (MDL) Principle}~\citep{grunwaldbook2007, grunwald2019, yamanishi23}. Such choices of $\bar{P}_1$ lead to e-variables that, while not exactly minimax-optimal, are typically close to optimal in terms of regret minimization, and therefore provide a practical and principled strategy for robust hypothesis testing. To clarify this connection, we take a brief detour to explain how e-value-based methods relate to the MDL Principle and its central concept, the universal distribution.

\subsection{GRO e-values and description lengths}
\label{sec:MDL}
The Minimum Description Length Principle provides a general framework for model selection: from a set of candidate models, it chooses the one that yields the shortest encoding of the observed data. In this approach, each model is represented by a single probability distribution, and models are compared via their \emph{description length}. The preferred model is the one with the smallest description length.

When comparing two models $\mathcal{M}_0$ and $\mathcal{M}_1$, the difference in description lengths is
\begin{align}\label{eq:mdl}
\Delta\mathrm{DL}(\mathbf{x})
&= \mathrm{DL}_1(\mathbf{x}) - \mathrm{DL}_0(\mathbf{x}) \nonumber\\
&= -\log \bar{P}_1(\mathbf{x}) + \log \bar{P}_0(\mathbf{x}),
\end{align}
where $\bar{P}_1$ and $\bar{P}_0$ are the representative distributions for $\mathcal{M}_1$ and $\mathcal{M}_0$.  
By Kraft's inequality~\cite{grunwaldbook2007, li2008}, the code length to describe $\mathbf{x}$, using a code that compresses optimally in expectation under $Q$, is (up to rounding) $-\log Q(\mathbf{x})$ bits; thus, $-\log \bar{P}_j(\mathbf{x})$ is the code length implied by $\bar{P}_j$.

\subsection*{Universal distributions and worst-case redundancy}
The key point is how to determine a single probability distribution $\bar{P}_j$ representing $\mathcal{M}_j$: it should perform well regardless of which specific distribution within $\mathcal{M}_j$ generated the data. In other words, if a distribution $P \in \mathcal{M}_j$ achieves a short expected code length $\mathbb{E}_P[-\log P(\mathbf{x})]$, then $\bar{P}_j$ should yield a similarly short one. Such $\bar{P}_j$ are called \emph{universal distributions} for the model $\mathcal{M}_j$~\cite{grunwaldbook2007}.

To be more precise, we can define the \emph{redundancy} of $\bar{P}_j$ relative to a parameter $\bm{\theta}_j$ as
\begin{equation}
\mathrm{RED}_j(\bm{\theta}_j;\bar{P}_j) := 
\mathbb{E}_{\bm{\theta}_j} \big[ - \log \bar{P}_j(\mathbf{x}) + \log P_{j}(\mathbf{x}; \bm{\theta}_j) \big]
\label{eq:redundancy}
\end{equation}
This quantity measures the expected extra bits needed when using $\bar{P}_j$ instead of the expected optimal code for $P_j(\cdot; \bm{\theta}_j)$. The latter is not available in practice, since the true $\bm{\theta}_j$ is typically unknown.
Thus, it is useful defining the \emph{worst-case redundancy}:
\begin{equation}
\label{eq:worst_case_redundancy}
\mathrm{RED}_j({\bm{\Theta}_j};\bar{P}_j) := \max_{\bm{\theta}_j \in \bm{\Theta}_j}
\mathrm{RED}_j(\bm{\theta}_j;\bar{P}_j).
\end{equation}
A distribution is universal if this quantity is small.

Ideally, one would like to find a $\bar{P}_j$ that minimizes the worst-case redundancy, but this is generally infeasible. However, for a $d_j$-dimensional parametric model under standard regularity conditions (satisfied by the Maximum Entropy models considered later), the Bayesian choice $\bar{P}_j = P_j^{w_j}$ with a regular prior $w_j$ attains near-optimal performance: its redundancy is within a constant of the optimal value as the sample size grows. This is formalized in the following result. 

\begin{definition}[INECCSI sets {\citep{grunwaldbook2007}}]\label{def:ineccsi} 
Let
\[
\mathcal{M}_j=\{P_{j}(\mathbf{x};\bm{\theta})\}_{\bm{\theta}\in\bm{\Theta}_j}, \quad j \in \{0,1\}.
\]
A subset $\bm{\Theta}_j' \subset \bm{\Theta}_j$ is an \emph{INECCSI subset} if its interior is a non-empty, convex, compact subset of the interior of $\bm{\Theta}_j$.
\end{definition}

INECCSI subsets exclude boundary effects and ensure regular asymptotics.  
For instance, in the Bernoulli model with $\Theta = [0,1]$, any $[\epsilon, 1-\epsilon]$ with $0 < \epsilon < 1/2$ is INECCSI.

Let $\mathcal{M}_j^{(m)}$ be the i.i.d. extension of $\mathcal{M}_j$ to $m$ observations, i.e., a model over $\mathbf{y}^m = (\mathbf{x}_1, \ldots, \mathbf{x}_m)$ where each $\mathbf{x}_i \in \mathcal{X}$ is independently sampled from $P_j(\cdot; \bm{\theta})$. Let $P_{j}^{(m)}$ be the i.i.d. extension of $P_j$ and $\bar{P}_j^{(m)}$ be a distribution on $\mathbf{y}^m$. Let $\mathrm{RED}^m({\bm{\Theta}_j};\bar{P}_j^{(m)})$ be the worst-case redundancy attained by $\bar{P}_j^{(m)}$. Since $\mathbb{E}_{{\bm{\theta_j}}}[- \log P_j^{(m)}(\mathbf{y}^m; \bm{\theta}_j)]$ grows linearly in $m$, universality of $\bar{P}_j^{(m)}$ requires $\mathrm{RED}^m$ to grow sub-linearly in $m$.

A standard result~\cite{grunwaldbook2007} states that for every INECCSI subset $\bm{\Theta}_j'$ and regular prior $w_j$, there exists $C > 0$ such that for all $m$:
\begin{multline}
\label{eq:bic}
 \frac{d_j}{2} \log m - C \;\le\;
 \inf_{\bar{P}_j^{(m)}} \mathrm{RED}^m({\bm{\Theta}'_j};\bar{P}_j^{(m)}) \\
 \le\; \mathrm{RED}^m({\bm{\Theta}'_j}; P_j^{w_j (m)}) \;
 \le\; \frac{d_j}{2} \log m + C,
\end{multline}
where the infimum is over all distributions on ${\cal X}^{(m)}$. The key implications of these results are:
\begin{itemize}
\item the minimum achievable worst-case redundancy grows as $(d_j/2) \log m$;
\item Bayesian marginal likelihoods are universal distributions, and their redundancy exceeds the minimum attainable by at most a constant --- in this sense, they are asymptotically almost optimal. 
\end{itemize}

\medskip

In a variation of the definition of universality, we may search for $\bar{P}_j$ such that $- \log \bar{P}_j(\mathbf{x}) - \log P_{\hat{\bm\theta}_j}(\mathbf{x})$ is small, in the worst-case over all possible data realizations $\mathbf{x}$; here $\hat{\bm{\theta}}_j(\mathbf{x})$ is the maximum likelihood estimator of $\bm{\theta}_j$ relative to $\mathbf{x}$. This leads to comparing $\bar{P}_j(\mathbf{x})$ to the best-fitting model \textit{a posteriori}, that is, the model that, with hindsight, would give the shortest code for the observed data.

Within the MDL literature, this more stringent criterion is often viewed as the ideal one. The distribution that achieves this is called the Normalized Maximum Likelihood (NML) distribution~\citep{grunwaldbook2007, yamanishi23, grunwald2019}. It assigns probabilities by maximizing the likelihood of the observed data while normalizing over all possible datasets of the same size:
\begin{equation}
\bar{P}_j^{\textrm{NML}}(\mathbf{x}) = \frac{P_j(\mathbf{x}; \hat{\bm{\theta}}_j(\mathbf{x}))}{\sum_{\mathbf{y} \in \mathcal{X}} P_j(\mathbf{y}; \hat{\bm{\theta}}_j(\mathbf{y}))}.
\end{equation}
 Importantly, the NML distribution does not require any prior, and achieves universality both in worst-case data and in worst-case expected regret. In fact, 
 for the MEM models we introduce below, inequality~\eqref{eq:bic} also holds when $P_j^{w_j(m)}$ is replaced by $\bar{P}_j^{\textrm{NML}(m)}$.

\subsection*{Universal distributions guarantee low-regret e-variables} We can now formalize the connection between e-values and MDL: we show that using a universal distribution $\bar{P}_1$ as the numerator in the e-variable construction \eqref{GRO_form} leads to small regret. This provides a principled justification for the use of Bayesian mixtures and NML in e-value methods.

To see this, notice that (minus) the log-ratio of any variable of the form
\begin{equation*}
S(\mathbf{x}) = \frac{\bar{P}_1(\mathbf{x})}{P_0^{w_0}(\mathbf{x})}
\end{equation*}
induces a difference in description lengths between models $\mathcal{M}_1$ and $\mathcal{M}_0$:
\begin{equation}\label{S_as_MDL}
- \log S(\mathbf{x}) = - \log \bar{P}_1(\mathbf{x}) - [ \log P_0^{w_0}(\mathbf{x})].
\end{equation}

This mirrors expression \eqref{eq:mdl}, but with one crucial difference: in order for $S$ to be an e-variable, the denominator $P_0^{w_0}$ cannot be chosen freely, as it must be the prior $w_0^*$ that ensures that $S$ qualifies as a GRO e-variable (i.e., satisfies the e-variable condition).

This formulation, however, brings a clear interpretative advantage: the description length difference expressed in \eqref{S_as_MDL} now has a direct statistical interpretation. Indeed, if $S$ is an e-variable, the corresponding code-length difference can be mapped to a statistical significance measure, since Type I error control is guaranteed. This grounds the MDL code-length difference in a frequentist hypothesis testing framework. In particular, smaller values of $-\log S(\mathbf{x})$ correspond to larger e-values and hence stronger evidence against the null model $\mathcal{M}_0$.
This observation addresses a longstanding issue in MDL: although it provides a principled model comparison method, it lacks explicit statistical guarantees such as Type I error control~\citep[Open Problem No. 9, page 413]{grunwaldbook2007}. Restricting attention to code-length differences that admit an e-value interpretation not only provides such guarantees but also makes it possible to assign a well-defined evidential value to differences in description lengths. This can be seen as the natural solution to the problem --- at least for the two-model comparison case~\citep{grunwald2019}. Extending this insight to multiple models remains an important open challenge.

\medskip

The connection between e-values and MDL becomes even more compelling when considering the regret of e-variables. Indeed, we now show that the worst-case regret of an e-variable using numerator $\bar{P}_1$ is never larger than the worst-case redundancy of $\bar{P}_1$. Let us restrict attention to an INECCSI subset $\bm{\Theta}_1' \subset \bm{\Theta}_1$. Moreover, for clarity, let's denote the regret relative to the GRO e-variable associated to $\bar{P}_1$, i.e. the regret obtained by putting $S_{\text{cand}} = \bar{P}_1(\mathbf{x})/P_0^{w_0^*}(\mathbf{x})$ in definition \ref{regret}, as $\text{REG}_1(\bm{\Theta}_1';\bar{P}_1)$. Then, for any distribution $\bar{P}_1$, according to definitions \eqref{regret} and \eqref{eq:gro_a_posteriori} (see section \ref{SM_red_reg}) it holds:

\begin{equation}
\text{REG}_1(\bm{\Theta}_1';\bar{P}_1)  \leq \text{RED}_1(\bm{\Theta}_1'; \bar{P}_1).
\label{eq:upperbound}
\end{equation}

This shows that, in the worst-case over $\bm{\theta}_1$, the regret of an e-variable built with numerator $\bar{P}_1$ is upper bounded by the redundancy of $\bar{P}_1$. Consequently, choosing a universal distribution $\bar{P}_1$, which by definition provides small redundancy, guarantees small regret.

This motivates our choices in the next sections: we will construct e-variables by setting $\bar{P}_1$ either to the Bayesian mixture $P_1^{w_1}$ (with a regular prior), or to the Normalized Maximum Likelihood $\bar{P}_1^{\textrm{NML}}$. Both yield small regret of order $(d_1/2) \log m + O(1)$.
These choices are also common in the literature. The NML distribution was used (implicitly) in~\cite{jang2023tighter} as an e-variable numerator, and Bayesian mixtures are widely adopted in the construction of e-variables~\citep{ramdas2023savi}.

\section{Application to maximum entropy models} \label{sec:application_to_MEMS}
Here, we provide explicit formulas for hypothesis tests that involve either microcanonical or canonical maximum entropy models. We focus on the case of discrete data. For a given choice of sufficient statistics $\mathbf{c}(\mathbf{x})$, we denote by $\mathcal{C}$ the discrete set of values of $\mathbf{c}(\mathbf{x})$ that are realizable by at least one $\mathbf{x} \in \mathcal{X}$; 
for mathematical convenience, we assume that $\mathcal{C}$ is a (finite or countable) subset of ${\mathbb R}^d$ for some $d \in {\mathbb N}$. 
Moreover, for any given value $\mathbf{c} \in \mathcal{C}$, let $\Omega(\mathbf{c})$ represent the number of configurations satisfying the constraint $\mathbf{c}(\mathbf{x})=\mathbf{c}$, formally defined as:
\begin{equation} \label{omega_def}
\Omega(\mathbf{c}):=\sum_{\mathbf{x}\,: \,\mathbf{c}(\mathbf{x}) = \mathbf{c}}1.
\end{equation}

Entropy maximization, when the hard constraints $\mathbf{c}(\mathbf{x}) = \mathbf{c}$ are enforced on each realizable configuration $\mathbf{x}$, yields a \emph{microcanonical} model whose functional form is a uniform distribution over data satisfying the constraints: 
\begin{equation}\label{p_mic}
P_\textrm{mic}(\mathbf{x};\mathbf{c})=
\begin{cases}
\frac{1}{\Omega(\mathbf{c})}, & \text{if}\:\mathbf{c}(\mathbf{x})=\mathbf{c};\\
0, & \text{else}.
\end{cases}
\end{equation}
The parameters of a microcanonical model correspond to the sufficient statistics themselves, with values in the discrete parameter space $\bm{\Theta}_{\textrm{mic}} = \mathcal{C}$.

When soft constraints $\mathbb{E}[\mathbf{c}(\mathbf{x})] = \mathbf{c}$ are enforced (that is, the value $\mathbf{c}$ of the sufficient statistic is to be met only as an ensemble average), the maximization of the entropy returns a \emph{canonical} model where the resulting functional form of the probability distribution is, instead, exponential, with positive probability for all possible data: 

\begin{equation}\label{p_can}
P_\textrm{can}(\mathbf{x};\bm{\theta})=\frac{e^{-\bm{\theta}\cdot\mathbf{c}(\mathbf{x})}}{Z(\bm{\theta})}
\end{equation}
where $Z(\bm{\theta})\equiv\sum_{\mathbf{x} \in \mathcal{X}}e^{-\bm{\theta}\cdot\mathbf{c}(\mathbf{x})}$ is a normalization term known as \textit{partition function}. Canonical models coincide with what is called  \textit{exponential families with uniform carrier function\/} in the statistics literature~\cite{Brown86}, and the formula above is generally referred to as canonical parametrization, where the parameters $\bm{\theta} \in \bm{\Theta}_{\text{can}}$ may be viewed as the Lagrange multipliers resulting from the entropy maximization.
For each value $\mathbf{c}$ defining  the microcanonical model in Eq.~\eqref{p_mic}, there is a corresponding value $\bm{\theta}$ such that $\mathbb{E}_{\bm{\theta}}[\mathbf{c}(\mathbf{x})]=\mathbf{c}$ under the canonical distribution in Eq.~\eqref{p_can}.

The above `duality' between canonical and microcanonical models implies that, alternatively, canonical models can also be parameterized using the expected value of the sufficient statistics. Given parameters $\bm{\theta}$, define the mean value vector:
\begin{equation}
\bm{\mu}(\bm{\theta}) := \mathbb{E}_{\bm{\theta}}[\mathbf{c}(\mathbf{x})].
\end{equation}
This defines a smooth, one-to-one mapping between the canonical parameter space $\bm{\Theta}_{\text{can}}$ and the set of realizable mean values, which we denote by $\tt M$. In exponential family theory, $\bm{\mu}$ is known as the \emph{mean value parameter}. For future reference, we refer to $P_{\bm{\mu}} = P_{\text{can}}(\cdot ; \bm{\theta}(\bm{\mu}))$ as the canonical distribution defined in its mean value parametrization, where $\bm{\theta}(\bm{\mu})$ is the mapping from mean-value parameters to corresponding canonical parameters, i.e. the inverse of $\bm{\mu}(\bm\theta)$.

A well-known result in this setting is that, if the set of possible constraint values $\mathcal{C}$ is finite, then:
\begin{itemize}
\item the canonical parameter space is the full space $\bm{\Theta}_{\text{can}} = \mathbb{R}^d$;
\item the corresponding space of mean values $\tt M$ coincides with the interior of the convex hull of $\mathcal{C}$.
\end{itemize}

This result ensures that the mapping $\bm{\theta} \mapsto \bm{\mu}$ is not only bijective but also covers all ``physically meaningful'' expected constraint values\footnote{It follows from the general theory of exponential families with finite support~\cite[Theorem 9.2]{BarndorffNielsen78}, under a technical condition known as \emph{steepness}, which holds when $\mathcal{C}$ is finite.}. In the rest of the paper, we will make use of this bijection and employ whichever parameterization is most convenient. In particular, when dealing with Bayesian marginal likelihoods and their priors, we will typically work in the mean-value space, bearing in mind that all results can be equivalently expressed in the canonical parameter space via the mapping $\bm{\theta} \mapsto \bm{\mu}(\bm{\theta})$.

In what follows, we define GRO e-variables for tests where both the null and the alternative hypotheses are two microcanonical or two canonical MEMs that differ in the choice of constraints. Our main theoretical results are presented in a general form, but to guide the reader through the derivations, we will use a running example throughout (Examples A, B, and C): a simple $2\times 2$ contingency table, representing two groups of binary data.

\subsection{Microcanonical test}
\label{subsec:micro_test}
Consider a test where the null $\mathcal{M}_{\text{mic}, 0}$ is a microcanonical model with sufficient statistics $\mathbf{c}_0$ taking values in set $\mathcal{C}_0$ and the alternative $\mathcal{M}_{\text{mic}, 1}$ is a microcanonical model with sufficient statistics $\mathbf{c}_1$ taking values in set $\mathcal{C}_1\neq \mathcal{C}_0$. The parameters of microcanonical models are discrete and correspond to their sufficient statistics. As shown in~\cite{micro_cano}, the NML microcanonical distribution is equivalent to a Bayesian distribution with a uniform prior over the sufficient statistics. Therefore, in this section, we restrict our analysis to the case of microcanonical Bayesian universal distributions for the alternative hypothesis, denoted by $P_{\text{mic},1}^{W_1}$, where $W_1$ is a probability mass function defined on $\mathcal{C}_1$. Thus, the microcanonical GRO e-variable reads
\begin{equation}\label{eq:emicro}
    S^\text{GRO}_{\text{mic}} = \frac{P_{\text{mic},1}^{W_1}(\mathbf{x})}{P_{\text{mic}, 0}^{W_0^*}(\mathbf{x})}
\end{equation}
and it solves the discrete version of the GRO optimization problem:
\begin{equation}\label{micro_DKL_problem}
W_0^* = \arg \min_{W\in \mathcal{W}_{\mathbf{c}_0}} D_{\text{KL}}(P^{W_1}_{\text{mic}, 1} \Vert P_{\text{mic},0}^{W_0})
\end{equation}
where instead of prior densities $w_0$, we need to consider prior probability mass functions $W_0$, and $\mathcal{W}_{\mathcal{C}_0}$ is the set of all such distributions on the parameter space $\mathcal{C}_0$. We solve the microcanonical GRO optimization problem explicitly and exactly (full derivation in SM; here we report only the main results) and find the optimal prior distribution on the null:
\begin{equation}\label{optimal_micro_W}
    W_0^*(\mathbf{c}_0) = \sum_{\mathbf{x} \, : \, \mathbf{c}_0(\mathbf{x}) = \mathbf{c}_0 } P^{W_1}_{\text{mic}, 1}(\mathbf{x}),
\end{equation}
i.e., $W_0^*(\mathbf{c}_0)$ is the marginal distribution of the null sufficient statistic $\mathbf{c}_0(\mathbf{x})$ induced by $P^{W_1}_{\text{mic}, 1}$. 
In the special case where the alternative sufficient statistics completely determine the value of the null, we say that {\em Condition A\/} holds:
\begin{align}
\label{condition_A}
& \textbf{Condition A:} \\ 
& \text{there exists a function $f: \mathcal{C}_1 \rightarrow \mathcal{C}_0$ s.t. }  \mathbf{c}_0(\mathbf{x}) = f(\mathbf{c}_1(\mathbf{x})). \nonumber
\end{align}
Under Condition A,  one can write:
\begin{equation}\label{W_gro_if_A}
    W_0^*(\mathbf{c}_0)  =   \sum_{\mathbf{c}_1\, : \, f(\mathbf{c_1}) = \mathbf{c}_0} W_1(\mathbf{c}_1),
\end{equation}
i.e., the GRO-optimal prior on the null is the distribution induced on the null sufficient statistics by the alternative prior, or equivalently, the marginal distribution of $\mathbf{c}_0$ induced by $W_1$.

Once that $W_0^*$ is computed, the microcanonical GRO-optimal e-variable can always be expressed as

 \begin{equation}\label{explicit_GRO_micro}
 S^{\text{GRO}}_{\text{mic}}(\mathbf{x}) = \frac{\Omega_0(\mathbf{c}_0(\mathbf{x}))}{\Omega_1(\mathbf{c}_1(\mathbf{x}))} \frac{W_1(\mathbf{c}_1(\mathbf{x}))}{W_0^*( \mathbf{c}_0(\mathbf{x}))}.
 \end{equation}
Finally, although the fact that $S^\text{GRO}_{\text{mic}}$ is an e-variable follows from a general theorem (Theorem~1 in~\cite{safe_testing}, as mentioned above Equation (\ref{eq:sgro})), we additionally provide a further, direct proof showing that its expected value under the null is exactly one: 
\begin{equation}
  \mathbb{E}_0[S^\text{GRO}_{\text{mic}}] = 1 \quad \forall P_{\text{mic}, 0} \in \mathcal{M}_{\text{mic}, 0}. 
\end{equation}
This direct proof, as well as the section's other detailed calculations and proofs, can be found in section \ref{SM_section_micro}. For clarity, we now provide a first example application. 

\subsection*{Example A} 
\label{ex:A}
\begin{figure}[H]
\includegraphics[width=0.5\textwidth
]{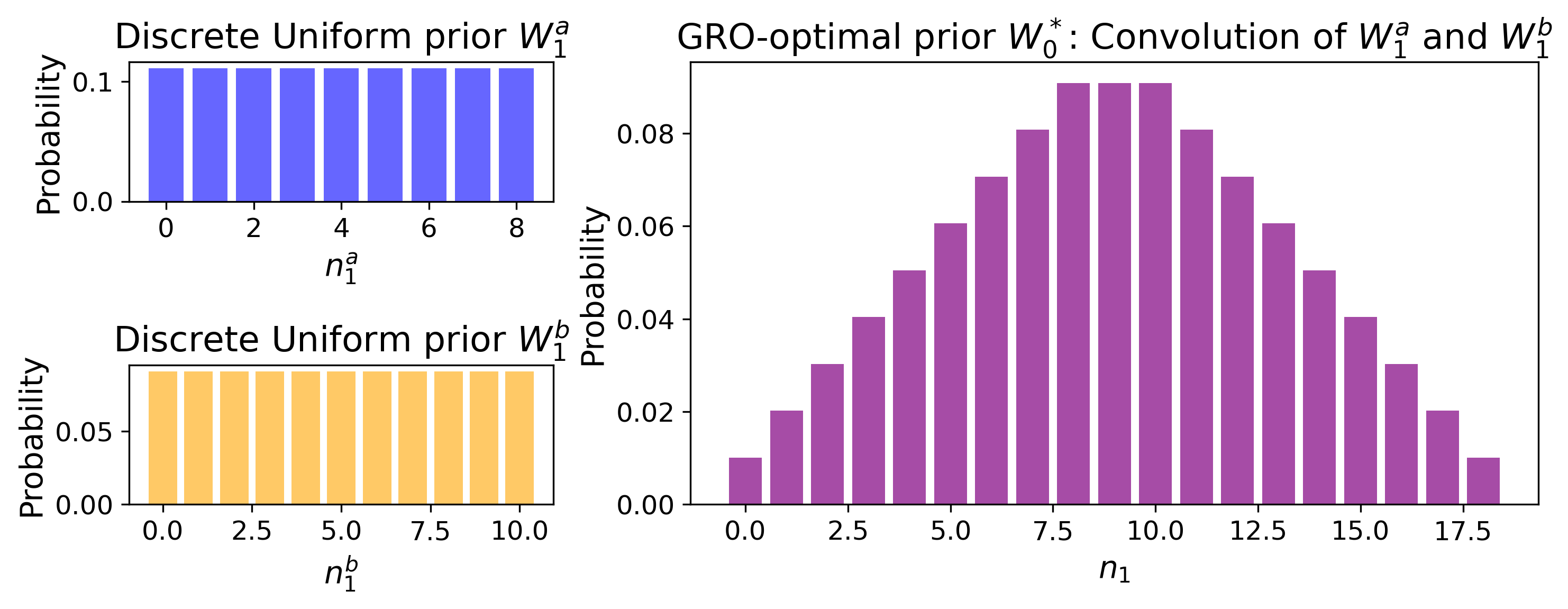}
\caption{In the microcanonical Example A, when the prior on the alternative sufficient statistics $n_1^a$ and $n_1^b$ are uniform distributions (on the left), the resulting GRO-optimal prior on the null sufficient statistic $n_1$ is the convolution of the two uniform distributions, which results in a triangular distribution (on the right). In this example, $n^a = 8$ and $n^b = 10$.}
\label{fig_two_uniforms}
\end{figure}

Let us consider the dataset $\mathbf{x} = (\mathbf{x}^a, \mathbf{x}^b)$ consisting of two groups of binary data, represented as $\mathbf{x}^a = (x_1^a, ..., x^a_{n^a})$ and $\mathbf{x}^b = (x_1^b, ..., x^b_{n^b})$, with $n^a$ and $n^b$ the respective group sizes. The total sample size is $n = n_a + n_b$. We denote by $n_1^a = \sum_{i=1}^{n_a} x^a_i$ and $n_1^b = \sum_{i=1}^{n_b} x^b_i$ the total number of $1$s in $\mathbf{x}^a$ and $\mathbf{x}^b$, and by $n_1 = n_1^a + n_1^b$ the total number of $1$s in $\mathbf{x}$. The aim is to build a microcanonical test to check whether the probability of observing $x = 1$ changes according to the different groups. To do so, we set the alternative sufficient statistics equal to the number of $1$s in each group, $
\mathbf{c}_1  = (n_1^a, n_1^b)$, and the null sufficient statistic equal to the total number of $1$s,  $
c_0  = n_1$. In the microcanonical formulation, these quantities are treated as fixed in the respective models. To find the microcanonical GRO e-variable, we apply formula \eqref{explicit_GRO_micro}, where:
\begin{itemize}
    \item $\Omega_0(n_1) = \binom{n}{n_1}$ is the number of permutations of $\mathbf{x}$ preserving the total number of $1$s;
    \item $\Omega_1(n_1^a, n_1^b) = \binom{n^a}{n_1^a}\binom{n^b}{n_1^b}$ is the number of permutations of $\mathbf{x}$ preserving the total number of $1$s in each group.  
\end{itemize}
For the sake of this example, we put independent, discrete uniform priors on the alternative parameters $n_1^a$ and $n_1^b$:
\begin{equation}
W_1(n_1^a, n_1^b) = \mathcal{U}_a(n_1^a)~\mathcal{U}_b(n_1^b) = \frac{1}{n^a +1} ~\frac{1}{n^b + 1}.
\end{equation} 
In this case, Condition A \eqref{condition_A} holds, as the null sufficient statistics can be written as a function of the alternative one: $n_1 = n_1^a + n_1^b$. Thus, the optimal prior on the null $W_0^*$ is the distribution of $n_1$ induced by $W_1$. In this case, that is simply the convolution of $\mathcal{U}_a$ and $\mathcal{U}_b$, which is a triangular discrete function:
\begin{equation}\label{uniform_convolution}
    W^*_0(n_1) =
    \begin{cases}
        \frac{n_1+1}{(n^a+1)(n^b+1)}, & \text{if } \scriptstyle 0 \leq n_1 \leq \min(n^a, n^b), \vspace{0.3em} \\
        \frac{\min(n^a, n^b) + 1}{(n^a+1)(n^b+1)}, & \text{if } \scriptstyle \min(n^a, n^b) < n_1 \leq \max(n^a, n^b), \vspace{0.3em} \\
        \frac{n^a+n^b+1 - n_1}{(n^a+1)(n^b+1)}, & \text{if } \scriptstyle \max(n^a, n^b) < n_1 \leq n^a+n^b, \vspace{0.3em} \\
        0, & \text{if } \scriptstyle \text{otherwise}.
    \end{cases}
\end{equation}
as shown in Figure~\ref{fig_two_uniforms}.

\subsection{Canonical test}\label{subsec:cano_test}
Consider a test where the null $\mathcal{M}_{\text{can}, 0}$ is a canonical model with sufficient statistics $\mathbf{c}_0$ taking values in set $\mathcal{C}_0$, and the alternative $\mathcal{M}_{\text{can}, 1}$ is a canonical model with sufficient statistics $\mathbf{c}_1$ taking values in set $\mathcal{C}_1\neq \mathcal{C}_0$. The goal is to find the canonical GRO e-variable:
\begin{equation}
 S^\text{GRO}_{\text{can}} = \frac{\bar{P}_{\text{can}, 1}(\mathbf{x})}{P_{\text{can}, 0}^{w_0^*}(\mathbf{x})}
\end{equation}
where $w_0^*$ is a prior density on the mean-value parameter space ${\tt M}_0$ and solves the optimization problem 
\begin{equation}
\label{canonical_optimization}
w_0^* = \arg \min_{w\in \mathcal{W}_{\bm\mu_0}}D_{\text{KL}}(\bar{P}_1\Vert P_0^{w_0}).
\end{equation}
No exact general solution is currently available for this problem. While in some cases it can be solved analytically or numerically, in the majority of cases, there is neither a known analytic solution nor a feasible numerical approach. Here, we propose two candidate approximations, the \textit{microcanonical approximation} and the \textit{pseudo approximation}. The first will serve as an actual approximation, and the second as a tool to assess whether the former approximation is good. 

The definition of the microcanonical approximation is based on two facts, proven in section \ref{SM_section_app_micro_cano}:
\begin{itemize}
\item A canonical universal distribution $\bar{P}_{\text{can}, 1}$ with sufficient statistics $\mathbf{c}_1$ can always be expressed as a microcanonical Bayesian marginal likelihood, i.e., there always exists a prior probability mass function $W_{\text{can},1}(\mathbf{c})$ such that 
\begin{equation}
	\label{eq:induced_prior}
   \bar{P}_{\text{can}, 1} = P_{\text{mic}, 1}^{W_{\text{can},1}}
\end{equation}
with $W_{\text{can},1}$ obtained by setting (for $j=1$): 
\begin{equation}\label{eq:fromPtoW}
W_{\text{can},j}(\mathbf{c}_j) = \sum_{\mathbf{x}\,:\, \mathbf{c}_j(\mathbf{x}) = \mathbf{c}_j} \bar{P}_{\text{can}, j} (\mathbf{x}),
\end{equation}
i.e.,  $W_{\text{can},j}(\mathbf{c}_j)$ is equal to the distribution of  $\mathbf{c}_j(\mathbf{x})$ induced by $\bar{P}_{\text{can},j}(\mathbf{x})$.
\item Given the canonical and microcanonical models $\mathcal{M}_{\text{can}}$ and $\mathcal{M}_{\text{mic}}$ built upon the same sufficient statistic $\mathbf{c}(\mathbf{x})$, a microcanonical e-variable $E$ is always a canonical e-variable:
\begin{multline}\label{eq:secondfact}
\quad \quad \mathbb{E}_{P}[\,E\,] \leq 1 \;\;  \forall \, P \in \mathcal{M}_{\text{mic}} \\ \quad \Rightarrow \quad \mathbb{E}_{P}[\,E\,] \leq 1  \;\; \forall P \, \in \mathcal{M}_{\text{can}}.
\end{multline}
\end{itemize}
Following the first fact, given $\bar{P}_{\text{can}, 1}$ and using the results of the previous section, we can build the approximating microcanonical GRO e-variable $S^{\text{GRO}}_{\text{mic}}$ for the microcanonical test based on the corresponding $P_{\text{mic}, 1}^{W_{\text{can},1}} = \bar{P}_{\text{can}, 1} $.
Thus \eqref{eq:emicro} becomes 
\begin{equation}\label{eq:approxmicro}
       S^\text{GRO}_{\text{mic}} = \frac{
       \bar{P}_{\text{can},1}(\mathbf{x})}{P_{\text{mic}, 0}^{W_0^*}(\mathbf{x})}
\end{equation}
where, readapting \eqref{optimal_micro_W} 
\begin{equation}
\label{eq:W_micro_for_cano}
W_0^*(\mathbf{c}_0) = \sum_{\mathbf{x}\, : \, \mathbf{c}_0(\mathbf{x}) = \mathbf{c}_0} \bar{P}_{\text{can}, 1}(\mathbf{x}).
\end{equation} 
Given the second fact \eqref{eq:secondfact}, the resulting microcanonical GRO e-variable is a valid canonical e-variable, i.e., it is an e-variable for the test between two canonical models, even if, for this test, it is not the GRO-optimal one. As such, from \eqref{GRO_problem}, it will have a smaller e-power than the canonical GRO one unless the two coincide:
\begin{equation}
\mathbb{E}_{\bar{P}_{\text{can}, 1}}\left[\log S^{\text{GRO}}_{\text{mic}}\right] \leq \mathbb{E}_{\bar{P}_{\text{can}, 1}}\left[\log S^{\text{GRO}}_{\text{can}} \right]
\end{equation}

The pseudo approximation is further built over the microcanonical one. The prior $W_0^*$ (used to build $S^{\text{GRO}}_{\text{mic}}$), defined on ${\cal C}_0$, is transformed into a smooth density $w_{\text{pseudo},0}$ over the corresponding (continuous) mean-value parameter space $\tt M_0$. This is obtained through a high resolution limit, by computing $W^*_0(\mathbf{c}_0)$ for a much higher dimension and by properly rescaling and normalizing it such that it is interpreted as a Riemann approximation of a continuous density on $\bm{\mu}_0$. A practical example of this procedure, which might seem abstract at this stage, is given in Examples B and C. Moreover, a pseudo-code is provided in section \ref{SM_section_pseudo}.
Given that, in general, $w_{\text{pseudo},0}$ is different from the GRO-optimal prior $w_0^*$, which in most cases remains unknown, the resulting variable 
\begin{equation}\label{eq:pseudo}
   S_\text{pseudo} = \frac{\bar{P}_{\text{can}, 1}(\mathbf{x})}{P_{\text{can}, 0}^{w_{\text{pseudo}, 0}}(\mathbf{x})}
\end{equation}
is not an e-variable, unless $w_{\text{pseudo},0} = w_0^* $. Indeed, from Theorem 1 of~\cite{safe_testing}, $S^{\text{GRO}}_{\text{can}}$ is the only e-variable of that form. Moreover, from \eqref{DKL_problem}, it holds: 
\begin{equation}
D_{\text{KL}}(\bar{P}_{\text{can}, 1} \Vert P_{\text{can}, 0}^{w_0^*}) \leq D_{\text{KL}}(\bar{P}_{\text{can}, 1} \Vert P_{\text{can}, 0}^{w_{\text{pseudo}, 0}})
\end{equation}
or, equivalently:
\begin{equation}
\mathbb{E}_{\bar{P}_{\text{can}, 1}}\left[ \log S^{\text{GRO}}_{\text{can}}\right] \leq \mathbb{E}_{\bar{P}_{\text{can}, 1}}\left[\log S_{\text{pseudo}} \right]
\end{equation}
Consequently, one has 
\begin{eqnarray}
\label{eq:sandwich}
\mathbb{E}_{\bar{P}_{\text{can}, 1}}\left[ \log S^{\text{GRO}}_{\text{mic}}\right] &\leq& \mathbb{E}_{\bar{P}_{\text{can}, 1}}\left[ \log S^{\text{GRO}}_{\text{can}}\right]\nonumber \\ &\leq& \mathbb{E}_{\bar{P}_{\text{can}, 1}}\left[\log S_{\text{pseudo}} \right],
\end{eqnarray}
i.e., the two approximations provide an upper and a lower bound for the e-power of the canonical GRO e-variable. In summary, when the canonical GRO e-variable is not available, we can follow a two-step procedure:
\begin{enumerate}\label{two_cano_steps}
\item We build the corresponding microcanonical approximation, knowing that it is a valid candidate e-variable. To build it, we first transform the canonical universal distribution into a microcanonical one, by finding $W_{\text{can}, 1}$ as in \eqref{eq:fromPtoW}. Then, we compute $S_{\text{mic}}^{\text{GRO}}$ according to the formulas expressed in the previous section (Equations \eqref{optimal_micro_W} and \eqref{explicit_GRO_micro}). 
\item The goodness of the microcanonical approximation can be evaluated by looking at the width of the interval
\begin{align}\label{eq:r}
r  = \mathbb{E}_{\bar{P}_{\text{can}, 1}}\left[\log S_{\text{pseudo}} \right]  - \mathbb{E}_{\bar{P}_{\text{can}, 1}}\left[\log S^{\text{GRO}}_{\text{mic}}\right] \geq 0
\end{align}
where, for future reference, it is useful to note that, using definitions  (\ref{eq:pseudo}) and \eqref{eq:approxmicro} and sufficiency, we can rewrite 
\begin{align}\label{eq:r2} 
r & = \mathbb{E}_{\bar{P}_{\text{can}, 1}}\left[\log P_{\text{mic}, 0}^{W_0^*}(\mathbf{x})
- \log P_{\text{can}, 0}^{w_{\text{pseudo}, 0}}(\mathbf{x})
\right] \nonumber \\ 
&= \mathbb{E}_{\bar{P}_{\text{can}, 1}}\left[\log W_0^*(\mathbf{c}_0(\bm{x})) 
- \log W_{\text{pseudo}, 0}(\mathbf{c}_0(\mathbf{x}))
\right]. 
\end{align}
where $W_{\text{pseudo}, 0}(\mathbf{c}_0(\mathbf{x}))$ is defined as in (\ref{eq:fromPtoW}). 
\end{enumerate} 
The evaluation above is under $\bar{P}_{\text{can}, 1}$-expectation; this makes sense if we use a Bayesian universal distribution $\bar{P}_{\text{can}, 1}= {P}^{w_1}_{\text{can}, 1}$ and the prior $w_1$ is a reasonable expression of our uncertainty. If we are not so sure about our priors, or if $\bar{P}_{\text{can}, 1}$ is non-Bayesian, we may be interested in a more stringent, worst-case measure for evaluating the performance of the microcanonical approximation. In analogy with the worst-case REG defined in \ref{regret}, we define an alternative version of $r$, denoted by $r'$, which can be defined both relatively to a single parameter $\bm{\theta}_1$ (equivalently and more conveniently relative to $\bm{\mu}_1 = {\bm{\mu}}(\bm{\theta}_1)$:
\begin{align}\label{eq:r'_theta}
r'(\bm{\mu}_1) & = \mathbb{E}_{\bm{\mu}_1} \left[\log S_{\text{pseudo}} - \log S^{\text{GRO}}_{\text{mic}}\right] \\
& =  \mathbb{E}_{\bm{\mu}_1} \left[\log W_0^*(\mathbf{c}_0(\bm{x})) 
- \log W_{\text{pseudo}, 0}(\mathbf{c}_0(\mathbf{x})) \right],\nonumber 
\end{align}
where $\mathbb{E}_{\bm\mu}$ denotes the expected value under $P_{\bm{\mu}}$, and in its worst-case version, which for clarity will be simply denoted by $r'$:
\begin{equation}
\label{eq:r'}
r' := \max_{\bm{\mu}_1\in \tt{M}_1} r'(\bm{\mu}_1).
\end{equation}
It can be easily argued that 
\begin{equation}
    r \geq 0 \Rightarrow r' \geq 0.
\end{equation}

In case $r$ (or $r'$) is small, we know that our easily computable microcanonical e-variable $S^{\text{GRO}}_{\text{mic}}$ is close to optimal according to the GRO criterion for the canonical problem, and hence can be used instead of the canonical  $S^{\text{GRO}}_{\text{can}}$. In the following example, which is a continuation of Example A, we show a practical case where this turns out to be true.

\subsection*{Example B (continued from Example A)}
We consider the same setting as in Example A, but in this case, we are interested in constructing a canonical test. In a canonical formulation, the observed number of $1$s is fixed only in expectation. As a result, the null model is a collection of $n$ i.i.d. Bernoulli variables, where the parameter is the probability $p_0 \in [0,1]$  of observing $x=1$, which is the same regardless of the group. The alternative model, instead, assumes that data in the two groups are independent Bernoulli variables, where the parameters are the probabilities $(p_a, p_b) \in [0,1]^2$ of observing $x=1$, which depend on the group. The aim of the tests is to assess whether $p_a$ and $p_b$ are the same or whether they are different. Again, for the sake of this example, we put independent, continuous uniform priors on the alternative parameters $p_a$ and $p_b$, $w_1(p_a, p_b) = u(p_a)u(p_b)$ where $u(p) = 1$ if $p\in[0,1]$ and $u(p) = 0$ else. In this simple case, the Bayesian marginal likelihood can be computed analytically, and it reads:
\begin{eqnarray}
P_{\text{can},1}^{w_1} &=& \!\int_0^1\!p_a^{n_1^a}(1-p_a)^{n^a - n_1^a} dp_a   \int_0^1\!p_b^{n_1^b}(1-p_b)^{n^b - n_1^b} dp_b\nonumber\\
&=&\binom{n^a}{n_1^a}^{-1}\frac{1}{n^a + 1} ~~\binom{n^b}{n_1^b}^{-1}\frac{1}{n^b + 1} 
\label{eq:evidence_with_uniform_prior}
\end{eqnarray}

\begin{figure}[ht]
\center
\includegraphics[width=0.5\textwidth
]{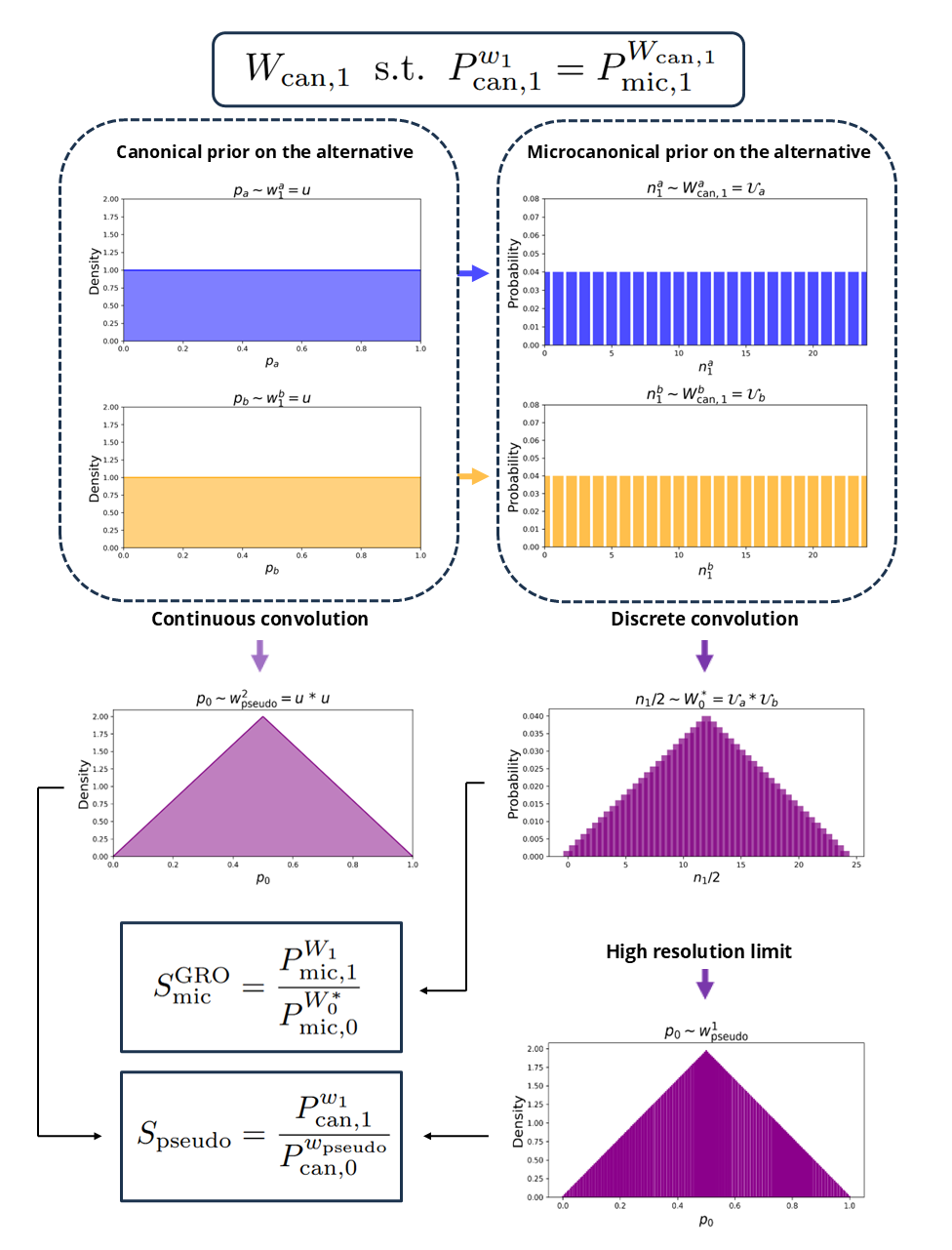}
\caption{\small 
Procedures to compute the microcanonical approximation $ S^{\text{GRO}}_{\text{mic}} $ and the pseudo approximation $ S_{\text{pseudo}} $ for testing between two binary data streams, under uniform priors (as in Examples~A, B, and C). Starting from two independent continuous uniform priors on the alternative (top left), we construct discrete microcanonical priors (top right) satisfying $ P_{\text{can},1}^{w_1} = P_{\text{mic},1}^{W_{\text{can},1}} $. The optimal discrete prior $ W_0^* $, used in $ S^{\text{GRO}}_{\text{mic}} $, is obtained by convolving the alternative priors. The continuous prior $ w_{\text{pseudo},0} $ for $ S_{\text{pseudo}} $ is derived either from $ W_0^* $ through a high resolution limit ($ w_{\text{pseudo},0}^1 $), or by directly convolving the original continuous priors ($ w_{\text{pseudo},0}^2 $).
}
\label{fig_procedure}
\end{figure}

Following the procedure described in this section, we first build the microcanonical approximation. To do so, we need to compute the probability $W_{\text{can}, 1}$ induced by $P_1^{w_1}$ on the alternative sufficient statistics, such that $P_{\text{can},1}^{w_1}  = P_{\text{mic},1}^{W_{\text{can},1}}$. By inspecting Eq. \eqref{eq:evidence_with_uniform_prior}, it is easy to observe that $W_{\text{can},1}$ is the uniform distribution: $W_{\text{can}, 1} = (n_a + 1)^{-1}(n_b + 1)^{-1} = \mathcal{U}_a(n_1^a)~\mathcal{U}_b(n_1^b)$. Thus, we can compute the microcanonical approximation $S^{\text{GRO}}_{\text{mic}}$ by using the results of Example A. 
As a second step, we check whether this microcanonical e-variable is a good approximation by studying the behavior of the interval width $r$ as the total size $n$ increases. To evaluate $S_{\text{pseudo}}$, we compute the prior $w_{\text{pseudo}, 0}$ as described above (denoted by $w^1_{\text{pseudo}, 0}$ in Figure~\ref{fig_procedure} to be distinct from $w^2_{\text{pseudo}, 0}$ of the following Example C). A schematic representation of how $S^{\text{GRO}}_{\text{mic}}$ and $S^{\text{pseudo}}$ are built is shown in Figure~\ref{fig_procedure}.

\begin{figure}[H]
\center
\includegraphics[width=0.45\textwidth
]{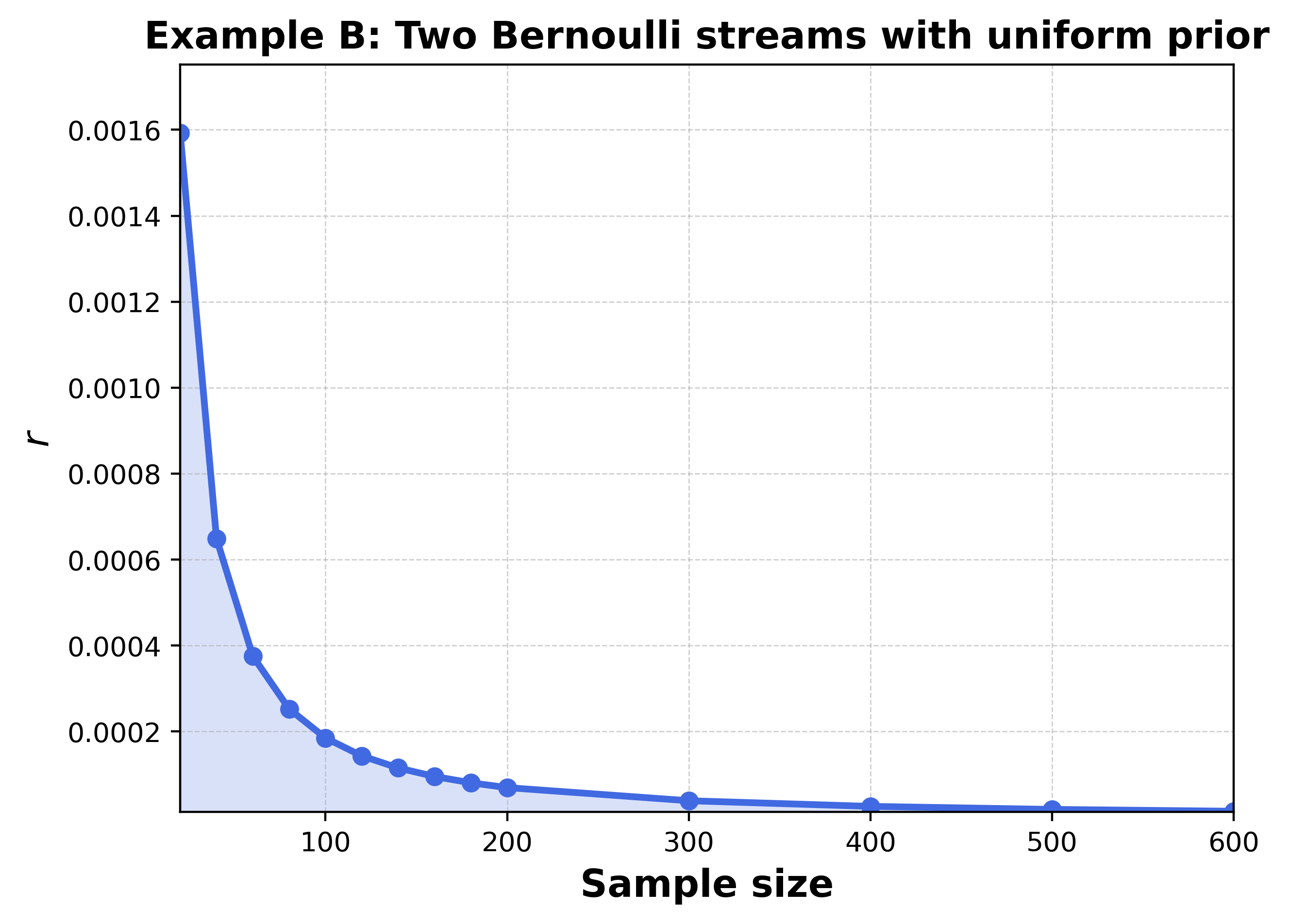}
\caption{Convergence of the interval width $r$ for a canonical test between two streams of binary data (as in Example B), for $n^a = n^b = m$, as the sample size $n = 2m$ grows.}
\label{fig_r_two_uniforms}
\end{figure}

Once $S^{\text{GRO}}_{\text{mic}}$ and $S^{\text{pseudo}}$ are computed, we can compute $r$ and show that the microcanonical approximation works very well in our simple example (see Figure~\ref{fig_r_two_uniforms}).

\subsection{Asymptotic justification for the microcanonical approximation}

We now provide a theoretical result that explains why the microcanonical approximation tends to perform very well in practice, even when the canonical GRO e-variable is not available. Specifically, it suggests that in many cases the gap $r$ defined in Equation~\eqref{eq:r} converges very fast to $0$ as the sample size increases.

First, let $\mathcal{M}$ be a canonical maximum entropy model with sufficient statistic $\mathbf{c}(\mathbf{x})$ taking values in a finite set $\mathcal{C} \subset \mathbb{R}^d$. The canonical distribution has an exponential form
\begin{equation}
P_{\text{can}}(\mathbf{x}; \bm{\theta}) = \frac{e^{-\bm{\theta} \cdot \mathbf{c}(\mathbf{x})}}{Z(\bm{\theta})},
\end{equation}
and induces the mean-value mapping
\begin{equation}
\bm{\mu}(\bm{\theta}) := \mathbb{E}_ {\bm{\theta}}[\mathbf{c}(\mathbf{x})],
\end{equation}
with $\bm{\mu}$ taking values in the mean-value parameter space $\tt M$.

Next, consider the i.i.d.\ extension $\mathcal{M}^{(m)}$ in which 
$\mathbf{y}^{(m)} = (\mathbf{x}_1, \dots, \mathbf{x}_m)$ are $m$ i.i.d.\ samples from $P_{\text{can}}(\cdot; \bm{\theta})$.
The sufficient statistic of $\mathbf{y}^{(m)}$ is the sum
\begin{equation}\label{eq:sumstats}
\mathbf{s}^{(m)}(\mathbf{y}^{(m)}) = \sum_{j=1}^m \mathbf{c}(\mathbf{x}_j).
\end{equation}

Let $w$ be a prior density over $\tt{M}$, and let $Q^w$ be the induced probability for any measurable ${\tt M}' \subseteq {\tt M}$:
\begin{equation}
Q^w(\bm{\mu} \in {\tt M}') := \int_{\tt M'}  w(\bm{\mu}) \, d\bm{\mu},
\end{equation}

The following theorem shows that the normalized sufficient statistic converges in distribution to $Q^w$. Convergence is quantified in terms of probabilities of subsets, with error decaying at rate $O(\log m / m)$.

\begin{theorem}\label{thm:one}
Let $w$ be any regular prior density on the mean value parameter space $\tt{M}\subset {\mathbb R}^d$. Then, for any INECCSI (Definition~\ref{def:ineccsi}) subset  ${\tt M}'$ of ${\tt M}$, we have
\begin{multline}\label{eq:weakconvergence}
\left| P^{w^{(m)}}_{\text{\rm can}}\left\{ \frac{\mathbf{s}^{(m)}(\mathbf{y}^{(m)})}{m}  \in {\tt M'} \right\} - Q^w\left\{ \bm{\mu} \in {\tt M'} \right\} \right| \\
= O\!\left( \frac{\log m}{m} \right).
\end{multline}
\end{theorem}

In words: under the Bayesian marginal likelihood $P^{w^{(m)}}_{\text{can}}$, the normalized sufficient statistic $\mathbf{s}^{(m)}/m$ becomes increasingly close to being distributed according to the prior over mean-value parameters.

\medskip
Assume we can extend both canonical models $\mathcal{M}_0$ and $\mathcal{M}_1$ to i.i.d. models $\mathcal{M}_0^{(m)}$ and $\mathcal{M}_1^{(m)}$ as above, where (\ref{eq:sumstats}) holds for both $\mathbf{c}= \mathbf{c}_0$  (sum denoted by $\mathbf{s}_0$) and $\mathbf{c}= \mathbf{c}_1$ (sum denoted by $\mathbf{s}_1$).
In such settings, Theorem~\ref{thm:one} supports the claim that the gap $r$ between the microcanonical and pseudo approximations vanishes for large $m$. This is best explained in terms of our running example. 

\subsection*{Example C (continued from Examples A and B)}
Suppose $n^a=n^b=m$, so that data can be grouped into $m$ i.i.d.\ \emph{blocks}, each consisting of one binary outcome from group $a$ and one from $b$. For each $m$ the sufficient statistics are:
\begin{itemize}
  \item $\mathbf{s}_1^{(m)} = (n_1^{a\,(m)}, n_1^{b\,(m)})$: number of ones in each group under the alternative,
  \item $\text{s}_0^{(m)} = \tfrac{1}{2}\big(n_1^{a\,(m)}+n_1^{b\,(m)}\big)$: average number of ones across both groups under the null. The division by $2$ is required to ensure that, for a single outcome, ${\tt M}_0 = [0,1]$, and can be interpreted, intuitively, as a set of probabilities. 
\end{itemize}
After normalization by $m$, $\mathbf{s}_1^{(m)}/m \in {\tt M}_1 = [0,1]^2$ and $\text{s}_0^{(m)}/m \in {\tt M}_0 = [0,1]$. Notice that every discrete distribution $W_j^{(m)}$ on the sufficient statistics of ${\cal M}_j^{(m)}$ induces a discrete distribution $V_j^{(m)}$ on the normalized sufficient statistics: $P_{\text{can},1}^{w_1(m)}$ induces a probability $W_{\text{can}, 1}^{(m)}$ on the alternative sufficient statistics $\mathbf{s}_1$, and a corresponding one, denoted here by $V_1^{(m)}$, on the normalized alternative sufficient statistics $\mathbf{s}_1/m$. Similarly, the microcanonical optimal prior $W^{*(m)}_0$ on the null sufficient statistic $\text{s}_0$ induces a distribution $V^*_0$ on $\text{s}_0/m$.  

In Example B, we used a prior $w_1$ under which $p_a$ and $p_b$ were independently and uniformly distributed, i.e., 
$w_1(p_a,p_b) = w^a_{1}(p_a) \cdot w^b_{1}(p_b)$ with $w^a_{1}= w^b_{1}= u$. The independent uniform prior has a remarkable property: $V_1^{(m)}$ coincides {\em exactly} with the product of two independent discrete uniforms, each defined on $\{0,1/m,\ldots,1\}$, corresponding to the components of $\mathbf{s}_1^{(m)}/m$.
Consequently, the distribution $V_0^{*(m)}$ is {\em exactly\/} equal to a triangular discrete distribution, which is the convolution of these two discrete uniforms (Figure~\ref{fig_two_uniforms}). Theorem~\ref{thm:one} indicates that something analogous, but now in an asymptotic sense, will happen for every regular prior $w_1(p_a,p_b) = w^a_{1}(p_a) w^b_{1}(p_b)$, as long as $p_a$ and $p_b$ are still independent under $w_1$. More in detail, even if $w^a_{1}$ and/or $w^b_{1}$ are not uniform, $V_1^{(m)}$ will converge to a distribution on ${\tt M}_1$ that is a discretized version of $w_1$, and $V_0^{*(m)}$ will still be the exact convolution of the two components of $V_1^{(m)}$, which are the (approximate) discretized versions of the components of $w_1$. To illustrate, in Example 3 below, $w^a_{1}$ and $w^b_{1}$ will be taken to be of general beta form rather than restricted to uniform, and then we will see Theorem~\ref{thm:one} in action, the correspondence becoming asymptotic rather than precise at each $m$. 
Still, $V_0^{*(m)}$ will converge, as $m$ grows, to a continuous, strictly positive density on ${\tt M}_0$, denoted by $w^1_{\text{pseudo},0}$. This distribution can be approximated by considering a very large $m$ and "smoothening" the corresponding $V_0^{*(m)}$ to $w^1_{\textrm{pseudo},0}$. This limiting procedure is precisely what we referred to earlier as the \emph{high resolution limit}. Once $w^1_{\textrm{pseudo},0}$ is obtained, $P^{w^1_{\textrm{pseudo},0} (m)}_{\textrm{can}}$ induces a probability distribution $W^{1 (m)}_{\text{pseudo},0}$ on the null sufficient statistics. As above, we can define 
\begin{equation}
V_{\text{pseudo},0}^{1 (m)}\!\left(\frac{\;\mathbf{s}_0^{(m)}}{m}\right) := W_{\text{pseudo},0}^{1 (m)}\!\big(\mathbf{s}_0^{(m)}\big).
\end{equation}
Now we invoke Theorem \ref{thm:one} again: it indicates that $V_{\text{pseudo},0}^{1 (m)}$ converges to 
$w^1_{\text{pseudo},0}$. 
Thus, one may expect $V_0^{*(m)}$ and $V_{\text{pseudo},0}^{1 (m)}$, and consequently $W_0^{*(m)}$ and $W_{\text{pseudo},0}^{1 (m)}$, to be close and $r$ to be small, according to Eq. \eqref{eq:r2}. The microcanonical e-variable becomes thus an excellent approximation of the canonical one --- which is what we set out to argue. 

In the current example, we can go further: let $w^2_{\textrm{pseudo},0}$ be the continuous convolution of the independent priors $w^a_{1}$ and $w^b_{1}$. Applying Theorem~\ref{thm:one} to $\mathcal{M}_0$ with this density shows that the induced distribution $V_{\text{pseudo},0}^{2 (m)}$ on $\mathbf{s}_0^{(m)}/m$ converges to a discretized version of $w^2_{\textrm{pseudo},0}$. At the same time, $V_0^{*(m)}$, being the convolution of a discretized $w_1$, converges to the discretized convolution of $w_1$. Thus, for large $m$, $w^1_{\textrm{pseudo},0}$ and $w^2_{\textrm{pseudo},0}$ become indistinguishable. In practice, one can compute $S_{\text{pseudo}}$ either by directly convolving the continuous components of $w_1$, or by taking the discrete convolution $W_0^{* (m)}$ and then its high-resolution limit: both approaches yield the same result (Figure~\ref{fig_procedure}).

\paragraph*{How precise and general is this?}
In the reasoning above, we invoked Theorem~\ref{thm:one} several times to go back and forth between prior distributions on mean-value parameters and marginal distributions on sufficient statistics. Specifically: (a) at the level of 
${\cal M}_1$ (blue and yellow arrows in Figure~\ref{fig_procedure}); and (b) at the level of ${\cal M}_0$, for relating $W^{*(m)}_0$ to $P_{\textrm{can}}^{w^1_{\textrm{pseudo},0}}$ (b1, bottom right arrow in Figure~\ref{fig_procedure}) and $P_{\textrm{can}}^{w^2_{\textrm{pseudo},0}}$ (b2, bottom left arrow). 

In step (a), the theorem is not really needed when $w^a_1$ and $w^b_1$ are uniform (as in the figure). Nevertheless, as long as the priors remain independent and regular, Theorem~\ref{thm:one} suggests that step (a) holds even if they are not uniform. More generally, moving from the binary 2-group case to a general MEM ${\cal M}_1$, Theorem~\ref{thm:one} still suggests that step (a) is valid whenever $w_1$ factorizes into independent regular priors, making the mean-value parameters independent. We write “suggest” rather than “prove” because the convergence in (\ref{eq:weakconvergence}) is too weak to formally imply $r \rightarrow 0$ (it concerns probabilities of sets, whereas \eqref{eq:r2} involves expectations of log densities). Nevertheless, it provides strong heuristic evidence, and we do observe convergence numerically (see Figure~\ref{fig_r_two_uniforms}). All reasoning based on Theorem~\ref{thm:one} should thus be understood as heuristic rather than fully formal.

Turning now to step (b) for general ${\cal M}_0$ and ${\cal M}_1$: as long as $\mathbf{s}_0^{(m)}$ is a linear function of $\mathbf{s}_1^{(m)}$, the use of Theorem~\ref{thm:one} in steps (b1) and (b2) remains heuristically justified, provided $w_1$ factorizes into regular independent priors as above. This linearity condition holds in all our examples (e.g.\ in Example C, $\mathbf{s}_0^{(m)}$ is the average of the components of $\mathbf{s}_1^{(m)}$). It guarantees that the limiting density $w^1_{\text{pseudo},0}$ exists, and Theorem~\ref{thm:one} then suggests that it coincides with $w^2_{\text{pseudo},0}$. 

In the more general case where $\mathbf{s}_0^{(m)}$ is a function (not necessarily linear) of $\textbf{s}_1^{(m)}$ --- i.e.\ Condition A holds --- then it may still be true that $V^{*(m)}_0$ converges to a high-resolution limiting density $w^1_{\text{pseudo},0}$. In that case, Theorem~\ref{thm:one} still suggests that step (b1) remains valid, so that $r$ becomes small with growing sample size, making the microcanonical approximation effective. However, in such settings, it is less clear whether the approach based on $w^2_{\text{pseudo},0}$ still makes sense.

We stress that this asymptotic justification does not rely on $\bar P_1$ being a Bayesian mixture with prior $w_1$. 
The construction leading to $w^1_{\text{pseudo},0}$ applies to any universal distribution $\bar P_1$ on the alternative (including the NML), since $\bar P_1$ always induces a discrete distribution on the alternative sufficient statistic; from this, one can derive $V_0^{*(m)}$ and then obtain $w^1_{\text{pseudo},0}$ via the high-resolution limit. 
By contrast, the alternative route based on $w^2_{\text{pseudo},0}$ explicitly requires a factorized regular prior on the alternative, and is therefore not directly available in the non-Bayesian case.

\section{Application to contingency tables and related models}
\label{sec:contingency_tables}
Contingency tables are a fundamental tool in statistical analysis for examining the relationship between categorical variables. Given a dataset where observations are classified according to categorical factors, a contingency table provides a structured way to summarize the frequencies of different category combinations.

Formally, a contingency table is an $l\times k$ matrix where each entry represents the count of occurrences for a particular combination of row and column categories. Such tables are widely used in fields where categorical data naturally arise, such as biostatistics, social sciences, and market analysis.

In network science, this approach plays a crucial role in link analysis, where the presence or absence of an edge ($ x = 1$ or $x=0$) in a network is studied across different subsets of nodes. For instance, in community detection, one may ask whether the probability of forming a link differs within and between predefined groups of nodes. This idea is closely related to the Stochastic Block Model (SBM), a generative model in which nodes are assigned to latent groups, and connection probabilities are determined by group memberships. Contingency tables provide a natural way to summarize and test the differences in connection probabilities across groups, helping to assess whether observed patterns deviate from a null model where edges are formed independently of group structure. See e.g., ~\cite{alves24,jerdee24} for connections between network modeling, and contingency tables and the discussion in  \ref{sec:application_to_NM} of this paper. 

In this work, we focus on binary categorical data, which corresponds to $l = 2$ in the general $l\times k$ contingency tables setting. We first apply our results to the simple case of two groups, i.e., $2\times 2$ contingency tables. We consider microcanonical and canonical tests. For canonical tests, our main focus will be that of finding the microcanonical approximation in practical cases; this translates into finding the induced prior on the alternative \eqref{eq:induced_prior} and then applying formula \eqref{eq:approxmicro}. We will finally verify the approximation validity by evaluating the interval width $r$, and show results on the regret. Later, we extend these results to the more general case of $2 \times k$ contingency tables. 

\subsection{\texorpdfstring{$\mathbf{2\times 2}$ contingency tables}{2x2 contingency tables}}
\label{subsec:2x2}
A $2 \times 2$ contingency table is a fundamental tool to assess whether the distribution of a binary outcome differs between two groups. Given a dataset where each observation consists of a binary variable $x \in \{0, 1\}$ and a categorical label indicating group membership, the data can be summarized in the following $2 \times 2$ table:

\[
\def\arraystretch{1.9}
\begin{array}{c|c c|c}
    & \text{Group A} & \text{Group B} & \text{Total} \\
\hline 
x = 1 & n^a_{1} & n^b_{1} & n_1 \\
x = 0 & n^a_{0} & n^b_{0} & n_0 \\ 
\hline
\text{Total} & n^a & n^b & n
\end{array}
\] 
The dataset $\mathbf{x}$ consists of two groups, represented as $\mathbf{x}_a = (x_1^a, ..., x^a_{n^a})$ and $\mathbf{x}_b = (x_1^b, ..., x^b_{n^b})$, where $n^a$ and $n^b$ are the respective group sizes. The table reports the number of ones ($n_1^a$ and $n_1^b$) and zeros ($n_0^a$ and $n_0^b$) in each group, along with their totals, $n_1$ and $n_0$.
The key question is whether the probability of observing $x = 1$ differs between the two groups. This problem translates into a hypothesis testing problem, where:
\begin{itemize} 
\item In the alternative hypothesis, the two groups are distinct, meaning the number of ones is constrained separately in each group:
\begin{equation}\label{eq:c1}
\mathbf{c}_1  = (n_1^a, n_1^b).
\end{equation}
\item In the null hypothesis, the groups are indistinguishable, so only the total number of ones is constrained:
\begin{equation} 
\text{c}_0 = n_1. \end{equation} 
\end{itemize}
These constraints define the sufficient statistics under each hypothesis and form the basis for the microcanonical and canonical tests discussed next. The reader may have noticed that this is exactly the setting of Examples A, B, and C in section \ref{sec:application_to_MEMS}. Nevertheless, for the sake of clarity, in this section all quantities will be defined again and in more detail, at the cost of repeating ourselves. \\

\subsection*{\texorpdfstring{$\mathbf{2\times 2}$ microcanonical test}{2x2 microcanonical test}}
 In the microcanonical formulation, we enforce hard constraints on the observed counts, treating them as fixed quantities. The null model with sufficient statistics $n_1$ reads 
\begin{equation}\label{micro_null}
    P_\textrm{mic, 0}(\mathbf{x}; n_1)=
    \begin{cases}
        \frac{1}{\Omega_0(n_1)}, & \text{if}\:n_1(\mathbf{x})=n_1;\\
        0, & \text{else};
    \end{cases}
\end{equation}
where 
\begin{equation}
\Omega_0(n_1) = \binom{n}{n_1}
\end{equation}

\noindent is the number of permutations of $\mathbf{x}$ preserving the total number of $1$s. 
The alternative model with sufficient statistics $(n_1^a, n_1^b)$ reads

\begin{equation}\label{micro_alternative_2x2}
    P_\textrm{mic, 1}(\mathbf{x}; n_1^a, n_1^b) =
    \begin{cases}
        \frac{1}{\Omega_1(n_1^a, n_1^b)}, & 
        \makebox[70pt][r]{if $(n_1^a(\mathbf{x}), n_1^b(\mathbf{x}))$} \\ 
        & \makebox[100pt][r]{\hspace{1.5em} $= (n_1^a,n_1^b)$;} \\
        0, & \text{else},
    \end{cases}
\end{equation}
where
\begin{equation}
\Omega_1(n_1^a, n_1^b)  = \binom{n^a}{n_1^a}\binom{n^b}{n_1^b}
\end{equation}

\noindent is the number of permutations of $\mathbf{x}$ preserving the total number of $1$s in each group. 

For any given prior $W_1$ on the alternative sufficient statistics, $S^\text{GRO}_{\text{mic}}$ is found exactly by computing $W_0^*$ and applying \eqref{explicit_GRO_micro}. In this case, Condition A \eqref{condition_A} is satisfied, as the null sufficient statistics can be written as a function of the alternative one:
\begin{equation}
    n_1 = n_1^a + n_1^b.
\end{equation}
Thus, following \eqref{W_gro_if_A}, the optimal prior on the null is the distribution of $n_0$ induced by $W_1(n_1^a, n_1^b)$. If $n_1^a$ and $n_1^b$ are independently distributed:
\begin{equation}
    W_1(n_1^a, n_1^b) = W_1^a(n_1^a) \cdot W_1^b(n_1^b),
\end{equation}
then $W_0^*$ is simply the convolution of $W_1^a$ and $W_1^b$:
\begin{equation}
W_0^* = W_1^a * W_1^b,
\end{equation}
where $f * g$ represents the convolution between functions $f$ and $g$.\\

\noindent \textbf{Example 1: Microcanonical test with NML} In the microcanonical case, resorting to the Normalized Maximum Likelihood approach is completely equivalent to putting a uniform prior on both parameters of the alternative model, as shown in~\cite{micro_cano}. Consequently, this case is reduced to Example A in \ref{subsec:micro_test}, and is not considered further.

\subsection*{\texorpdfstring{$\mathbf{2\times 2}$ canonical test}{2x2 canonical test}}
The null canonical model obtained by constraining the average number of $1$s, i.e., the expected value of $n_1$, is represented by the exponential distribution

\begin{equation}\label{cano_null}
P_{\text{can}}(\mathbf{x};\theta_0)=\frac{e^{-\theta_0 \cdot n_1(\mathbf{x})}}{(1+e^{-\theta_0})^{n}},
\end{equation}
which can be rewritten in the mean-value parametrization:
\begin{equation}
P_{\text{can}}(\mathbf{x};p_0)=p_0^{n_1(\mathbf{x})}(1-p_0)^{n-n_1(\mathbf{x})}
\end{equation}
upon defining 
\begin{equation}
p_0=\frac{e^{-\theta_0}}{1+e^{-\theta_0}}.
\end{equation}
The null model is the distribution of a collection of $n$ i.i.d. Bernoulli variables, where the occurrence of $x = 1$ has the same probability $p_0$ regardless of the group. 

The alternative model, obtained by constraining the expected values of $n_1^a$ and $n_1^b$, reads:
\begin{equation}\label{cano_alternative_cano}
P_{\text{can}}(\mathbf{x};\theta_a, \theta_b)=\frac{e^{-\theta_a \cdot n_1^a(\mathbf{x})-\theta_b \cdot n_1^b(\mathbf{x})}}{(1+e^{-\theta_a})^{n^a}(1+e^{-\theta_b})^{n^b}},
\end{equation}
or, equivalently, 
\begin{multline}\label{cano_alternative_mean}
P_{\text{can}}(\mathbf{x};p_a, p_b)=p_a^{n_1^a(\mathbf{x})}(1-p_a)^{n^a-n_1^a(\mathbf{x})} \\ \times p_b^{n_1^b(\mathbf{x})}(1-p_b)^{n^b-n_1^b(\mathbf{x})}
\end{multline}
upon defining the mean-value parameters:
\begin{equation}
p_a=\frac{e^{-\theta_a}}{1+e^{-\theta_a}} \quad \text{and} \quad p_b=\frac{e^{-\theta_b}}{1+e^{-\theta_b}} .
\end{equation}
The alternative model assumes that data in group A and group B are independent Bernoulli variables, where the probability of $x = 1$ is different according to the group. In this scenario, the aim of the test is to assess whether $p_a$ and $p_b$ are the same or whether they are different.   
In what follows, we explicitly apply the procedure described in section \ref{subsec:cano_test} to different choices of $\bar{P}_{\text{can}, 1}$.\\

\noindent \textbf{Example 2: Canonical test with NML.}
 First, we focus on the test between two canonical models with an NML approach, i.e., $\bar{P}_{\text{can},1} = P_{\text{can}, 1}^{\text{NML}}$. For a model with two independent Bernoulli distributions, the exact expression of the NML distributions reads~\cite{staniczenko2014, micro_cano}:
\begin{equation}
P_{\text{can}, 1}^{\text{NML}}\label{nml_2}(\mathbf{x})=P_{\text{can}, 1}^{\text{a, NML}}(\mathbf{x})\cdot P_{\text{can}, 1}^{\text{b, NML}}(\mathbf{x}) 
\end{equation}
with
\begin{equation}\label{nml_bernoulli}
P_{\text{can}, 1}^{\text{i, NML}}(\mathbf{x}) = \frac{ \left(\frac{n_1^i(\mathbf{x})}{n^i}\right)^{n_1^i(\mathbf{x})} \left(1 - \frac{n_1^i(\mathbf{x})}{n^i}\right)^{n^i - n_1^i(\mathbf{x})}}{\frac{e^{n^i}\Gamma(n^i, n^i)}{(n^i )^{n^i-1}}+1}
\end{equation}
for $i \in \{a, b\}$. In the formula above, $\Gamma(s,t)$ is the upper incomplete gamma function. 
According to the procedure described in \ref{two_cano_steps}, a good candidate e-variable in this case is the microcanonical approximation, i.e., the GRO e-variable of the corresponding microcanonical test. To build it, we need the distribution of the alternative sufficient statistics $(n_1^a, n_1^b)$ induced by $P_{\text{can}, 1}^{\text{i, NML}}$, which is 
\begin{equation}
    W_{\text{can},1}(n_1^a, n_1^b) = W_{\text{can},1}^a(n_1^a)\cdot W_{\text{can},1}^b(n_1^b)
\end{equation}
with 
\begin{equation}
W_{\text{can},1}^a(n_1^a) = \Omega_1^a(n_1^a)  \cdot \frac{ \left(\frac{n_1^a(\mathbf{x})}{n^a}\right)^{n_1^a(\mathbf{x})} \left(1 - \frac{n_1^a(\mathbf{x})}{n^a}\right)^{n^a - n_1^a(\mathbf{x})}}{\frac{e^{n^a}\Gamma(n^a, n^a)}{(n^a)^{n^a-1}}+1} 
\end{equation}
and
\begin{equation}
W_{\text{can},1}^b(n_1^b) =\Omega_1^b(n_1^b) \cdot \frac{ \left(\frac{n_1^b(\mathbf{x})}{n^b}\right)^{n_1^b(\mathbf{x})} \left(1 - \frac{n_1^b(\mathbf{x})}{n^b}\right)^{n^b - n_1^b(\mathbf{x})}}{\frac{e^{n^b}\Gamma(n^b, n^b)}{(n^b)^{n^b-1}}+1}. 
\end{equation}
Given that $n_1^a$ and $n_1^b$ are independently distributed, $W_0^*(n_1)$ is the convolution of $W_{\text{can},1}^a(n_1^a)$ and $W_{\text{can},1}^b(n_1^b)$, which can be computed numerically.\\

\noindent \textbf{Example 3: Independent beta priors.} The beta probability distribution reads: 
\begin{equation}
    \text{Beta}(y ; \alpha, \beta) =
    \frac{y^{\alpha - 1} (1 - y)^{\beta - 1}}{B(\alpha, \beta)},
    \quad \text{for } y \in (0,1),
\end{equation}
where $B(\alpha,\beta)$ is the beta function, defined as
\begin{equation}
    B(\alpha, \beta) = \int_0^1 t^{\alpha - 1} (1 - t)^{\beta - 1} dt.
\end{equation}
The beta prior represents a popular choice because it is flexible enough to encompass several cases of interest (Table \ref{SM_tab_beta_distribution} of the Supplementary Material). Here, we put two independent beta priors $w_1^a$ and $w_1^b$ with parameters, respectively, $(\alpha_a, \beta_a)$ and $(\alpha_b, \beta_b)$, on $p_a$ and  $p_b$. The Bayesian marginal likelihood resulting from this choice can be written explicitly as

\begin{equation}
    \bar{P}_{\text{can}, 1}(\mathbf{x}) =\frac{B(\bar{\alpha}^a, \bar{\beta}^a)}{B(\alpha^a, \beta^a)}
    \cdot
    \frac{B(\bar{\alpha}^b, \bar{\beta}^b)}{B(\alpha^b, \beta^b)}
\end{equation}
where
\begin{align*}
    \bar{\alpha}^a &= n_1^a + \alpha^a, & \bar{\beta}^a &= n^a - n_1^a + \beta^a, \\
    \bar{\alpha}^b &= n_1^b + \alpha^b, & \bar{\beta}^b &= n^b - n_1^b + \beta^b.
\end{align*}
As in the previous example, to obtain the microcanonical approximation for this problem, we look for the probability mass function induced by $\bar{P}_{\text{can}, 1}$ on the alternative sufficient statistics, which reads:
\begin{equation}
    W_{\text{can}, 1}(n_1^a, n_1^b) = W_{\text{can}, 1}^a(n_1^a)\cdot W_{\text{can}, 1}^b(n_1^b)
\end{equation}
with 
\begin{equation}
W_{\text{can}, 1}^a(n_1^a) = \Omega_1^a(n_1^a)  \cdot \frac{B(\bar{\alpha}^a, \bar{\beta}^a)}{B(\alpha^a, \beta^a)}
\end{equation}
and
\begin{equation}
W_{\text{can}, 1}^b(n_1^b) =\Omega_1^b(n_1^b) \cdot \frac{B(\bar{\alpha}^b, \bar{\beta}^b)}{B(\alpha^b, \beta^b)}.
\end{equation}
With this choice, $W_1^a$ and $W_1^b$ are \textit{beta-binomial distributions}. 
Given that $n_1^a$ and $n_1^b$ are independently distributed, as we expected because we put independent priors on $p_a$ and $p_b$, $W_0^*(n_1)$ is the convolution of $W_{\text{can}, 1}^a(n_1^a)$ and $W_{\text{can}, 1}^b(n_1^b)$.
Whether this expression can be written in closed form depends on the specific values of the beta parameters chosen. For example, if all beta parameters are equal to $1$, $W^*_0$ reduces to the convolution between two discrete uniform distributions \eqref{uniform_convolution}. When no closed form is available, the convolution can be computed numerically. 

In Figure~\ref{SM_fig_beta_convolutions_2x2} we show $W_1^a$, $W_1^b$, $W_0^*$, $w^1_{\text{pseudo},0}$ and $w^2_{\text{pseudo},0}$ for different choices of the beta parameters. As expected, in all cases where $w_1^a$ and $w_1^b$ are well defined in the whole parameter space, $w^1_{\text{pseudo},0}$ and $w^2_{\text{pseudo},0}$ are almost indistinguishable. This is a consequence of Theorem 1: the distribution of the mean value parameter ($w^2_{\text{pseudo},0}$) and that of the sufficient statistic ($w^1_{\text{pseudo},0}$) resemble each other when the sample size is big (high resolution limit). \\

\noindent In the next section, we show results obtained by numerical simulations for what concerns the optimality of the microcanonical approximation and the regret, measured in the examples reported in this section. For simplicity, when necessary, we will assume that the two groups have the same sample size, i.e., $n^a = n^b = m$, and that the independent beta priors on the alternative, denoted by $w_1^a$ and $w_1^b$, have all parameters equal to a certain value $\gamma > 0$.

\subsection*{Evaluating the microcanonical approximation}

In order to evaluate the goodness of the microcanonical approximation, we employ two approaches: a direct comparison and a comparison through $r$.

In the first case, we directly compare the e-power of the microcanonical approximation to the GRO-optimal canonical one, where the latter is computed by numerically solving the optimization problem \eqref{canonical_optimization}. We find that the e-power of the microcanonical approximation converges to that of the canonical GRO e-variable as the total size grows (Figure~\ref{fig:epowerdifference_2x2}). The e-power of the pseudo approximation converges as well, even though the convergence is slower compared to that of the microcanonical one. From these plots, we can already conclude that the microcanonical one works as a good approximation of $S_{\text{can}}^{\text{GRO}}$.

Notice that, if $n_1^a$ and $n_1^b$ are both big enough, using $P_{\text{can},1}^\text{NML}$ is asymptotically equivalent to using a Bayesian universal distribution with a {\em Jeffreys prior}~\cite{grunwaldbook2007} on the alternative parameters, which in our case is equivalent to a beta prior with parameters all equal to $\gamma = 0.5$. More precisely, let us, for simplicity, set $n^a = n^b = m$. Then, for any INECCSI subset ${\tt M}'_1 \subset {\tt M}_1$, as $m \rightarrow \infty$, with $w_1$ equal to the density of Jeffreys prior,
\begin{equation}
\sup_{\bm{\mu}_1 \in {\tt M}_1}  
\mathbb{E}_{\bm{\mu}_1} 
\left[  - \log {P}_{\text{can},1}^{w_1}(\mathbf{x}^m) + 
\log P_{\text{can},1}^\text{NML}
\right] = o(1),
\end{equation} 
where $o(1)$ denotes a quantity that goes to $0$ as $m \to +\infty$. Nevertheless, a difference persists at the boundaries (outside ${\tt M}_1'$), where Jeffreys prior diverges and the NML induced priors do not. This difference becomes even more important when convoluting the independent Jeffreys priors to compute $W^*_0$. This explains why the first and third pictures in Figure~\ref{fig:epowerdifference_2x2} are quite different. For this reason, in all simulations, we implement the exact NML formula instead of its Jeffreys approximation.\\

The numerical approach to directly compare the e-power is feasible in a few simple cases and only for relatively small sample sizes. Conversely, the value of $r$ can be easily evaluated, even for very large system sizes. In Figure~\ref{fig:r_2x2}, we show the plot of $r$ as defined earlier to evaluate the effectiveness of the microcanonical approximation in different scenarios. The results confirm those of Figure~\ref{fig:epowerdifference_2x2}, as in all cases considered, $r$ converges to $0$. In conclusion, we argue that the microcanonical approximation is a perfect candidate in this case.

\subsection*{Results on regret}
Let's again consider the $m$-dimensional i.i.d. extension of our models. In section \ref{SM_bound_on_regret} of the Supplementary Materials, we show that, if the error $r'(\bm\mu_1)$ in \eqref{eq:r'_theta} vanishes as the sample size $m$ increases, both the canonical growth-optimal e-variable $S^{\text{GRO}}_{\text{can}}$ and its microcanonical approximation $S^{\text{GRO}}_{\text{mic}}$ satisfy:
\begin{equation}
\label{eq:upperboundb}
\text{REG}_1(\bm{\mu}_1, \cdot) = \frac{d_1 - d_0}{2} \cdot \log m + O(1).
\end{equation}
In the $2 \times 2$ case, where $d_1 = 2$ and $d_0 = 1$, this becomes:
\[
\text{REG}_1(\bm{\mu}_1; \cdot) = \frac{1}{2} \log m + O(1).
\]

This result holds uniformly over all $\bm\mu_1 \in {\tt M}'_1$, provided that ${\tt M}'_1$ is an INECCSI set (i.e., excluding regions near the boundary of the parameter space). However, it does not extend to the full parameter space ${\tt M}$, where the asymptotic form \eqref{eq:bic} may fail to hold even in well-specified cases.

\begin{figure}[ht]
\includegraphics[width=0.4\textwidth
]{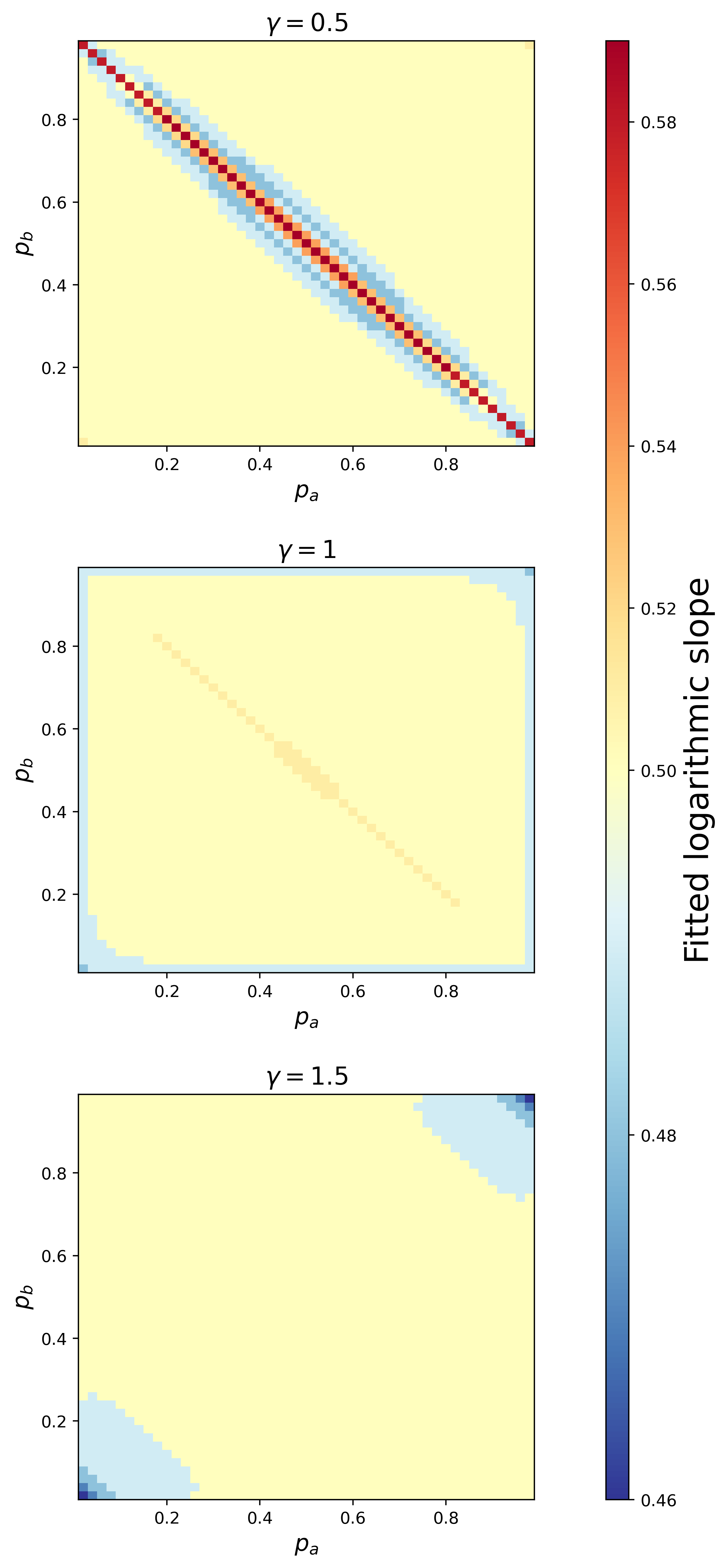}
\label{asymptotic_slope}
\caption{Fitted slope of the logarithmic growth $a\log m + b$ of the microcanonical approximation regret \ref{regret_with_theta1} in the $2\times2$ case ($n^a = n^b = m$), shown for different combinations of the alternative parameters $(p_a, p_b)$. The expected asymptotic slope is $0.5$ (yellow). Three alternative beta priors are considered: $\alpha = \beta = 0.5$, $\alpha = \beta = 1$ and $\alpha = \beta = 1.5$  . The sample sizes used for fitting are $m \in \{600, 800, 1000, 1200, 1400, 1600, 1800\} $. $p_a$ and $p_b$ vary in the interval $[0.02, 0.98]$, with a grid step of $0.02$. Values at the boundaries are excluded to improve the readability of the plots.\label{fig_slope} }
\end{figure}

Our experiments confirm these insights. We evaluated worst-case regret in the $2 \times 2$ setting for different values of the beta prior parameter $\gamma$. Notice that in what follows we apply our reasoning to the mean value parameter spaces, $(p_a, p_b) \in {\tt M}_1 = [0,1]^2$ and $p_0 \in {\tt M}_0 = [0,1]$, and that we consider INECCSI sets with respect to ${\tt M}_1$. From experimental results, collected in Figure~\ref{fig_slope}, we observe a clear dichotomy:
\begin{itemize}
    \item For $\gamma < 1$, the convolution of the beta priors $w_1^a$ and $w_1^b$ is non-differentiable at $p_0 = 1/2$, as shown in Figure~\ref{SM_fig_beta_convolutions_2x2} (e.g., for $\gamma=0.5$). Consequently, the convergence of $V^{*(m)}_0$ to a density over the mean-value space ${\tt M}_0$ (as discussed under Theorem~\ref{thm:one}) may be very slow or fail altogether. In this case, $S^{\text{pseudo}}$ becomes incomparable to $S^{\text{GRO}}_{\text{can}}$ and $S^{\text{GRO}}_{\text{mic}}$, and \eqref{eq:upperboundb} no longer holds (see Figure~\ref{SM_fig: regret_r'}). Indeed, in Figure~\ref{fig_slope}, we see that even on small INECCSI sets, the regret grows like $a \log m + b$ for some $a > 1/2$.
    
    \item For $\gamma = 1$, convergence is moderate. Although $r'$ decays quickly (Figure~\ref{SM_fig: regret_r'}), the experimental values of Figure~\ref{fig_slope} shows areas (e.g., the yellow counter-diagonal) where regret exceeds the expected rate. These may still belong to an INECCSI set, but convergence has not yet been reached at the sample sizes considered ($m \leq 1800$).
    
  \item For $\gamma > 1$, the convolution is differentiable, and convergence of $V^{*(m)}_0$ is fast. The asymptotic behavior $(1/2)\log m + O(1)$ is observed on INECCSI sets (see again Figure~\ref{fig_slope})
\end{itemize}
 These findings imply that, from a minimax perspective, using priors with $\gamma < 1$ is generally suboptimal. Such priors fail to achieve the expected regret rate of $(1/2)\log m + O(1)$ even when the true parameters lie well inside the parameter space.

This has implications for default prior choices. In both the Bayesian and MDL literatures, Jeffreys prior~\cite{grunwaldbook2007} is often recommended as a default when no prior knowledge is available, and is justified in the MDL framework because it achieves asymptotically minimax optimal redundancy (i.e., the middle inequality in (\ref{eq:bic}) becomes an equality~\cite{ClarkeB94}). However, in our setting, Jeffreys prior corresponds to $\gamma = 1/2$, which, despite its MDL-optimality, is \emph{not} optimal with respect to worst-case regret under e-values.

\subsection{\texorpdfstring{$\mathbf{2\times k}$ contingency tables}{2xk contingency tables}}
\label{subsec:2xk}
A  $2 \times k$ contingency table is a natural extension of the $2 \times 2 $ case, allowing us to assess whether the distribution of a binary outcome differs across multiple ($k $) groups. Given a dataset where each observation consists of a binary variable $x \in \{0, 1\} $ and a categorical label indicating group membership (among $k $ different groups), the data can be summarized in the following $2 \times k $ table:

\[
\def\arraystretch{1.9}
\begin{array}{c|c c c c|c}
    & \text{Group 1} & \text{Group 2} & \cdots & \text{Group } k & \text{Total} \\
\hline 
x = 1 & n^1_{1} & n^2_{1} & \cdots & n^k_{1} & n_1 \\
x = 0 & n^1_{0} & n^2_{0} & \cdots & n^k_{0} & n_0 \\ 
\hline
\text{Total} & n^1 & n^2 & \cdots & n^k & n
\end{array}
\] 
\vspace{0.3cm}

The dataset consists of $k $ groups, represented as $\mathbf{x}_i = (x_1^i, ..., x^i_{n^i}) $ for $i = 1, ..., k $, where $n^i $ denotes the size of group $i $. The table reports the number of ones ($n_1^i $) and zeros ($n_0^i $) in each group, along with their respective totals, $n_1 $ and $n_0 $.  

The key question remains whether the probability of observing $x = 1 $ differs between groups. This problem again translates into a hypothesis testing problem, where:

\begin{itemize} 
    \item Under the alternative hypothesis, the groups are distinct, meaning the number of ones is constrained separately in each group:
    \begin{equation}
        \mathbf{c}_1  = (n_1^1, n_1^2, \dots, n_1^k).
    \end{equation}
    
    \item Under the null hypothesis, the groups are indistinguishable, meaning only the total number of ones is constrained:
    \begin{equation} 
        \text{c}_0 = n_1.
    \end{equation}  
\end{itemize}

These constraints define the sufficient statistics under each hypothesis of the microcanonical and canonical tests discussed next. As the examples will illustrate, most of the results in this section naturally extend from the $2 \times 2$ case. The key distinction is that, in the latter case, the only relevant asymptotic behavior is as the total sample size $n$ grows large. In contrast, in the present setting, both $n$ and $k$ can grow large, with different scenarios arising depending on the application (see  \ref{sec:application_to_NM}). In all cases, the asymptotic behavior of e-variables plays a crucial role, particularly in the canonical test, where only asymptotic approximations are available, and we need to assess whether the microcanonical approximation can be used. 

\subsection*{\texorpdfstring{$\mathbf{2\times k}$ microcanonical test}{2xk microcanonical test}}
While the null model stays the same (Eq.    \eqref{micro_null}, the alternative model is simply the extension of \eqref{micro_alternative_2x2} from 2 to $k$ groups:
\begin{equation}\label{micro_alternative_2xk}
    P_\textrm{mic, 1}(\mathbf{x}; \{n_1^i\}) =
    \begin{cases}
        \frac{1}{\Omega_1(\{n_1^i\}),} & 
        \makebox[83pt][r]{if $(n_1^1(\mathbf{x}), ..., n_1^k(\mathbf{x}))$} \\ 
        & \makebox[100pt][r]{\hspace{1.5em} $= (n_1^1, ..., n_1^k)$,} \\
        0, & \text{else},
    \end{cases}
\end{equation}
where
\begin{equation}
\Omega_1(\{n_1^i\}) = \prod_{i=1} ^k \binom{n^i}{n_1^i}.
\end{equation}
As in the $2\times2$ case, Condition A \eqref{condition_A} is satisfied:
\begin{equation}
n_1 = \sum_{i=1}^k n_1^i 
\end{equation}
and the optimal prior on the null is the marginal distribution of $n_0$ induced by $W_1(\{n_1^i\})$. If all $n_1^i$ are independently distributed:
\begin{equation}
    W_1(\{n_1^i\}) = \prod_{i=1}^k W_1^i(n_1^i)
\end{equation}
then $W_0^*$ is simply the convolution of the individual alternative priors:
\begin{equation}
W_0^* = W_1^1 * ... * W_1^k.
\end{equation}
Interestingly, when the number of groups $k$ is large, and the priors are regular enough, a Central Limit Theorem holds; thus, $W_0^*$ is well approximated by a \textit{discrete Gaussian distribution}, i.e., if $k \gg 1$:
\begin{equation}\label{discrete_gaussian}
W_0^*(n_1) \approx \frac{1}{N(\mu_k, \sigma_k)} \exp\left( -\frac{(n_1 - \mu_k)^2}{2\sigma_k^2} \right)
\end{equation}
where $N$ is the normalization constant and
\begin{align*}
&\mu_k = \sum_{i=1}^{k} \mathbb{E}_{W_1^i}[n_1^i]\\
&\sigma_k^2 = \sum_{i=1}^{k} \text{Var}_{W_1^i}(n_1^i).
\end{align*}

This result is particularly convenient: when $k$ is big enough, the only effect of the choice of priors on the alternative, as long as they are independent and regular enough, is in determining the average and the variance of the optimal (approximated) Gaussian prior on the null. \\

\begin{figure}[H]
\includegraphics[width=0.5\textwidth
]{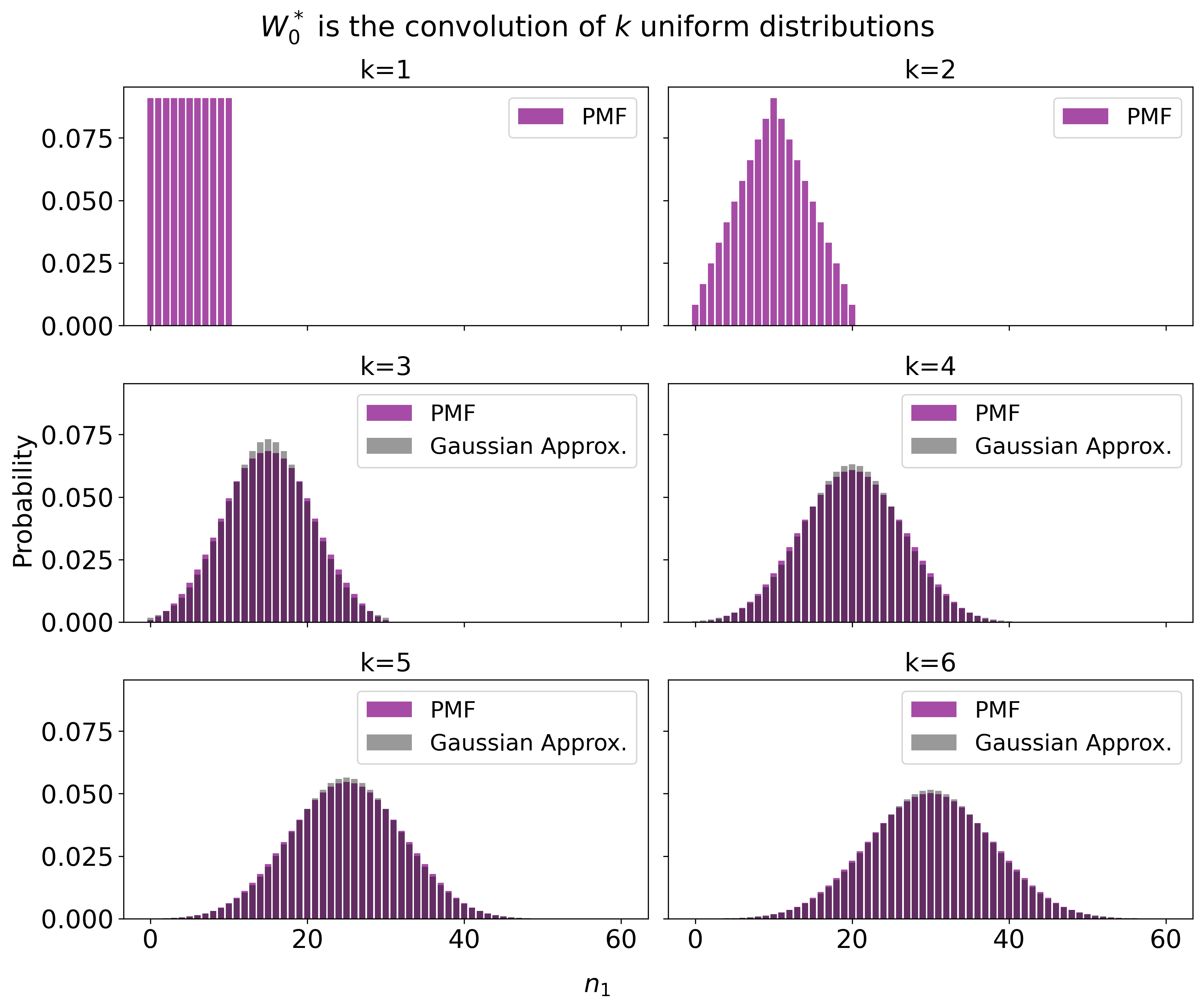}
\caption{The microcanonical GRO-optimal prior on the null $W_0^*$ for testing $2\times k$ tables is obtained as the convolution of the $k$ independent priors on the alternative, which are discrete uniform priors in the case shown in this picture. Each convolution, for $k>2$, is superposed to its discrete Gaussian approximation.}
\label{fig_many_uniforms}
\end{figure}

\noindent \textbf{Example 4: Independent uniform priors.}
Here, we extend Example 1, i.e., Example A, to the case of $k$ groups. When a uniform discrete prior $\mathcal{U}$ is put on each parameter of the alternative:
\begin{equation}
W_1(\{n_1^i\}) = \prod_{i = 1}^k \mathcal{U}_i(n_1^i) = \prod_{i=1}^k \frac{1}{n^i + 1},
\end{equation} 
the GRO null prior is again the convolution of all the individual priors, i.e., the convolution of $k$ discrete uniform distributions, which reads~\cite{stars_bars}:
\begin{multline}\label{uniform_many_convolutions}
    W^*_0(n_1) =\\= \sum_{S  \subseteq \{1, \, 
    \dots\, k \}} (-1)^{\mid S \mid }\binom{n_1 + k -1 -\sum_{j \in S} (n^j- n) }{n_1 - 1} \\ \times\left[ \prod_{i=1}^k n^i + 1\right]^{-1},
\end{multline}
where the sum runs over all possible subsets of $\{1, \, \dots\, k \}$ and $\mid S \mid$ is the number of elements of set $S$.
In the formula, the first factor stands for the number of ways in which a set of $k$ non-negative numbers ($\{n^i_1\}$) can be chosen uniformly such that their sum is equal to $n_1$, with the constraint that for each $i$, $n_1^i$ must be smaller than or equal to $n^i$. The second factor represents a normalization constant. If all $n^i$ are equal to a certain value $m$, the formula simplifies and reads:
\begin{multline}
    W^*_0(n_1) = \sum_{j =1}^{\lfloor n_1 /(m+1) \rfloor} (-1)^{j}\binom{n}{j}\binom{n_1 - j(m+1) + k-1 }{k - 1} \\ \times\left[ \prod_{i=1}^k n^i + 1\right]^{-1}.
\end{multline}
where $\lfloor x \rfloor$ is the floor function of $x$. This is the formula used to generate Figure~\ref{fig_many_uniforms}, where we show $W_0^*$, along with its Gaussian approximation, for increasing values of $k$.
\subsection*{\texorpdfstring{$\mathbf{2\times k}$ canonical test}{2xk canonical test}}
The null canonical model is the same as in the $2 \times 2$ case, Eq. \eqref{cano_null}, i.e., a collection of $n$ i.i.d. Bernoulli trials, where the probability of observing $x = 1$ is the same across all groups. The alternative model extends Eq. \eqref{cano_alternative_cano} and \eqref{cano_alternative_mean} to the case of $k$ groups, by constraining the expected values of $n_1^i$ separately for each group $i$, leading to the expression:
\begin{equation}
P_{\text{can}}(\mathbf{x}; \bm{\theta}) = \frac{e^{-\sum_{i=1}^{k} \theta_i n_1^i(\mathbf{x})}}{\prod_{i=1}^{k} (1+e^{-\theta_i})^{n^i}},
\end{equation}

or, equivalently, in the mean-value parametrization:

\begin{equation}
P_{\text{can}}(\mathbf{x}; \bm{p}) = \prod_{i=1}^{k} p_i^{n_1^i(\mathbf{x})} (1 - p_i)^{n^i - n_1^i(\mathbf{x})},
\end{equation}

where we define the group-specific probabilities as:

\begin{equation}
p_i = \frac{e^{-\theta_i}}{1+e^{-\theta_i}}, \quad \text{for each } i \in \{1, \dots, k\}.
\end{equation}

In this formulation, the alternative model assumes that data in each group are independent Bernoulli variables, where the probability of $x = 1$ depends on the group. The goal of the hypothesis test is to determine whether these probabilities are equal across all groups ($p_1 = p_2 = \dots = p_k$) or whether they differ, indicating that the probability of observing $x = 1$ is group-dependent.

In the following sections, we apply the procedure described in section \ref{subsec:cano_test} to different choices of $\bar{P}_{\text{can},1}$.\\

\noindent \textbf{Example 5: Canonical test with NML.}
Extending results from Example 2, for a model with $k$ independent Bernoulli distributions, the NML distribution reads~\cite{micro_cano}: 
\begin{equation}\label{nml_k}
P_{\text{can}, 1}^{\text{NML}} (\mathbf{x})= \prod_{i=1}^k P_{\text{can}, 1}^{\text{i, NML}}(\mathbf{x})
\end{equation}
where $P_{\text{can}, 1}^{\text{i, NML}}$ is that of Eq. \eqref{nml_bernoulli}. The microcanonical approximation is obtained by defining
\begin{equation}
    W_{\text{can},1}(\{n_1^i\}) = \prod_{i=1}^k W_{\text{can},1}^i(n_1^i)
\end{equation}
with 
\begin{equation}
W_{\text{can},1}^i(n_1^i) = \Omega_1^i(n_1^i)  \cdot \frac{ \left(\frac{n_1^i(\mathbf{x})}{n^i}\right)^{n_1^i(\mathbf{x})} \left(1 - \frac{n_1^i(\mathbf{x})}{n^i}\right)^{n^i - n_1^i(\mathbf{x})}}{\frac{e^{n^i}\Gamma(n^i, n^i)}{(n^i)^{n^i-1}}+1}.
\end{equation}
$W_0^*$ is then the convolution of all $W_{\text{can},1}^i(n_1^i)$, which again can be computed numerically or by resorting to the Gaussian approximation \eqref{discrete_gaussian}. \\

\noindent \textbf{Example 6: Independent beta priors.}
Here, we extend Example 3 to the case of $k$ groups.
We assume that each $p_i$ is independently distributed according to a beta prior with parameters $(\alpha^i, \beta^i)$. The Bayesian marginal likelihood reads:
\begin{equation}
    \bar{P}_{\text{can}, 1}(\mathbf{x}) = \prod_{i=1}^k
    \frac{B(\bar{\alpha}^i, \bar{\beta}^i)}{B(\alpha^i, \beta^i)}
\end{equation}
where
\begin{align*}
    &\bar{\alpha}^i = n_1^i + \alpha^i \\
    &\bar{\beta}^i = n^i - n_1^i + \beta^i \quad \text{for each } i \in \{1, \dots, k\}.
\end{align*}
To derive the microcanonical approximation, we compute the probability mass function induced by  $\bar{P}_{\text{can}, 1}(\mathbf{x})$ on $\{n_1^i\}$, which reads: 
\begin{equation}
    W_{\text{can},1}(\{n_1^i\}) = \prod_{i=1}^k W_{\text{can},1}^i(n_1^i)
\end{equation}
with
\begin{equation}
W_{\text{can},1}^i(n_1^i) = \Omega_1^i(n_1^i)  \cdot \frac{B(\bar{\alpha}^i, \bar{\beta}^i)}{B(\alpha^i, \beta^i)} \quad \text{for each } i \in \{1, \dots, k\}. 
\end{equation}
$W_0^*(n_1)$ is, then, the convolution of all $W_{\text{can},1}^i(n_1^i)$. If all beta parameters are equal to $1$, $W_0^*$ reduces to the convolution between $k$ discrete uniform distributions \eqref{uniform_many_convolutions}. When the beta parameters are such that no closed form is available, the convolution must be computed numerically. Alternatively, if $k$ is big enough, one can resort to the discrete Gaussian approximation \eqref{discrete_gaussian}. Analogously, a continuous Gaussian approximation can be used to approximate $w_{\text{pseudo}}$.   Figure~(\ref{SM_fig_beta_convolutions_2x10}), we show $W_{\text{can},1}^i$, $W_0^*$, and $w_{\text{pseudo}}$ for $k=10$.

\subsection*{Evaluating the microcanonical approximation}
To assess the effectiveness of the microcanonical approximation, we study the behavior of the interval width $r$ in different cases. To simplify the problem, all beta priors considered in our results have parameters equal to the same number, $\gamma$, and all groups share the same size, i.e., $n_i = m $ for all $i\in \{1,\dots, k\}$. In this scenario, we have that $n = m \cdot k$. We consider three cases: $m$ increases and $k$ is fixed; $n$ is fixed, and $m$ and $k$ change accordingly;  finally, $m$ and $k$ grow together according to a certain law. We evaluate $S_{\text{pseudo}}$, and consequently $r$, by using $w^1_\text{pseudo, 0}$, according to the procedure described in Example C, which is easily extended to the case of $k$ groups. Our experiments (Figure~\ref{fig:r_2x2_k_and_m}) show that:
\begin{enumerate}
\item $r$ converges quickly to $0$ for fixed $k$ as the m increases (or, equivalently, the total sample size increases);
\item $r$ grows slowly for $n$ fixed and $k$ getting bigger;
\item $r$ converges quickly to $0$ whenever $k$ and $m$ grow together according to different power laws. 
\end{enumerate}
The only case where $r$ does not converge to $0$ corresponds to a decreasing $m$ as $O(1/k)$. Our conclusion is that our microcanonical approximation $S_{\text{mic}}^{\text{GRO}}$ is an optimal candidate as long as $m$, i.e., the data size of each group, is big enough. 

\subsection{Connection to models of networks  and time series}
\label{sec:application_to_NM}

maximum entropy models are widely used to construct null models of complex systems that preserve specific structural or temporal features, while remaining otherwise random~\cite{jaynes,cimini2019, squartini2017, squartini2018,golan1996maximum}. 

For instance, when applied to networks, maximum entropy models in their canonical formulations are known as \textit{exponential random graph models}~\cite{holland1981,hunter2006,cimini2019}. Examples of commonly used maximum entropy network models are the Erd\H{o}s--R\'enyi model, Configuration Models, and Stochastic Block Models~\cite{cimini2019,squartini2017}. The framework presented here is fully general and can be applied to build and compute e-values when testing between general maximum entropy network models with different sufficient statistics, in both their canonical and microcanonical formulations. Moreover, section \ref{sec:MDL} establishes a link between e-values and the Minimum Description Length principle --- a framework increasingly used in recent years for network inference and model selection~\cite{peixoto17, peixoto25, vallescatala18,micro_cano}.

In particular, the hypothesis tests for contingency tables developed here have a direct correspondence with hypothesis tests between common network models. 
This mapping arises because the sufficient statistics in our contingency tables capture the same structural constraints as those imposed in standard network ensembles~\cite{squartini2017,cimini2019}. Indeed, a binary network is represented by a binary adjacency matrix, which is a (structured) collection of $1$s and $0$s, corresponding to the presence or absence of a link between two nodes.

The null model considered here, in both its canonical and microcanonical formulation, corresponds to the well-known \emph{Erd\H{o}s--R\'enyi} model (ER), where the sufficient statistic is the total number of links, equal to (half, if the network is undirected) the total number of $1$s observed in the adjacency matrix.

In the \emph{Stochastic Block Model} (SBM), nodes are partitioned into groups and the adjacency matrix of a network is structured in $k$ blocks, corresponding to the presence of inter- and intra-group links. For instance, in models of networks with community structure, intra-group link probabilities are larger than inter-group ones. The sufficient statistics are the number of links in each block. Testing an SBM against an ER model corresponds exactly to testing whether connection probabilities are identical across all blocks (i.e., communities are absent), and this SBM vs ER problem reduces to our canonical or microcanonical contingency table $2\times k$ test.

In the \emph{Partial Configuration Model} (PCM) for bipartite networks~\cite{zhang2022strong}, the degree of each node in one layer is constrained, while connections to the other layer are otherwise random. The (bi-)adjacency matrix is a $k \times m$ rectangular binary matrix, and the sufficient statistics are the number of links connected to each node in the constrained layer, i.e., the number of $1$s in each row. 
Testing a PCM against a bipartite ER model corresponds to testing whether all nodes in the constrained layer have the same connection probability (and therefore the same expected degree), i.e., testing for homogeneity of node properties in the graph. This again maps to a $2\times k$ contingency table, where each constrained node represents a ``group'' and each group size equals the number $m$ of nodes in the unconstrained layer.  

Besides network models, a final connection worth mentioning is the one between binary contingency tables and multivariate time series data describing, e.g., a system of units being active ($1$) or inactive ($0$) at discrete time steps (such as spiking neurons data). The PCM can, in this case, represent a model enforcing, for each time step, a different activation probability of the various units. Therefore, testing the PCM against a bipartite ER model corresponds in this case to testing non-stationarity vs stationarity of the observed process over time. 

We therefore conclude that our microcanonical e-variable for contingency tables can be directly applied to a wide range of problems, both exactly in the microcanonical case and as an approximation for the canonical case. Moreover, our results on the behavior of $r$ show that the microcanonical approximation works very well in both scenarios, as long as the size of each group is large enough. 
This circumstance is particularly convenient when studying models of large complex systems with a growing number of heterogeneous features, such as PCMs where the number of nodes in both layers can diverge in the ``thermodynamic limit'' of infinitely large graphs, SBMs used to model networks with a growing number of communities, and models of high-dimensional multivariate (nonstationary) time series.
As we mentioned, the growing number of features (and parameters) in these models is generally needed to replicate the heterogeneous properties of real-world networks and time series more closely. 
At the same time, it makes the study of these models more challenging because of the breakdown of various useful approximations valid for a finite number of parameters ---and even of the asymptotic equivalence between canonical and microcanonical versions of the resulting ensembles~\cite{squartini2015breaking,micro_cano}. Despite these complications, the results derived here nicely apply to those regimes.

\section{Conclusion}
In this work, we have developed a general framework for constructing optimal e-values for hypothesis testing between maximum entropy models with different constraints, in both microcanonical and canonical formulations. Our main theoretical contribution is the exact derivation of the microcanonical GRO e-variable and its use as a valid approximation to the canonical GRO e-variable when the latter is intractable. We provided analytical and numerical evidence that this approximation becomes asymptotically exact in many relevant regimes. 

We illustrated our results through applications to $2 \times 2$ contingency tables, in both Bayesian and non-Bayesian (NML) settings, showing numerically that the microcanonical approximation provides a good proxy for the canonical solution, confirming our theoretical results. We then extended the analysis to general $2 \times k$ tables, where numerical results suggest that the microcanonical approximation works and remains asymptotically optimal for different interplays between $k$ and the group sizes, as long as the latter are sufficiently large. Interestingly, when $k$ becomes large, the microcanonical e-variable is itself well approximated by choosing a discrete Gaussian prior on the null. We highlighted that this framework can be naturally translated into network-science terms, where many important models can be derived as maximum entropy models.

A central role in our construction is played by universal distributions. These are the same distributions that underlie the Minimum Description Length (MDL) principle, where they achieve minimax redundancy. Our results show that such universal distributions (including Bayesian and NML ones) can be conveniently used to build GRO e-variables as well, thus providing a direct and convenient connection between description lengths and e-variables. A possible direction to explore in future work is to extend this connection beyond pairwise model comparisons and investigate how GRO e-variables and MDL can be combined to design tests involving multiple models at once.

\bibliography{biblio}{}

\bibliographystyle{unsrt}

\onecolumngrid

\newpage
\begin{center}
\widetext
\textbf{\large Supplementary Materials}
\end{center}

\setcounter{equation}{0}
\setcounter{figure}{0}
\setcounter{table}{0}
\setcounter{page}{1}
\setcounter{section}{0}
\makeatletter

% numerazione visibile
\renewcommand{\theequation}{S\arabic{equation}}
\renewcommand{\thesection}{S\arabic{section}}
\renewcommand{\thefigure}{S\arabic{figure}}
\renewcommand{\thetable}{S\arabic{table}}

% ancore per hyperref (fondamentale!)
\renewcommand{\theHequation}{S\arabic{equation}}
\renewcommand{\theHsection}{S\arabic{section}}
\renewcommand{\theHfigure}{S\arabic{figure}}
\renewcommand{\theHtable}{S\arabic{table}}

\section{Redundancy and regret}
{\label{SM_red_reg}}
Here we show that, given the null and alternative models $
\mathcal{M}_0=\{P_{0}(\mathbf{x};\bm{\theta})\}_{\bm{\theta}\in\bm{\Theta}_0}
$ and $
\mathcal{M}_1=\{P_{1}(\mathbf{x};\bm{\theta})\}_{\bm{\theta}\in\bm{\Theta}_1}
$, the regret \eqref{regret} of the GRO e-variable \eqref{GRO_form}, i.e.,
\begin{equation}
S^{\text{GRO}}(\mathbf{x}) = \frac{\bar{P}_1(\mathbf{x})}{P_0^{w^*}(\mathbf{x})}
\end{equation}
for a given distribution $\bar{P}_1$, is bounded by the redundancy \eqref{eq:redundancy} of $\bar{P}_1$. Indeed, for any INECCSI (Def. \ref{def:ineccsi}) subset $\bm{\Theta}'$ of $\bm{\Theta}$:
\begin{align}
&\text{REG}(\bm{\Theta_1}';\bar{P}_1) 
= \max_{\bm{\theta}_1 \in \bm{\Theta}_1'} 
\mathbb{E}_{\bm{\theta}_1} \left[ \log S^{\text{GRO}(\bm{\theta}_1)} - \log \frac{\bar{P}_1(\mathbf{x})}{P_0^{w_0^*}(\mathbf{x})} \right] \nonumber \\
&= \max_{\bm{\theta}_1 \in \bm{\Theta}_1'} \min_{w'_0 \in \mathcal{W}_{\bm{\theta}_0}}
\mathbb{E}_{\bm{\theta}_1} \left[ 
\log \frac{P_1(\mathbf{x}; \bm{\theta}_1)}{P_0^{w'_0}(\mathbf{x})} 
- \log \frac{\bar{P}_1(\mathbf{x})}{P_0^{w_0^*}(\mathbf{x})} \right] \nonumber \\
&= \max_{\bm{\theta}_1 \in \bm{\Theta}_1'} 
\left( \min_{w'_0 \in \mathcal{W}_{\bm{\theta}_0}} \mathbb{E}_{\bm{\theta}_1} 
\left[ \log \frac{P_0^{w_0^*}(\mathbf{x})}{P_0^{w'_0}(\mathbf{x})} \right] 
+ \text{RED}_1(\bm{\theta}_1; \bar{P}_1) \right) \nonumber \\
& \leq \text{RED}_1(\bm{\Theta}_1'; \bar{P}_1).
\label{eq:SM_upperbound}
\end{align}

\section{Microcanonical test}\label{SM_section_micro}
In this section, the subscript ``mic" is omitted for the sake of clarity, as all the models considered are microcanonical models.
\subsection{Exact solution of the optimization problem}
  We consider a test between a microcanonical alternative $\mathcal{M}_1$ and a microcanonical null $\mathcal{M}_0$. Given a universal microcanonical distribution $\bar{P}_1$ (either NML or Bayesian) on the alternative, the GRO-optimal microcanonical e-variable
 \begin{equation}
 S^{\text{GRO}} = \frac{\bar{P}_1}{P_{0}^{W_0^*}}
 \end{equation}
is found by solving the optimization problem

\begin{align}\label{SM_DKL_problem_1}
    W_0^* &= \arg \min_{W\in \mathcal{W}_{\mathbf{c}_0}}D_{\text{KL}}(\bar{P}_1 \Vert P_0^{W}) \\
    &= \nonumber \arg \min_{W\in \mathcal{W}_{\mathbf{c}_0}} \sum_{\mathbf{x} \in \mathcal{X}} \bar{P}_{1}(\mathbf{x}) \log \frac{\bar{P}_{1}(\mathbf{x})}{\bar{P}^{W}_{0}(\mathbf{x})} 
\\ \nonumber &= \arg \max_{W\in \mathcal{W}_{\mathbf{c}_0}} \sum_{\mathbf{x} \in \mathcal{X}} \bar{P}_{1}(\mathbf{x}) \log \bar{P}^{W}_{0}(\mathbf{x}).
\end{align}
For a microcanonical model with sufficient statistics $\mathbf{c}_i$, the Bayesian marginal likelihood reads~\cite{micro_cano}
\begin{equation}\label{SM_bayesian_evidence}
P_i^{W_i}(\mathbf{x}) = \sum_{\mathbf{c}_i \in \mathcal{C}_i} P_i(\mathbf{x}; \mathbf{c}_i) W_i(\mathbf{c}_i) = P_i(\mathbf{x}; \mathbf{c}_i(\mathbf{x}))W_i(\mathbf{c}_i(\mathbf{x})) = \frac{W_i(\mathbf{c}_i(\mathbf{x}))}{\Omega_i(\mathbf{c}_i(\mathbf{x}))}
\end{equation}
where the latter equality is due to the definition of the microcanonical model, which assigns a positive probability only if $\mathbf{c}_i(\mathbf{x}) = \mathbf{c}_i $. By putting this result in \eqref{SM_DKL_problem_1}, one gets

\begin{align}
    W_0^* &= \arg \max_{W\in \mathcal{W}_{\mathbf{c}_0}} \sum_{\mathbf{x} \in \mathcal{X}} \bar{P}_{1}(\mathbf{x})[\log W_0(\mathbf{c}_0(\mathbf{x})) - \log \Omega_0(\mathbf{c}_0(\mathbf{x})) ]\\
    &= \nonumber  \arg \max_{W\in \mathcal{W}_{\mathbf{c}_0}} \sum_{\mathbf{x} \in \mathcal{X}} \bar{P}_{1}(\mathbf{x})\log W_0(\mathbf{c}_0(\mathbf{x})).   
\end{align}
The latter expression can be written as a sum over the values of $\mathbf{c}_0$: 
\begin{align}
    W_0^* = \arg \max_{W\in\mathcal{W}_{\mathbf{c}_0}}\sum_{\mathbf{c}_0\in \mathcal{C}_0} \left[\sum_{\mathbf{x} \, : \, \mathbf{c}(\mathbf{x}) =  \mathbf{c}_0}\bar{P}_{1}(\mathbf{x})\right]\log W(\mathbf{c}_0).
\end{align}
According to Gibbs inequality, the distribution maximizing the quantity above is 
\begin{align}\label{SM_GRO_micro}
    W_0^*(\mathbf{c}_0) = \sum_{\mathbf{x} \, : \, \mathbf{c}_0(\mathbf{x}) =  \mathbf{c}_0}\bar{P}_{1}(\mathbf{x})
\end{align}
i.e., the marginal distribution of the null sufficient statistic $\mathbf{c}_0(\mathbf{x})$ induced by $\bar{P}_{1}$, hereby denoted, for simplicity, by $\bar{P}^{\mathbf{c}_0}_{1}$. 

In the special case where the alternative sufficient statistics completely determine the value of the null one, we can write:
\begin{align}\label{SM_condition_A}
& \textbf{Condition A:} \\ 
& \text{there exists a function $f: \mathcal{C}_1 \rightarrow \mathcal{C}_0$ s.t.}  \mathbf{c}_0(\mathbf{x}) = f(\mathbf{c_1}(\mathbf{x})), \nonumber 
\end{align}

and thus 
\begin{align}\label{SM_GRO_micro_if_cond_A}
    W_0^*(\mathbf{c}_0) &= \sum_{\mathbf{x} \, : \, \mathbf{c}_0(\mathbf{x}) =  \mathbf{c}_0}\bar{P}_{1}(\mathbf{x})\\
    \nonumber &= \sum_{\mathbf{x} \, : \, \mathbf{c}_0(\mathbf{x}) =  \mathbf{c}_0} \frac{W_1(\mathbf{c}_1(\mathbf{x}))}{\Omega_1(\mathbf{c}_1(\mathbf{x}))} \\
    \nonumber &= \sum_{\mathbf{c}_1\, : \, f(\mathbf{c_1}) = \mathbf{c}_0} \Omega_1(\mathbf{c}_1) \frac{W_1(\mathbf{c}_1)}{\Omega_1(\mathbf{c}_1)} \\
     \nonumber &=  \sum_{\mathbf{c}_1\, : \, f(\mathbf{c_1}) = \mathbf{c}_0} W_1(\mathbf{c}_1).
\end{align}
In other words, the GRO-optimal prior on the null is the distribution induced on the null sufficient statistics by the alternative prior, or, equivalently, the marginal distribution of $\mathbf{c}_0$ induced by $W_1$, hereby denoted by $W_1^{\mathbf{c}_0}$.

Once that $W_0^*$ is computed, given that all universal microcanonical distributions considered here are, in fact, Bayesian marginal likelihoods, the microcanonical GRO-optimal e-variable can be expressed as

 \begin{align}
 S^{\text{GRO}}(\mathbf{x})  &= \frac{P_1^{W_1}(\mathbf{x})}{P_{0}^{W_0^*}(\mathbf{x})} \\
 \nonumber &= \frac{P_1(\mathbf{x}; \mathbf{c}_1(\mathbf{x}))}{P_0(\mathbf{x}; \mathbf{c}_0(\mathbf{x}))} \frac{W_1(\mathbf{c}_1(\mathbf{x}))}{W_0^*( \mathbf{c}_0(\mathbf{x}))} \\
 \nonumber &= \frac{\Omega_0(\mathbf{c}_0(\mathbf{x}))}{\Omega_1(\mathbf{c}_1(\mathbf{x}))} \frac{W_1(\mathbf{c}_1(\mathbf{x}))}{W_0^*( \mathbf{c}_0(\mathbf{x}))}.
 \end{align}

\subsection{The average under the null of the GRO e-variable is always unitary}
Here, we show that $
  \mathbb{E}_0[\,S^\text{GRO}\,] = 1 \quad \forall P_0 \in \mathcal{M}_{0}.
$ \\

We start by picking a generic distribution inside the null model:
\begin{equation}
    P_0(\mathbf{x}; \bar{\mathbf{c}}_0) = 
    \begin{dcases}
        \frac{1}{\Omega_0(\bar{\mathbf{c}}_0)} & \text{if }\mathbf{c}_0(\mathbf{x}) = \bar{\mathbf{c}}_0 \\
        0 & \text{else}.
    \end{dcases}
\end{equation}
Then, we compute the average under $P_0(\mathbf{x}; \bar{\mathbf{c}}_0)$ of $S^\text{GRO}$:
\begin{align}
        \mathbb{E}_0[S^\text{GRO}] &= \mathbb{E}_0\left[\frac{\bar{P}_1}{P_0^{W_0^*}}\right] = \sum_{\mathbf{x}\in \mathcal{X}} P_0({\mathbf{x}; \bar{\mathbf{c}}}_0) \frac{\bar{P}_1(\mathbf{x})}{P_0^{W_0^*}(\mathbf{x})} \\
        &= \frac{1}{\Omega_0(\bar{\mathbf{c}}_0)} \sum_{\mathbf{x} \,:\, \mathbf{c}_0(\mathbf{x}) = \bar{\mathbf{c}}_0}    \frac{\bar{P}_1(\mathbf{x})}{P_0^{W_0^*}(\mathbf{x})}
\end{align}
In what follows, we use the explicit expression of the Bayesian marginal likelihood \eqref{SM_bayesian_evidence}, i.e.,  $P_0^{W_0}(\mathbf{x}) = \frac{W_0(\mathbf{c}_0(\mathbf{x}))}{\Omega_0(\mathbf{c}_0(\mathbf{x}))}$, and that of the GRO optimal prior \eqref{SM_GRO_micro}
\begin{align}
        \mathbb{E}_0[\,S^\text{GRO}\,] &= \frac{1}{\Omega_0(\bar{\mathbf{c}}_0)} \sum_{\mathbf{x} \,:\, \mathbf{c}_0(\mathbf{x}) =  \bar{\mathbf{c}}_0}   \frac{\Omega_0(\mathbf{c}_0(\mathbf{x})) \bar{P}_1(\mathbf{x})}{W^*_0(\mathbf{c}_0(\mathbf{x}))} \\
        &= \nonumber \frac{1}{\Omega_0(\bar{\mathbf{c}}_0)} \frac{\Omega_0(\bar{\mathbf{c}}_0)}{W^*_0(\bar{\mathbf{c}}_0)}\sum_{\mathbf{x} \,:\, \mathbf{c}_0(\mathbf{x}) =  \bar{\mathbf{c}}_0} \bar{P}_1(\mathbf{x}) = 1.     
\end{align}

\section{Microcanonical approximation}
\label{SM_section_app_micro_cano}
\subsection{Every microcanonical e-variable is a canonical e-variable}
Given a sufficient statistics $\mathbf{c}(\mathbf{x})$, the microcanonical probability distribution can be expressed as a conditional canonical one:
\begin{equation}
P_{\text{mic}} (\mathbf{x}; \mathbf{c})  = P_{\text{can}} (\mathbf{x}; \bm{\theta} \mid \mathbf{c}).
\end{equation}
where the probability of $\mathbf{x}$ is conditioned on a certain value of $\mathbf{c}(\mathbf{x})$.
According to the law of total expectation, for every random variable $S(\mathbf{x})$ defined on $\mathcal{X}$:
\begin{equation}\label{SM_law_total}
\mathbb{E}_{{\text{can}}}[\,S\,] = \mathbb{E}_{\text{can},}\left[\, \mathbb{E}_{\text{can}} [\,S \mid \mathbf{c} \,]\, \right] = \mathbb{E}_{\text{can}}\left[\,\mathbb{E}_{\text{mic}} [\,S\,]\,\right].
\end{equation}
It follows that if $E_{\text{mic}}$ is a microcanonical e-variable, it is also a canonical one:  
\begin{equation}
    \mathbb{E}_{\text{mic}}[\,E_{\text{mic}}\,] \leq 1 \quad \forall P_{\text{mic}} \in \mathcal{M}_{\text{mic}}
    \quad \Rightarrow \quad
    \mathbb{E}_{\text{can}}[\,E_{\text{mic}}\,] \leq 1 \quad \forall P_{\text{can}} \in \mathcal{M}_{\text{can}}
\end{equation}
Moreover, as proven in the last section, it holds:
\begin{equation}
\mathbb{E}_{\text{mic}} [\, S^{\text{GRO}}_{\text{mic}}\,] = 1.
\end{equation}
Thus, from \eqref{SM_law_total}, it follows that
\begin{equation}
\mathbb{E}_{\text{can}} [\, S^{\text{GRO}}_{\text{mic}}\,] = 1.
\end{equation}
\subsection{A canonical universal distribution can always be expressed as microcanonical Bayesian marginal likelihood}
\label{SM_subsection_app_micro_cano_B}
In what follows, we show that: 
\begin{enumerate}
    \item Given a canonical universal distribution $\bar{P}_{\text{can}}$ relative to a sufficient statistic $\mathbf{c}$, we can always define a prior distribution $W_{\text{can}}(\mathbf{c})$ on the sufficient statistic such that
\begin{equation}
    \bar{P}_{\text{can}} = \bar{P}^{W_{\text{can}}}_{\text{mic}}.
\end{equation}
\item The opposite is not true: for some $\bar{P}^W_{\text{mic}}$, there is no choice of prior density $w(\bm{\theta})$ such that $\bar{P}^W_{\text{mic}} =\bar{P}^w_{\text{can}} $. 
\end{enumerate}
\subsubsection*{Proof 1}
By construction, a canonical universal distribution $\bar{P}_{\text{can}}(\mathbf{x})$ relative to the sufficient statistics $\mathbf{c}$ always assigns the same probability mass to configurations sharing the same value of the sufficient statistic, just as the corresponding $P_{\text{can}}(\mathbf{x}; \bm{\theta})$ does. Consequently, the following holds:
\begin{equation}
 \bar{P}_{\text{can}}(\mathbf{x}\,|\,\mathbf{c}) = P_{\text{mic}}(\mathbf{x};\mathbf{c}),
\end{equation}
i.e., when conditioned on the sufficient statistic, the universal canonical distribution reduces to a uniform distribution over the number of configurations corresponding to the observed value, which is the microcanonical distribution.  
Moreover, we can always write
\begin{equation}\label{SM_decomposition}
\bar{P}_{\text{can}}(\mathbf{x}) = \bar{P}_{\text{can}}(\mathbf{x}\;|\;\mathbf{c}(\mathbf{x})) \bar{P}_{\text{can}}^{\mathbf{c}}(\mathbf{c}(\mathbf{x})) = P_{\text{mic}}(\mathbf{x};\mathbf{c}(\mathbf{x})) \bar{P}_{\text{can}}^\mathbf{c}(\mathbf{c}(\mathbf{x}))
\end{equation}
where $\bar{P}_{\text{can}}^{\mathbf{c}}(\mathbf{c})$ is the distribution of $\mathbf{c}$ induced by $\bar{P}_{\text{can}}$. The proof follows by comparing the expression above with the general expression of the microcanonical Bayesian marginal likelihood (\ref{SM_bayesian_evidence}) and by setting $W_{\text{can}}(\mathbf{c}) = \bar{P}_{\text{can}}^{\mathbf{c}}(\mathbf{c})$.

\subsubsection*{Proof 2}
The canonical Bayesian marginal likelihood $\bar{P}_{can}^{w}$
\begin{equation}
P_{\text{can}}^{w}(\mathbf{x}) = \int_{\bm{\Theta}} P_{\text{can}}(\mathbf{x}; \bm{\theta}) w(\bm{\theta}) d\bm{\theta}
\end{equation}
is the weighted sum of positive functions; indeed, $P_{\text{can}}(\mathbf{x}; \bm{\theta})$ is an exponential function that assigns positive probability to all $\mathbf{x} \in \mathcal{X}$.  As such, for all proper choices of the prior density $w(\bm{\theta})$, $\bar{P}_{\text{can}}^{w}$ is strictly positive everywhere:
\begin{equation}
    \bar{P}_{can}^w(x) > 0 \quad \forall x \in \mathcal{X}.
\end{equation}
Consider a microcanonical prior $W(\mathbf{c})$ such that $W(\mathbf{c}^*) = 0$ for a certain value $\mathbf{c}^*$ of the sufficient statistics. Consequently, 
\begin{equation}
\bar{P}_{\text{mic}}^{W}(x^*) = 0\quad \forall \; x :  \mathbf{c}(\mathbf{x}) = \mathbf{c}^*.
\end{equation}
Thus, there is no choice of canonical prior $w(\bm{\theta})$ s.t. $\bar{P}_{\text{mic}}^{W} = \bar{P}_{\text{can}}^{w}$. 

\section{Proof of Theorem~\ref{thm:one}}
\label{app:proofthmone}
In the proof, we freely use well-known properties of exponential families as described by, e.g.,~\cite{BarndorffNielsen78}. 
Set $B := {\tt M}'$. 
Fix a constant $a$ and consider the sets, for $m=1,2, \ldots$: 
\begin{align*}
B^+_m  = \left\{\bm{\mu}: \inf_{\bm{\mu}' \in B}\| \bm{\mu} - \bm{\mu}'\ \|_2^2 \leq \frac{a \log m}{m} \right\}, 
B^-_m  = \left\{\bm{\mu} \in B: \inf_{\bm{\mu}' \not \in B}\| \bm{\mu} - \bm{\mu}'\ \|_2^2 \geq \frac{a \log m}{m} \right\}.
\end{align*}
$B^+_m$ is a superset of $B$, including a small region (whose volume tends to $0$ with sample size) just outside $B$'s boundary; similarly $B^-_m$ excludes a small region just inside $B$'s boundary. 
\newcommand{\mle}{\bm{\hat{\mu}}}
To shorten notation we write $\mle:= \mathbf{s}(\mathbf{y}^{(m)})/m$ and $P^w := P^{w\,(m)}_{\text{can}}$ and, for any measurable subset $B' \subset {\tt M}$,  $Q^w(B') := Q^w(\bm{\mu} \in B')$, and $P_{\bm{\mu}} := P_{\text{can}}(\cdot ; \bm{\theta}(\bm{\mu}))$ where $\bm{\theta}(\bm{\mu})$ is the mapping from mean-value parameters to corresponding canonical parameters, i.e. the inverse of the (1-to-1) mapping $\bm{\mu}(\bm\theta)$ defined in the main text.  We have: 
\begin{align}\label{eq:sanovA}
  P^{w}(  \mle\in B)  & 
  = \int_{\bm\mu \in {\tt M}} P_{\bm\mu}(\mle \in B) w(\bm\mu) d\bm\mu 
  \geq \int_{\bm\mu \in B^-_m} P_{\bm\mu}(\mle \in B) w(\bm\mu) d\bm \mu  
  \nonumber \\ & 
  \geq    Q^w(B^-_m) \inf_{\bm{\mu} \in B^-_m} P_{\bm{\mu}}(\mle \in B) \geq  Q^w(B^-_m) \inf_{\bm{\mu} \in B^-_m} (1- P_{\bm{\mu}}(\mle \not \in B))   \nonumber \\
  & \geq 
   Q^w(B^-_m)  - \sup_{\bm{\mu} \in B^-_m} P_{\bm{\mu}}(\mle \not \in B)
Q^w(B) + O \left(\frac{\log m}{m}\right) - \sup_{\bm{\mu} \in B^-_m} P_{\bm{\mu}}(\mle \not \in B).
\end{align}
and also 
\begin{align}\label{eq:sanovB}
  P^{w}(  \mle\in B) &  =  
  \int_{\bm{\mu}\in B^+_{m}} P_{\bm\mu}(\mle \in B) w(\bm\mu) d\bm\mu  + 
  \int_{\bm{\mu} \not \in B^+_{m}} P_{\bm\mu}(\mle \in B) w(\bm\mu) d\bm\mu 
 \nonumber \\
  & \leq Q^w(B^+_m) + \sup_{\bm{\mu} \in {\tt M}, \bm{\mu}  \not \in B^+_m}
  P_{\bm{\mu}}(\mle \in B) = Q^w(B) + O\left(\frac{\log m}{m}\right) +  \sup_{\bm{\mu} \in {\tt M} \setminus  B^+_m}
  P_{\bm{\mu}}(\mle \in B). 
\end{align}
We will now further bound the supremum terms in the above two formulas, showing that they are both of order $O(m^{-a})$. The result then follows by plugging in $a=1$. 
\paragraph{Supremum in (\ref{eq:sanovB})}
To bound the supremum in (\ref{eq:sanovB}), note first that $B$ is a convex set. We can therefore use Csisz\'ar's~\cite{Csiszar84} multivariate generalization of Chernoff's concentration inequality (see~\cite{GrunwaldHB24} for general discussion) to get that
\begin{align}\label{eq:suppie}
&     \sup_{\bm{\mu} \in {\tt M} \setminus  B^+_m}
  P_{\bm{\mu}}(\mle \in B) \leq  \sup_{\bm{\mu} \in {\tt M} \setminus  B^+_m, \bm{\mu}'\in B} e^{- m D_{\textrm{KL}}(P_{\bm{\mu}'} \| P_{\bm{\mu}}
  )} = e^{- m \inf_{\bm{\mu} \in {\tt M} \setminus  B^+_m, \bm{\mu}'\in B} D_{\textrm{KL}}(P_{\bm{\mu}'} \| P_{\bm{\mu}}
  )} \leq e^{- m \inf_{\bm{\mu} \in \partial B^+_m, \bm{\mu}'\in B} D_{\textrm{KL}}(P_{\bm{\mu}'} \| P_{\bm{\mu}}
  )},
\end{align}
where $D_{\textrm{KL}}(P_{\bm{\mu}'} \| P_{\bm{\mu}})$ is the Kullback-Leibler divergence between $P_{\bm{\mu}'}$ and $P_{\bm{\mu}}$ defined at a single outcome, and in the final inequality we used the standard fact that the  KL divergence between members of an exponential family is strictly convex in its first argument.  

Now since $B$ is an INECCSI subset of ${\tt M}$, clearly there exists another INECCSI subset of ${\tt M}$, say $\bar{B}$, and a finite $m_0$ such that $B^+_m \subset \bar{B} $ for all $m \geq m_0$. Since $\bar{B}$ is INECCSI, there exist $c, C$ with $0 < c < C < \infty$ such that all eigenvalues of the Fisher information matrix $I(\bm{\mu})$ in the mean-value parameterization are in between $c$ and $C$ for all $\bm{\mu} \in \bar{B}$. This means that the KL divergence between any $\bm{\mu}, \bm{\mu}' \in \bar{B}$ satisfies 
\begin{equation}\label{eq:KLsquare}
(1/2) c^k \frac{a \log m}{m} \leq 
\frac{1}{2} c^k \| \bm{\mu} - \bm{\mu}' \|_2^2  \leq  D_{\textrm{KL}}(P_{\bm{{\mu}}'} \| P_{\bm\mu}) \leq  \frac{1}{2} C^k \| \bm{\mu} - \bm{\mu}' \|_2^2
\leq (1/2) C^k \frac{a \log m}{m}.
\end{equation}
The result now follows by plugging in the lower bound on the KL divergence implied by the above into (\ref{eq:suppie}) and then setting $a=1$ and plugging further into (\ref{eq:sanovB}). 
\paragraph{Supremum in (\ref{eq:sanovA})}
To bound the supremum in (\ref{eq:sanovA}), fix any $\bm{\mu} \in B^-_m$ and let $R_{m,\bm{\mu}}$ be a hyper-rectangle centered at $\bm{\mu}$ that is a subset of $B$ and that has side-length $2 \epsilon_m$. By construction there is $c' > 0$ such that, for all $m$, we can take $\epsilon_m = c' \cdot \sqrt{( a \log m / m)}$. Noting that we can write $\bm{\mu} = (\bm{\mu}_1, \ldots, \bm{\mu}_d)$, with $d$ the dimensionality of both the canonical space $\bm{\Theta}$ and the mean-value space ${\tt M}$, we set $H^{\geq}_{\bm{\mu},j,\epsilon}:= \{\bm{\mu}' \in {\tt M}: \bm{\mu}'_j\geq \bm{\mu}_j + \epsilon\}$ and  $H^{\leq}_{\bm{\mu},j,\epsilon}:= \{\bm{\mu}' \in {\tt M}: \bm{\mu}'_j\leq \bm{\mu}_j - \epsilon\}$. We have: 
\begin{align}\label{eq:infie}
&     \sup_{\bm{\mu} \in B^-_m}
  P_{\bm{\mu}}(\mle \not \in B) \leq 
  \sup_{\bm{\mu} \in B^-_m}
  P_{\bm{\mu}}(\mle \not \in R_{m,\bm{\mu}}) \leq 
  \sum_{j=1}^d   
\left( P_{\bm{\mu}}(\mle \in H^{\leq}_{\bm{\mu},j,\epsilon_m} )
+  P_{\bm{\mu}}( 
\mle \in H^{\geq}_{\bm{\mu},j,\epsilon_m} 
) \right) \nonumber \\ &  \leq
\sum_{j=1}^d \left(  e^{-m \inf_{\bm{\mu}' \in
H^{\leq}_{\bm{\mu},j,\epsilon_m}} D_{\textrm{KL}}(P_{\bm{\mu}' } \| P_{\bm{\mu}}) \| )} + e^{-m 
\inf_{\bm{\mu}' \in
H^{\geq}_{\bm{\mu},j, \epsilon_m}} D_{\textrm{KL}}(P_{\bm{\mu}' } \| P_{\bm{\mu}}) \| )
}
\right) = O\left( m^{-a}\right).
\end{align}
Here the second inequality is the union bound and the third is once again Csisz\'ar's~\cite{Csiszar84} multivariate generalization of Chernoff's concentration inequality (in   (\ref{eq:suppie}),  we bounded $P_{\bm{\mu}}(\mle \in B)$ where $B$ was a convex set, allowing us to use Csisz\'ar's result directly; but in (\ref{eq:infie}), we need to bound $P_{\bm{\mu}}(\mle \not \in B)= P_{\bm{\mu}}(\mle \in {\tt M} \setminus B)$; since ${\tt M} \setminus B$ is not a convex set, yet convexity is required by Csisz\'r's result, we now first need to cover it by $2d$ convex sets $H^{\cdot}_{\bm{\mu},\cdot,\epsilon_m}$, for each of which we then use Csisz\'ar's result). The final inequality follows by using (\ref{eq:KLsquare}) again.

\newpage
\section{Pseudo approximation through the high resolution limit}
\label{SM_section_pseudo}

Below, we provide pseudocode to compute the density $w_{\text{pseudo},0}^1$ for a canonical test on $2\times 2$ contingency tables, under the alternative hypothesis with independent beta priors on the mean-value parameters. Notice that for $\alpha_a = \beta_a = \alpha_b = \beta_b = 1$, we retrieve the uniform prior of Example B. 

The procedure starts from the induced distribution on the sufficient statistics under the alternative, denoted $W_{\text{can},1}^{a}$ and $W_{\text{can},1}^{b}$ in the main text, which follow a beta-binomial form. To approximate the continuous limit, we increase the resolution by multiplying the original group sizes by a scaling factor (Steps 1–2). The higher the scaling factor, the better the approximation --- at the cost of greater computational effort.

The two high-resolution distributions are then convolved to obtain an approximation of the GRO-optimal prior on the null, $W_0^*$ (Step 3). Since we want a density over $p_0 \in [0,1]$, we define a pseudo-continuous support for $p_0$ accordingly (Step 4). Finally, the convolved distribution is normalized to form a proper density over this support (Step 5).

\begin{verbatim}
Input: 
    n_a, n_b               // group sizes
    scaling_factor         // resolution multiplier
    alpha_a, beta_a        // beta-binomial parameters for group a
    alpha_b, beta_b        // beta-binomial parameters for group b

Step 1: Define high-resolution support
    x_high_a = 0 to scaling_factor * n_a
    x_high_b = 0 to scaling_factor * n_b

Step 2: Compute high-resolution beta-binomial PMFs
    For each i in x_high_a:
        p_high_a[i] = BetaBinomialPMF(i, scaling_factor * n_a, alpha_a, beta_a)
    For each i in x_high_b:
        p_high_b[i] = BetaBinomialPMF(i, scaling_factor * n_b, alpha_b, beta_b)

Step 3: Convolve the two distributions
    conv_high = Convolve(p_high_a, p_high_b)

Step 4: Define pseudo-continuous support over [0, 1]
    p_0_fine = [0, 1, ..., len(conv_high)-1] / (scaling_factor * (n_a + n_b))

Step 5: Normalize the convolved distribution
    step = p_0_fine[1] - p_0_fine[0]
    conv_high = conv_high / (sum(conv_high) * step)

Output:
    conv_high   // approximate density over [0, 1]
    p_fine      // corresponding support
\end{verbatim}

\newpage
\section{Bound on regret}
\label{SM_bound_on_regret}
Here we provide a bound for the regret of a maximum entropy model \eqref{regret}.   

Let ${\cal M}_0$ be a maximum entropy model. We begin by recalling an extension of the classical $(d/2) \log m$ asymptotic redundancy result (see Equation~\eqref{eq:bic}) that remains valid even when the model ${\cal M}_0$ is misspecified --- i.e., it does not contain the true distribution $P^*$. This generalization appears in~\cite{grunwaldbook2007}, and is formalized in~\cite[Proposition 3]{hao2024evaluesexponentialfamiliesgeneral}.

Let $\textbf{x}^m$ be an i.i.d.\ sample from a distribution $P^*$ that may lie outside ${\cal M}_0$. Suppose that there exists a distribution $P_{\tilde{\bm{\theta}}_0} \in {\cal M}_0$ minimizing the Kullback-Leibler divergence to $P^*$:
\[
P_{\tilde{\bm{\theta}}_0} = \arg\min_{\bm{\theta}_0 \in \bm{\Theta}_0} D_{\textrm{KL}}(P^* \| P_{\bm{\theta}_0}).
\]
Then, for any regular prior density $w_0$, we have:
\begin{multline}
\label{eq:bicb}
\text{RED}_0(P^*; P_0^{w_0}) := \mathbb{E}_{P^*} 
\left[  - \log {P}_0^{w_0}(\mathbf{x}^m) + \log P_{\tilde{\bm{\theta}}_0}(\mathbf{x}^m) \right] 
\\ =  \frac{d_0}{2} \log m +O(1).
\end{multline}
In the well-specified case where $P^* \in {\cal M}_0$, it holds that $P^* = P_{\tilde{\bm{\theta}}_0}$, and $\text{RED}_0(P^*; P_1^{w_0})$ coincides with our earlier definition $\text{RED}_0(\tilde{\bm{\theta}}_0, P_1^{w_0})$ (up to notation), thus recovering the classical result \eqref{eq:bic}.

We now apply this general result in the setting where $P^* = P_{\bm{\theta}_1} \in {\cal M}_1$, in order to bound the regret $\text{REG}(\bm{\theta}_1; P^{w_1}_1 )$ of the Bayesian marginal $P^{w_1}_1$ for a regular prior $w_1$. Consider the pseudo-e-variable $S_{\text{pseudo}}$, used as a proxy for either $S^{\text{GRO}}_{\text{mic}}$ or $S^{\text{GRO}}_{\text{can}}$. Using the previous result to refine equation~\eqref{eq:upperbound}, we obtain:
\begin{align}
\text{REG}&(\bm\theta_1; S_{\text{pseudo}}) 
= 
\mathbb{E}_{\bm{\theta}_1} \left[ \log S^{\text{GRO}(\bm{\theta}_1)} - \log \frac{{P}^{w_1}_1(\mathbf{x})}{P_0^{w_{\text{pseudo},0}}(\mathbf{x})} \right] \nonumber \\
&= 
\mathbb{E}_{\bm{\theta}_1} \left[ \log \frac{
P^{w_{\text{pseudo},0}}_0(\mathbf{x})}{P_0^{\tilde{w}'_0}(\mathbf{x})} \right]+ \text{RED}_1(\bm{\theta}_1; P_1^{w_1}) \nonumber \\
&= 
\mathbb{E}_{\bm{\theta}_1} \left[ \log \frac{P^{w_{\text{pseudo},0}}_0(\mathbf{x})}{P_{\bm{\tilde\theta}_0}(\mathbf{x})} \right] + \text{RED}_1(\bm{\theta}_1; P_1^{w_1})   \nonumber \\ 
&= \text{RED}_1(\bm{\theta}_1; P_1^{w_1}) - \text{RED}_0(P_{\bm{\theta}_1}; P_0^{w_\text{pseudo, 0}}) + O(1) \nonumber \\ 
\label{SM_eq:upperboundb}
& \overset{(a)}{=}  \frac{d_1 - d_0}{2} \cdot \log m + O(1).
\end{align} 
The first two equalities are direct, and the third follows from~\cite[Theorem 3 (Parts 3,4)]{hao2024evaluesexponentialfamiliesgeneral}. Equality $(a)$ holds provided that $w_{\text{pseudo},0}$ is regular, in particular, that it has a continuous density.

\section{Supplementary tables and figures}
\begin{table}[ht]
    \centering
    \renewcommand{\arraystretch}{1.3}
    \begin{tabular}{|c|c|c|c|}
        \hline
        \textbf{Case} & \textbf{Parameter Condition} & \textbf{Behavior}  \\
        \hline
        \textbf{Uniform} & $\alpha = \beta = 1$ & Flat distribution \\
        \hline
        \textbf{Jeffreys Prior} & $\alpha = \beta = 0.5$ & U-shaped  \\
        \hline
        \textbf{Bimodal (U-shaped)} & $\alpha, \beta < 1$ & Peaks at 0 and 1 \\
        \hline
        \textbf{Left-skewed} & $\alpha > 1, \beta < 1$ & Peak near 1  \\
        \hline
        \textbf{Right-skewed} & $\alpha < 1, \beta > 1$ & Peak near 0  \\
        \hline
        \textbf{Bell-shaped} & $\alpha, \beta > 1$ & Normal-like \\
        \hline
        \textbf{Highly concentrated} & $\alpha = \beta \gg 1$ & Sharp peak  \\
        \hline
        \textbf{Degenerate (Dirac Delta)} & $\alpha, \beta \to \infty$ & Point mass at $\frac{\alpha}{\alpha + \beta}$ \\
        \hline
    \end{tabular}
    \caption{Behaviors of the beta distribution for different parameter values.}
    \label{SM_tab_beta_distribution}
\end{table}

\begin{figure}[ht]
\includegraphics[width=\textwidth
]{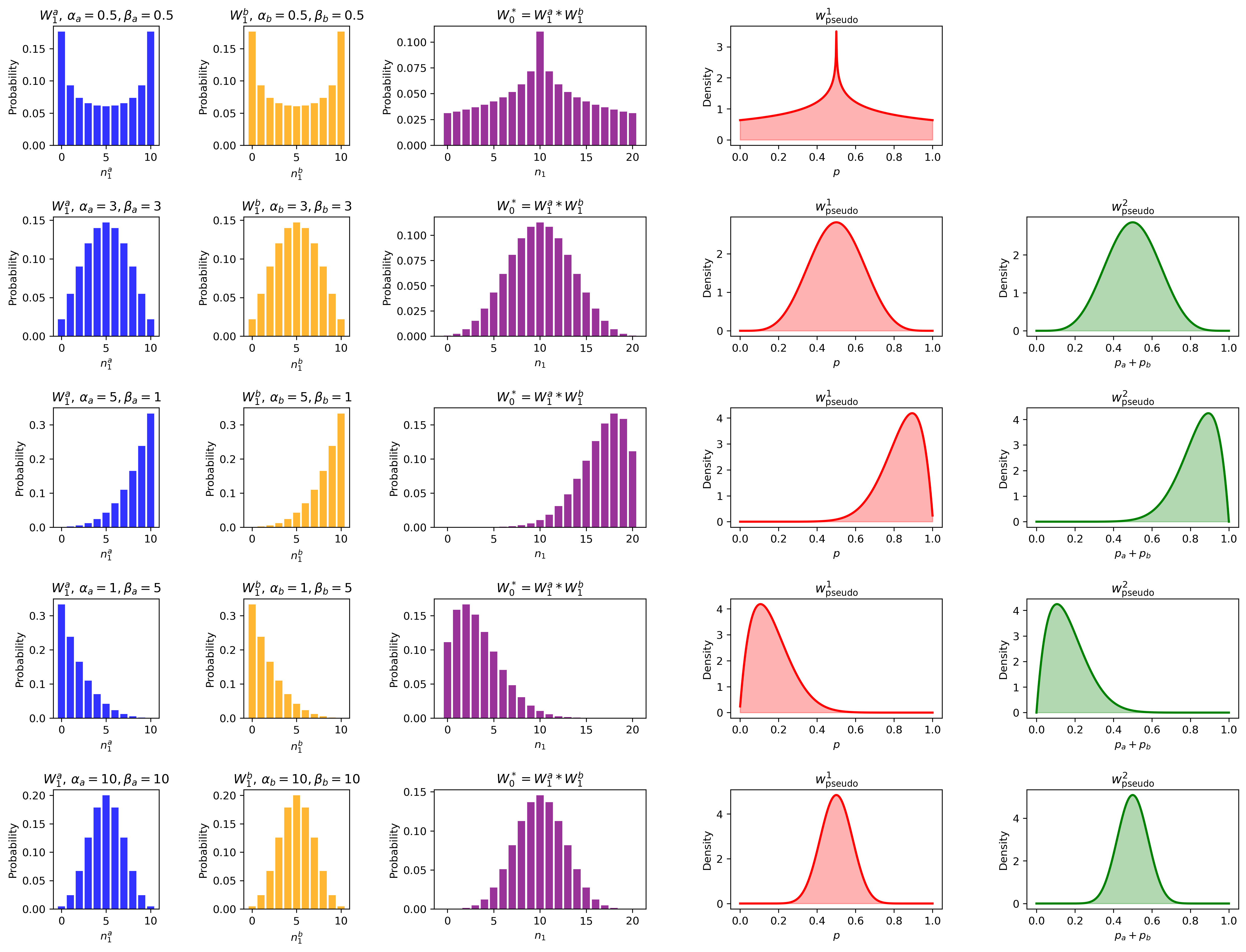}
\caption{Construction of the microcanonical optimal prior $W_0^*$ (third column) and the pseudo-prior approximations $w^1_{\text{pseudo},0}$ (fourth column) and $w^2_{\text{pseudo},0}$ (fifth column), for $2\times 2$ contingency tables with independent beta priors on the alternative. Starting from the induced independent beta-binomial distribution on the alternative sufficient statistics (first and second columns), the microcanonical GRO-optimal prior $W_0^*$ is obtained as their convolution. The pseudo prior approximation $w^1_{\text{pseudo},0}$ is obtained starting from $W_0^*$ through a high-resolution limit. The pseudo prior approximation $w^2_{\text{pseudo},0}$ is obtained by directly convoluting the original continuous beta priors, when well defined on the whole parameter space (i.e., for $\gamma \geq 1)$.}
\label{SM_fig_beta_convolutions_2x2}
\end{figure}

\begin{figure}[ht]
    \centering
    \includegraphics[width=0.6\textwidth]{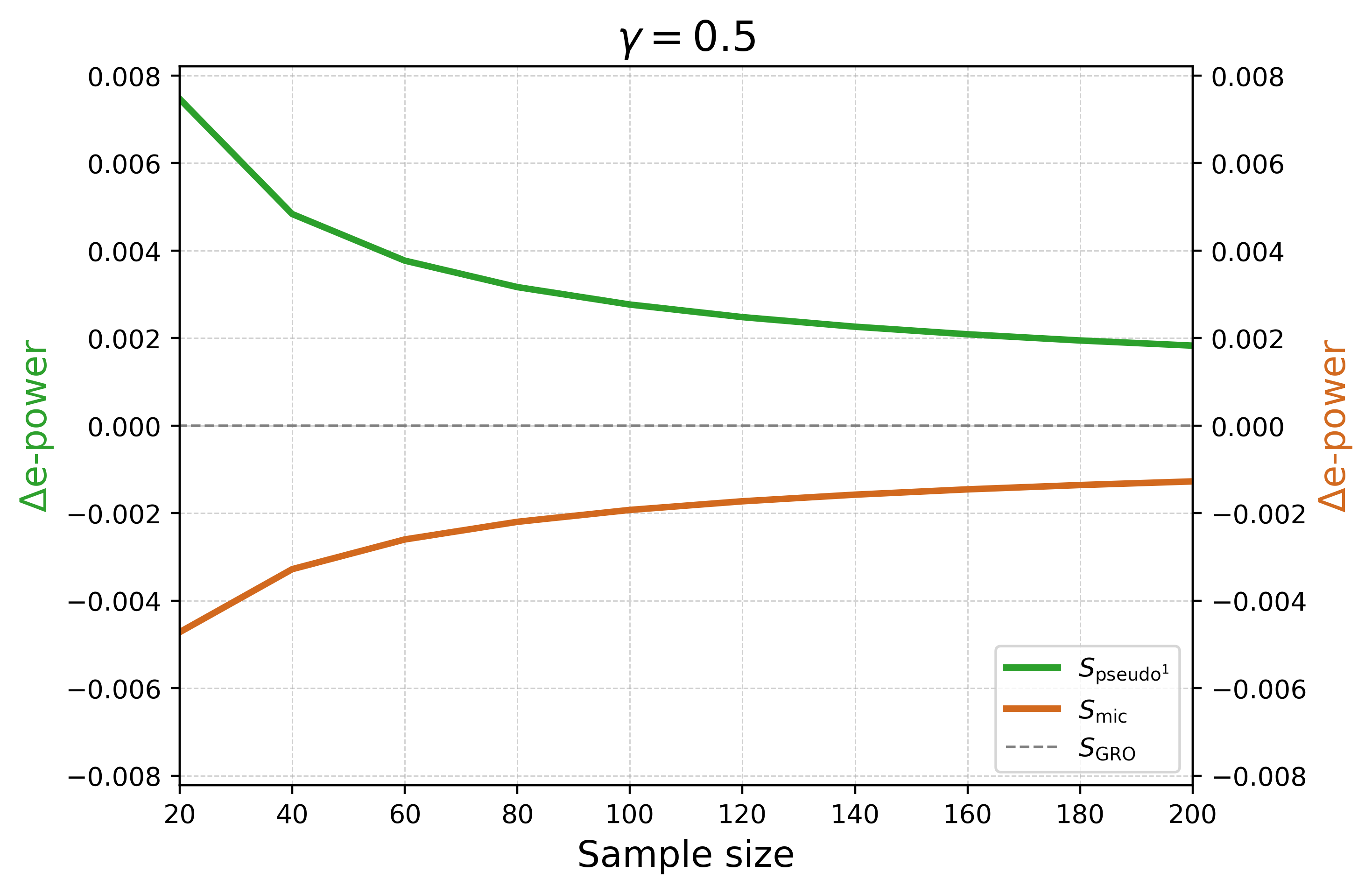} \\[2pt]
    \includegraphics[width=0.6\textwidth]{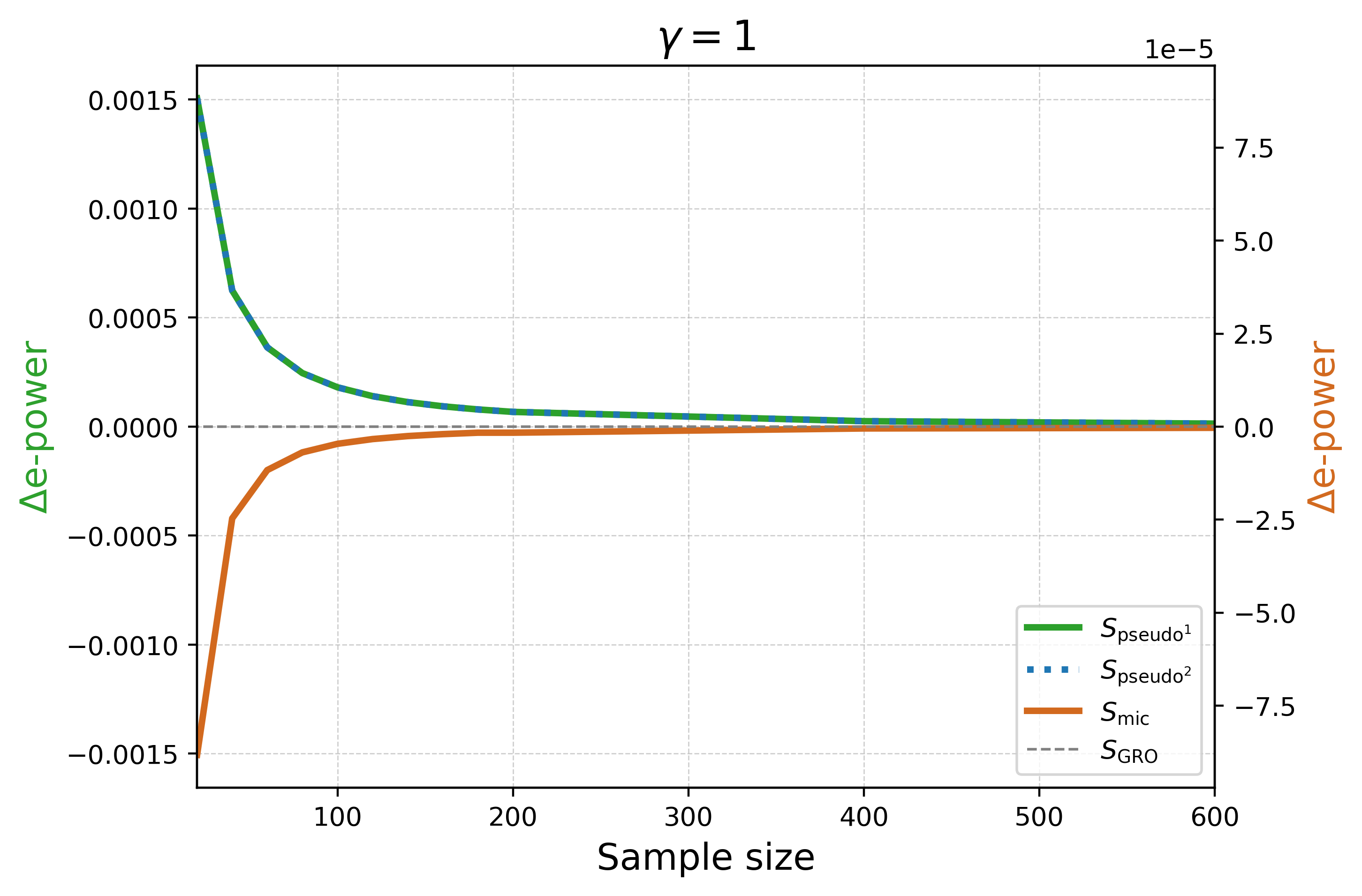} \\[2pt]
    \includegraphics[width=0.6\textwidth]{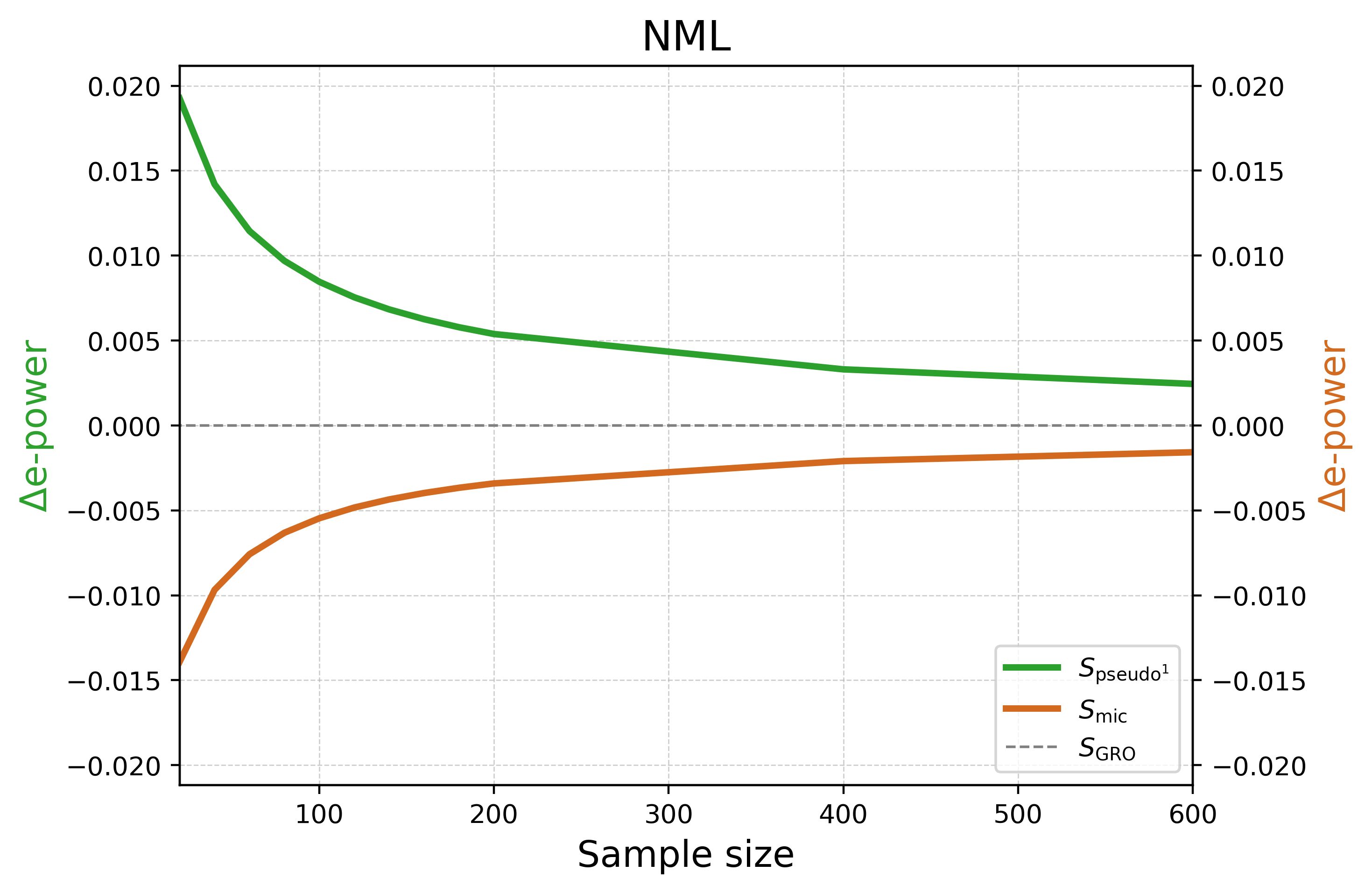} \\[2pt]
    \caption{E-power difference between the canonical GRO e-variable (computed numerically), its microcanonical approximation (orange curve), and the pseudo approximation (green curve), across sample sizes and different choices on the alternative (NML and beta with all parameters equal to $\gamma$). The microcanonical and pseudo approximations provide a lower and upper bound for the canonical GRO e-power, converging to it as the sample size grows. Results are shown for the $2\times 2$ contingency tables canonical test with $n^a = n^b = m$ and sample size equal to $2m$.}
    \label{fig:epowerdifference_2x2}
\end{figure}

\begin{figure}[ht]
    \centering
    \begin{minipage}{0.45\textwidth}
        \centering
        \includegraphics[width=\textwidth]{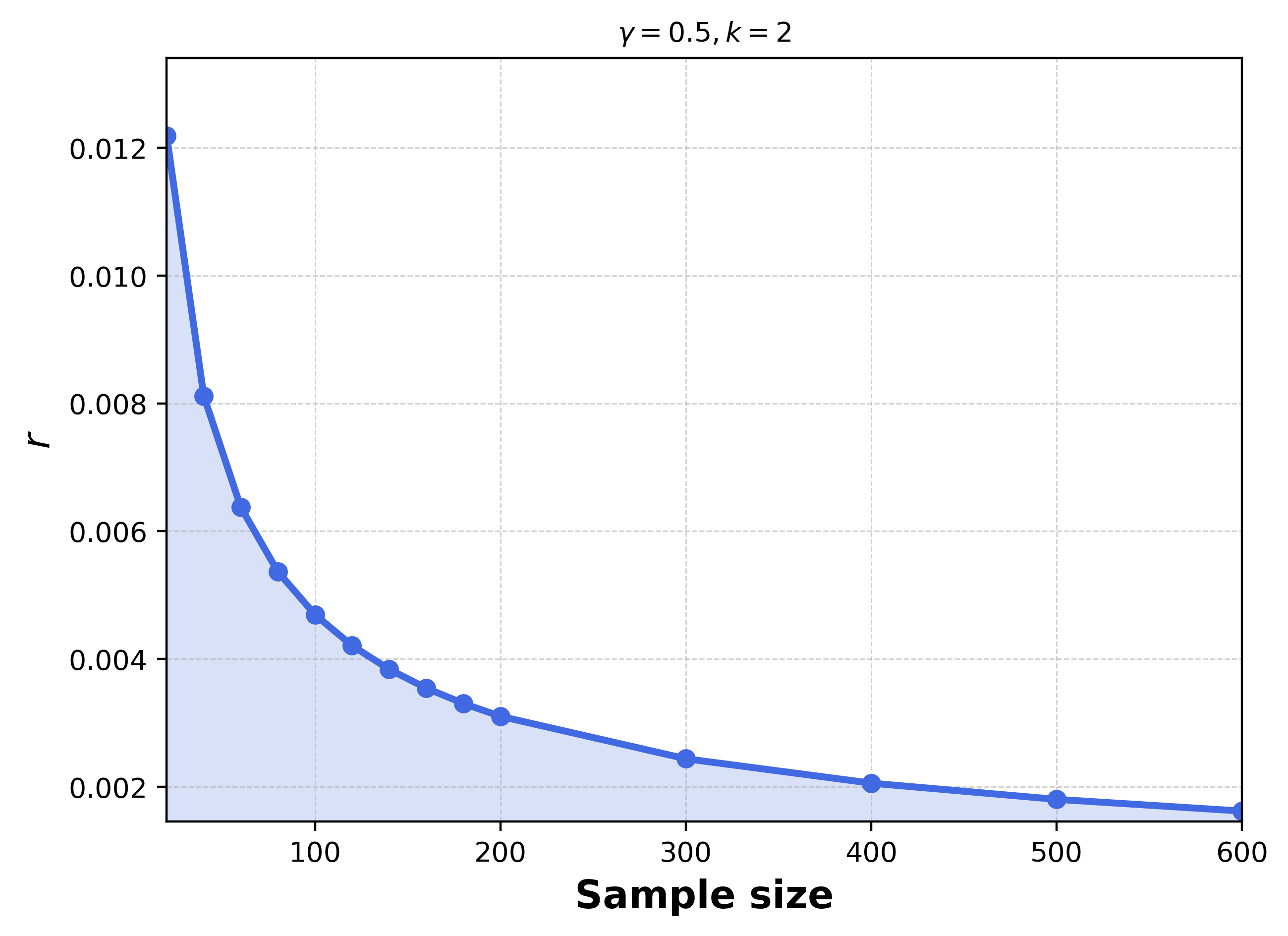} \\[2pt]
        \includegraphics[width=\textwidth]{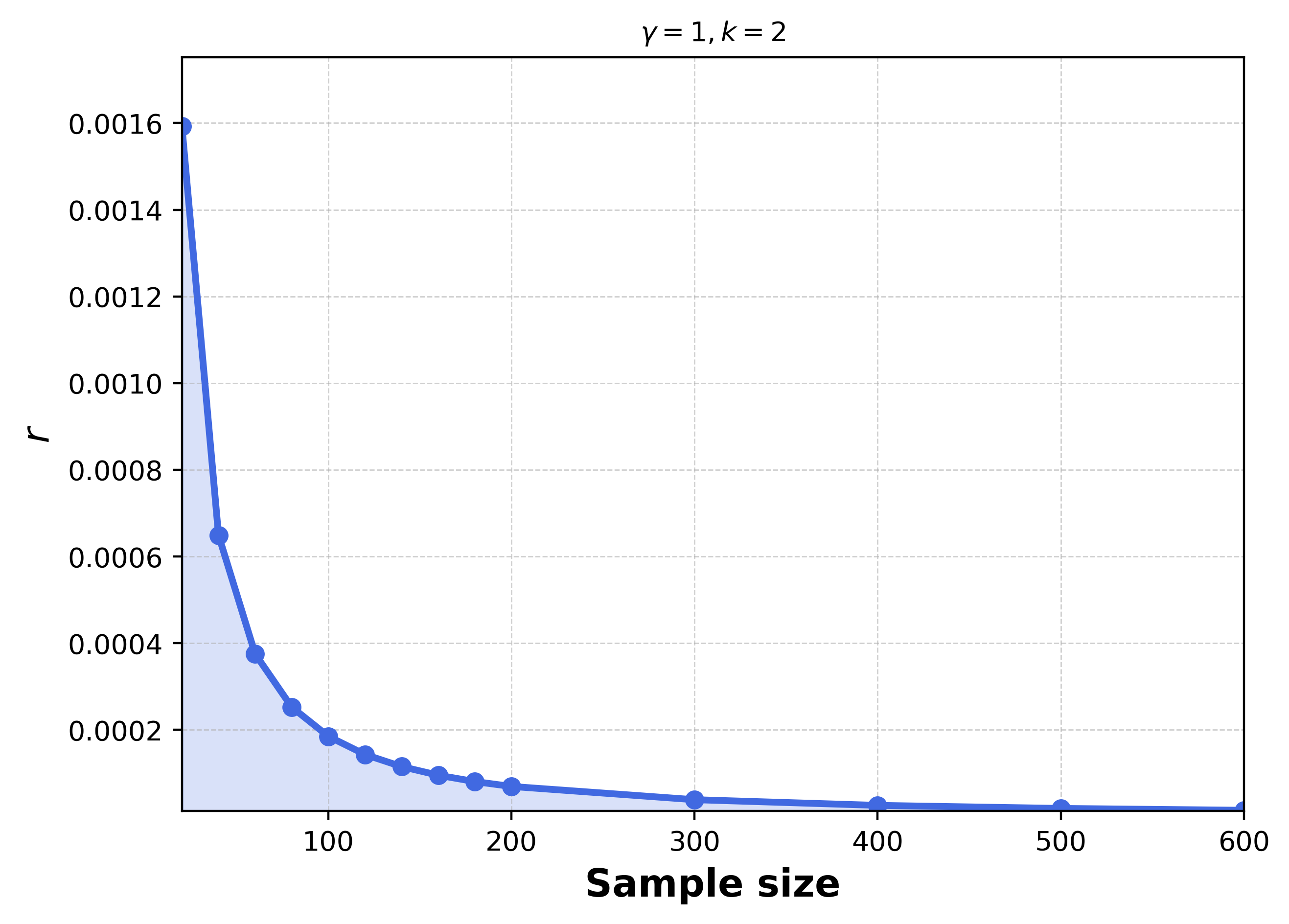} \\[2pt]
        \includegraphics[width=\textwidth]{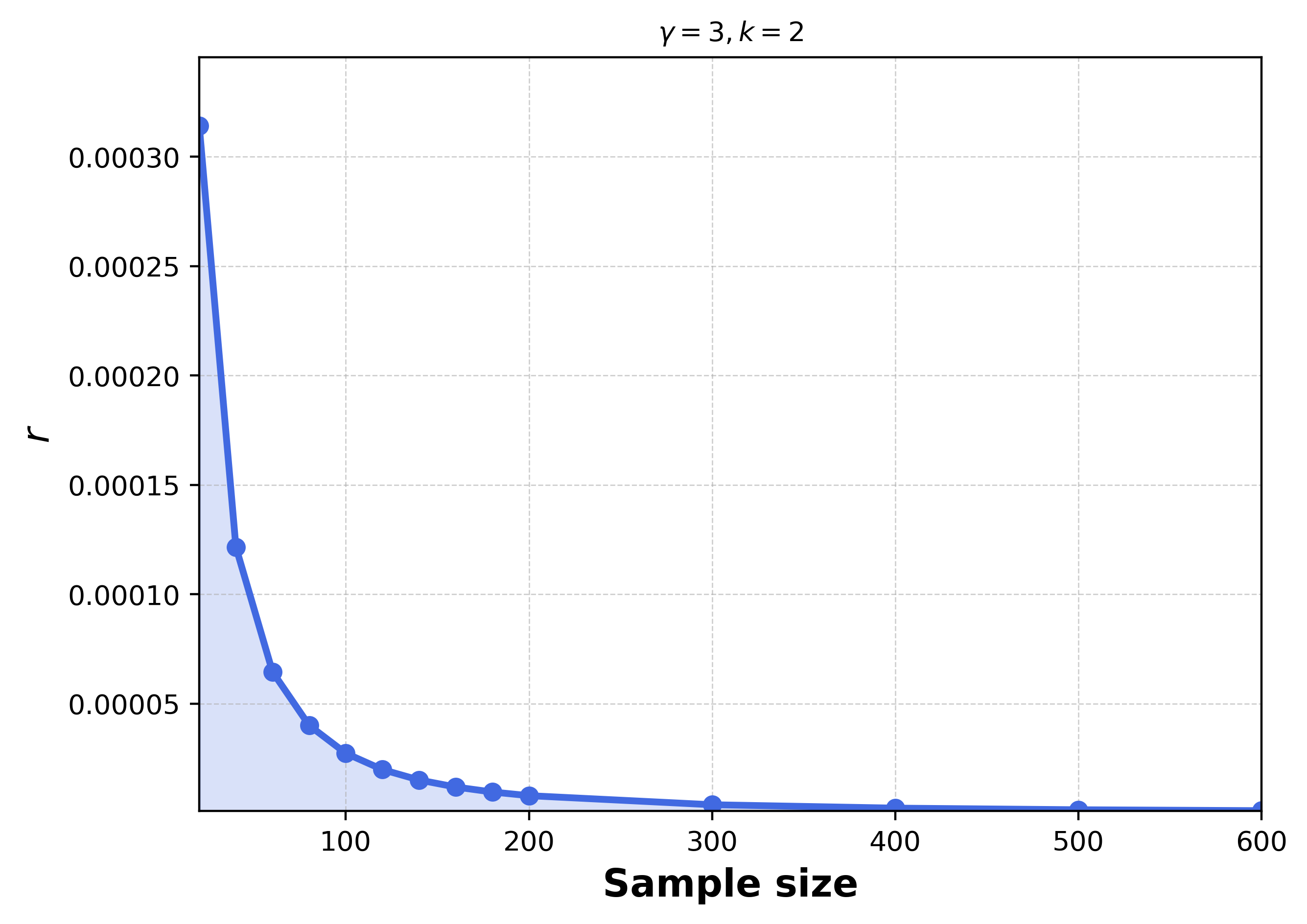} \\[2pt]
        \includegraphics[width=\textwidth]{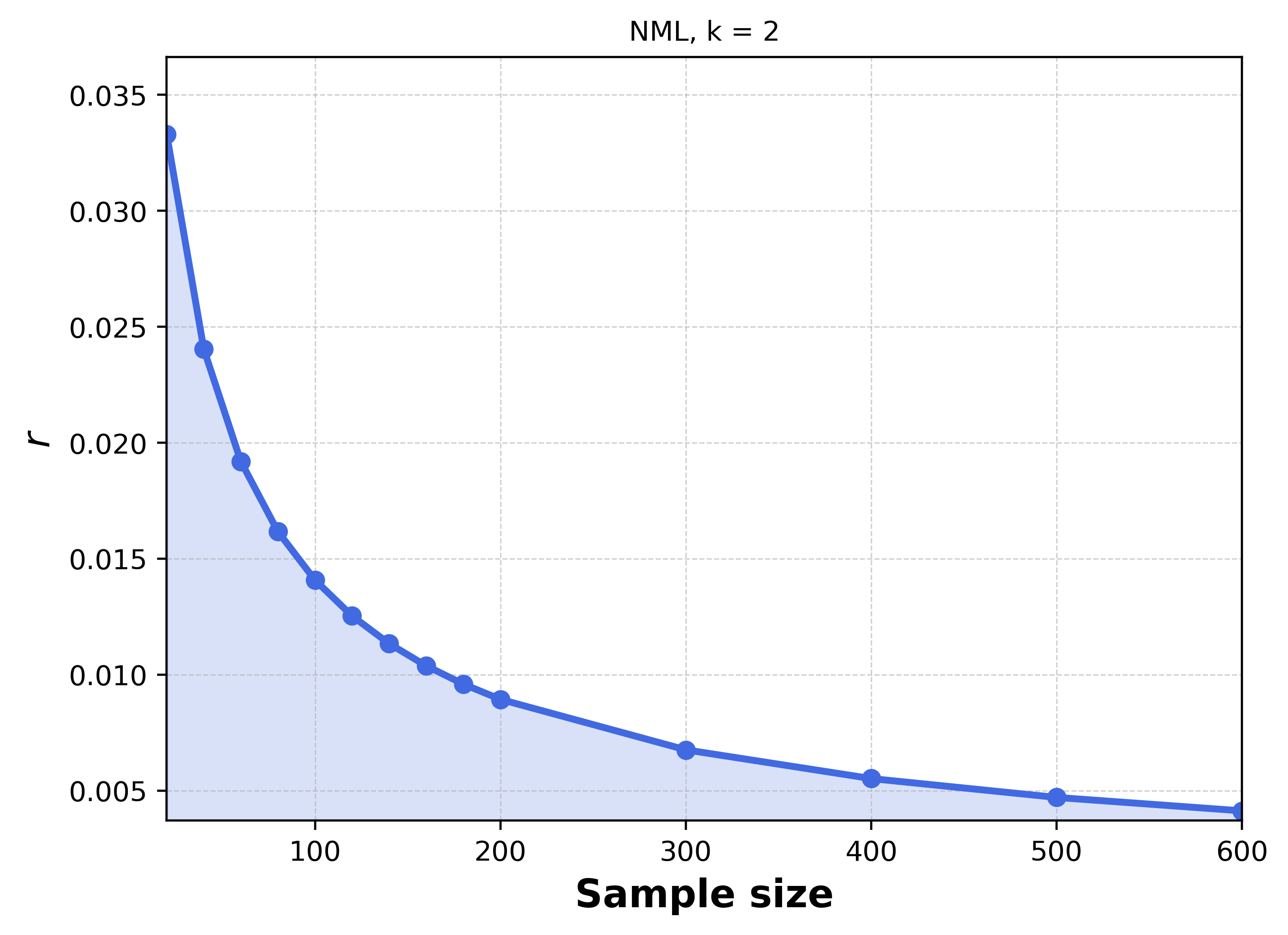}
    \end{minipage}
    \hfill
    \begin{minipage}{0.45\textwidth}
        \centering
        \includegraphics[width=\textwidth]{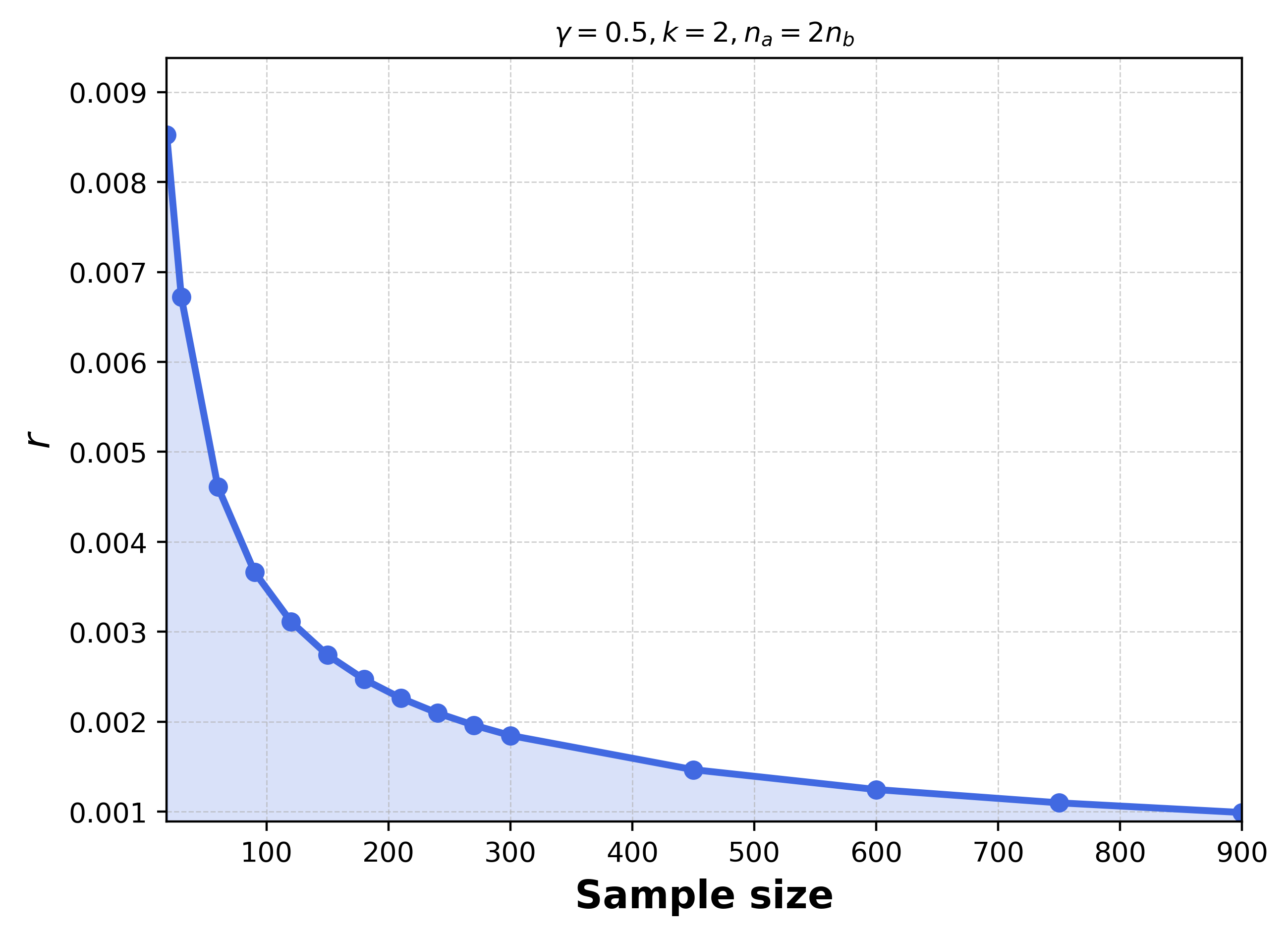} \\[2pt]
        \includegraphics[width=\textwidth]{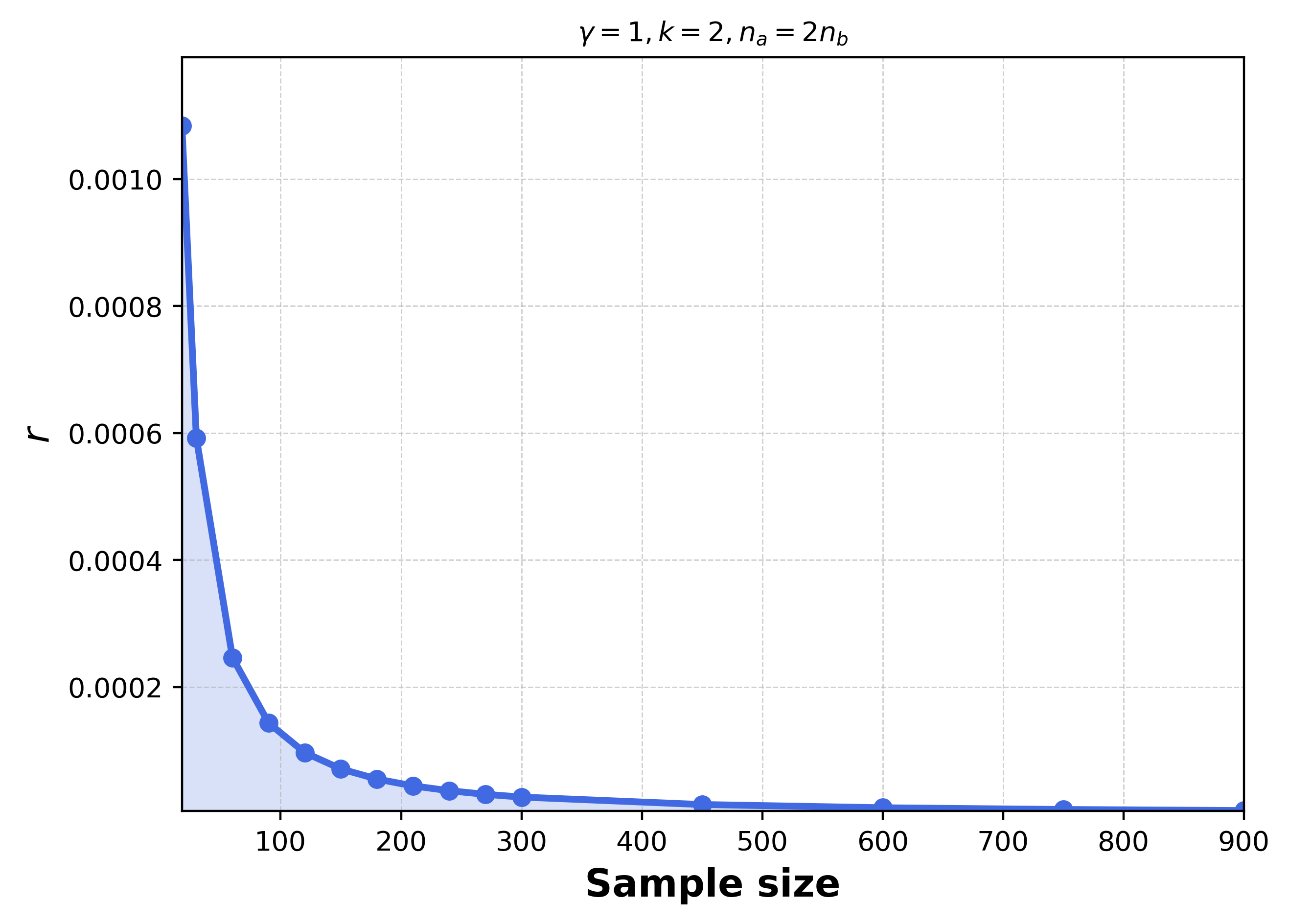} \\[2pt]
        \includegraphics[width=\textwidth]{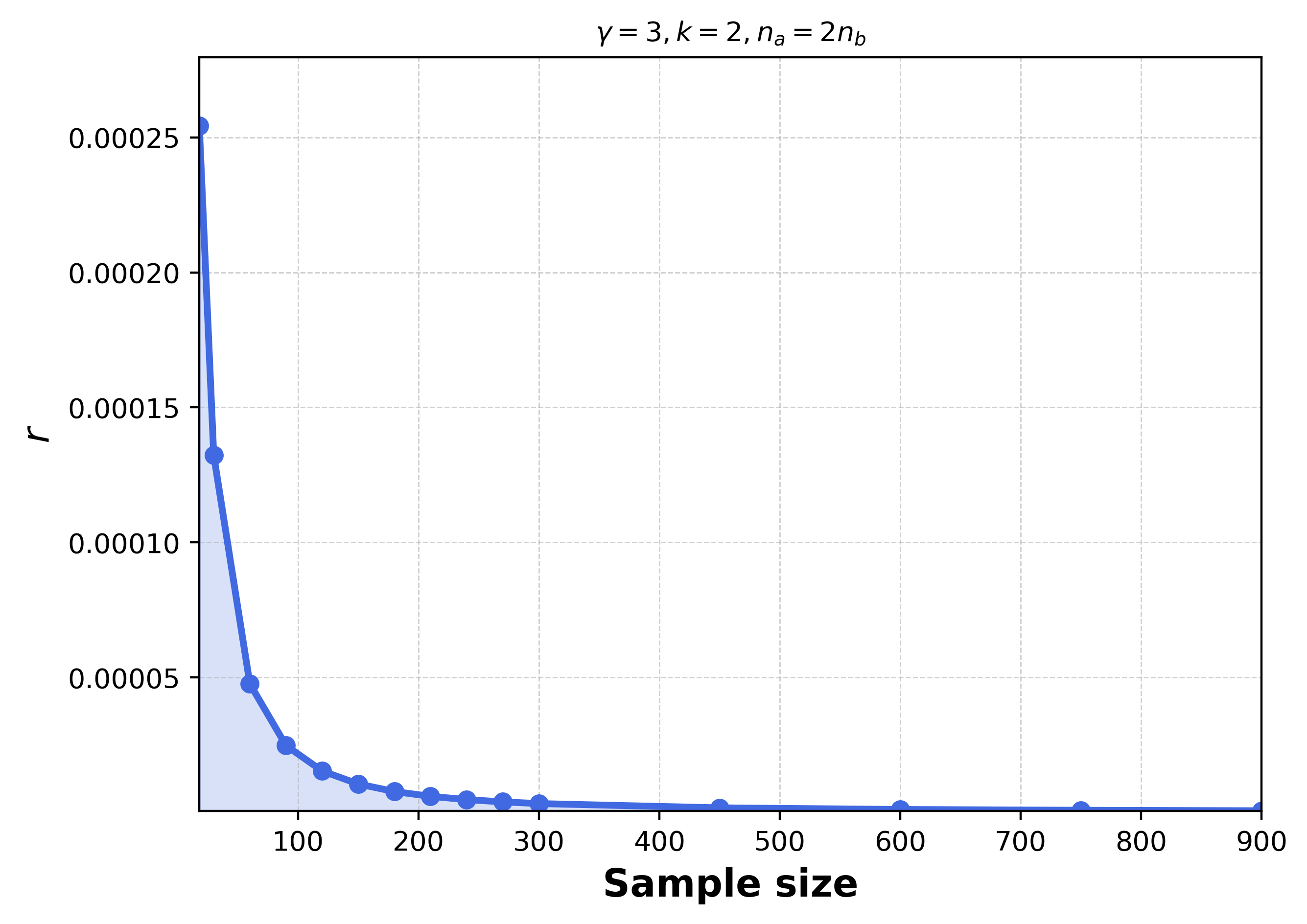} \\[2pt]
        \includegraphics[width=\textwidth]{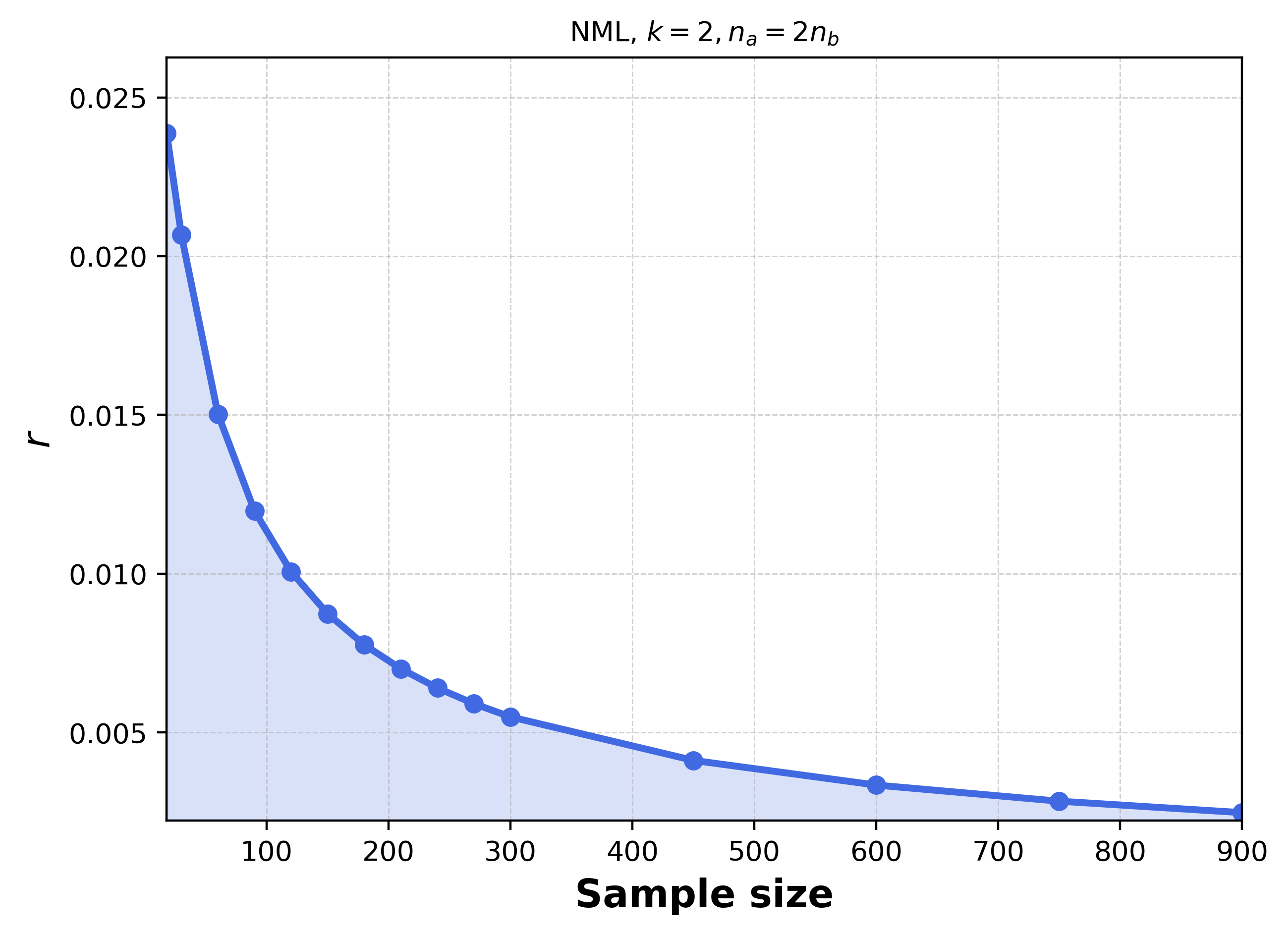}
    \end{minipage}
    \caption{Convergence of $r$ in $2\times 2$ contingency tables, for $n^a = n^b$ (left column) and $n^a = 2n^b$ (right column), as the sample size $n = n^a + n^b$ grows. Results are shown for different choices of $\bar{P}_{\text{can},1}$: beta independent priors with all parameters equal to $\gamma = 0.5$ (first row), $1$ (second row), $3$ (third row), and NML (fourth row).}
    \label{fig:r_2x2}
\end{figure}

\begin{figure}[ht]
    \centering
    \begin{minipage}{0.45\textwidth}
        \centering
        \includegraphics[width=0.9\textwidth]{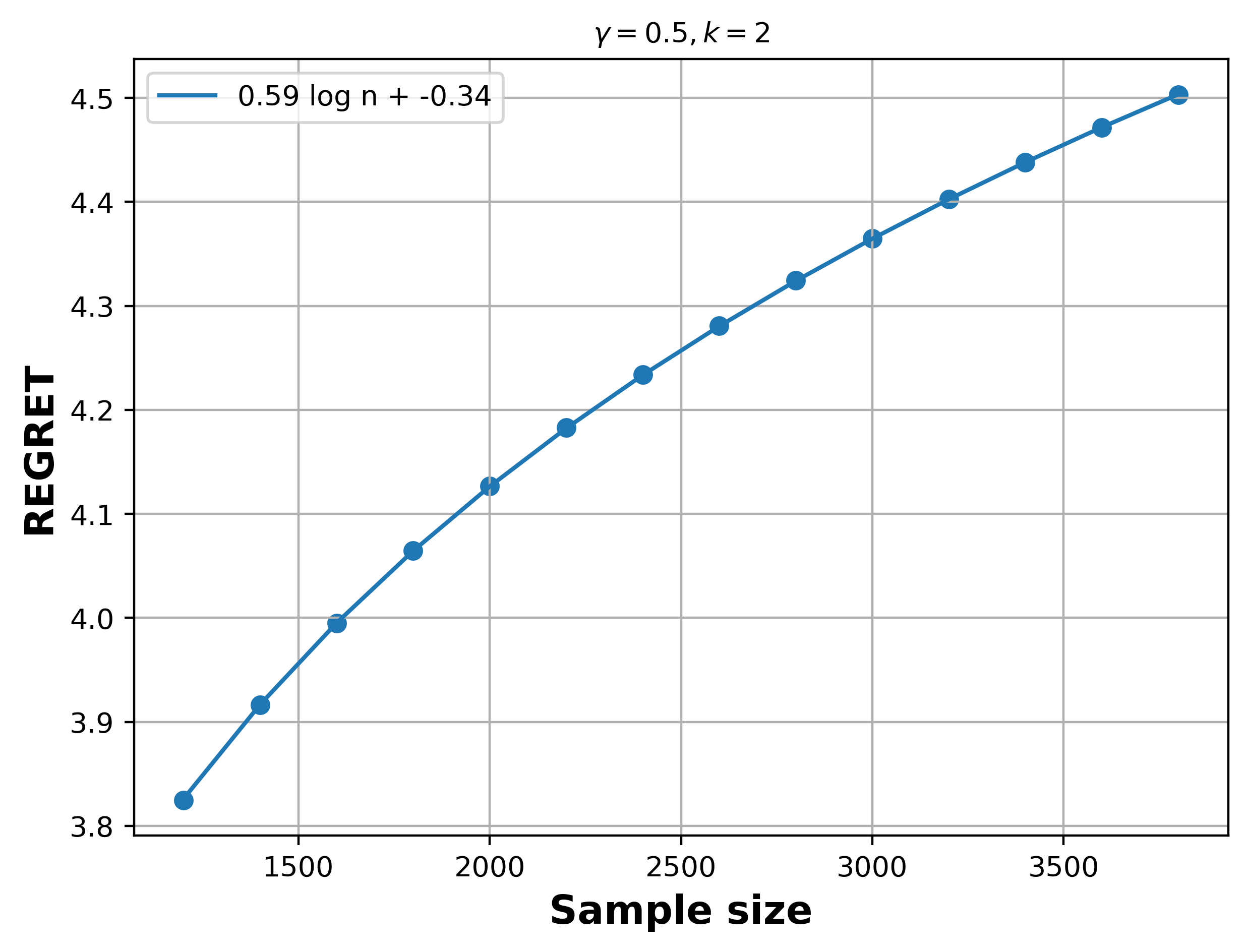} \\[2pt]
        \includegraphics[width=0.9\textwidth]{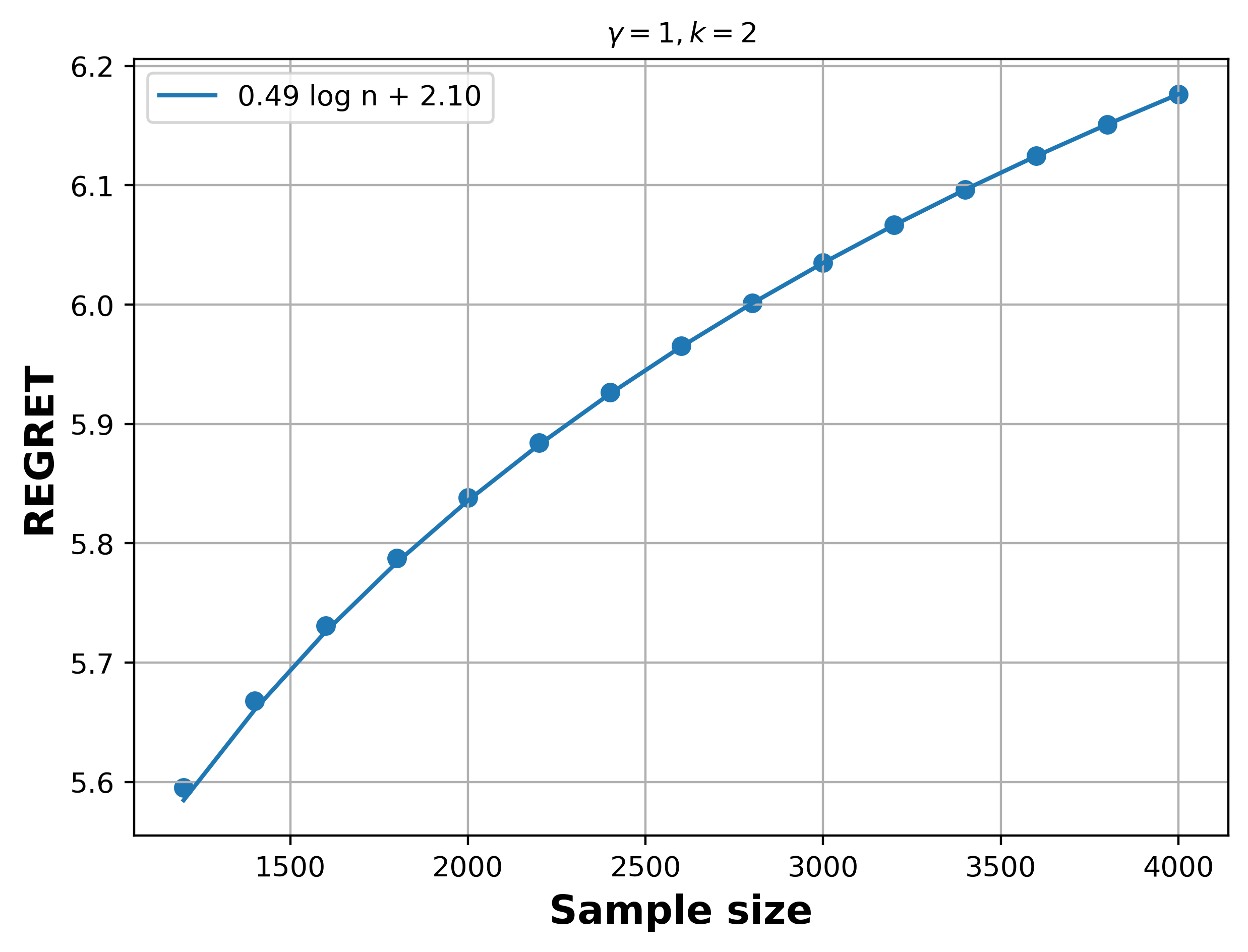} \\[2pt]
        \includegraphics[width=0.9\textwidth]{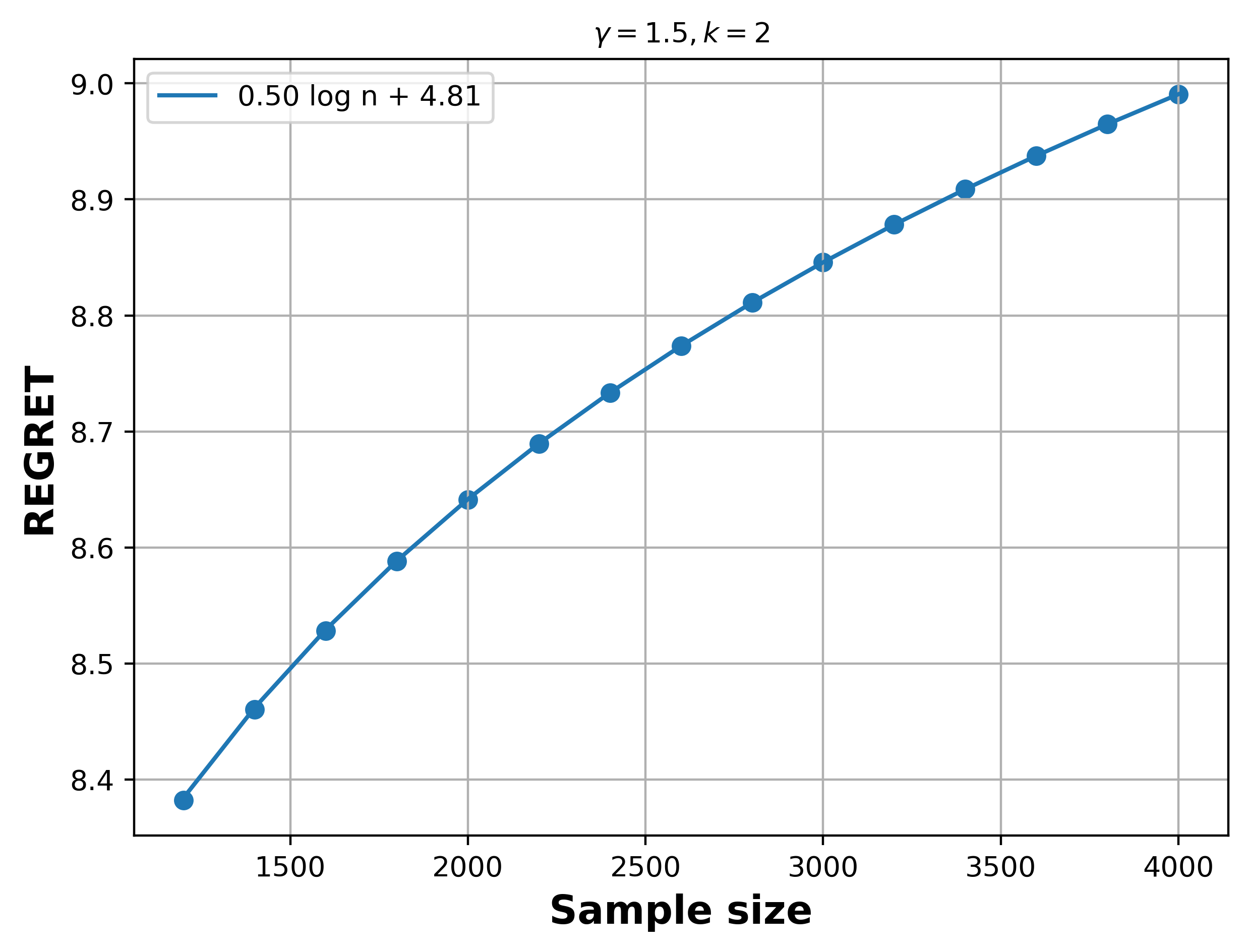} \\[2pt]
        \includegraphics[width=0.9\textwidth]{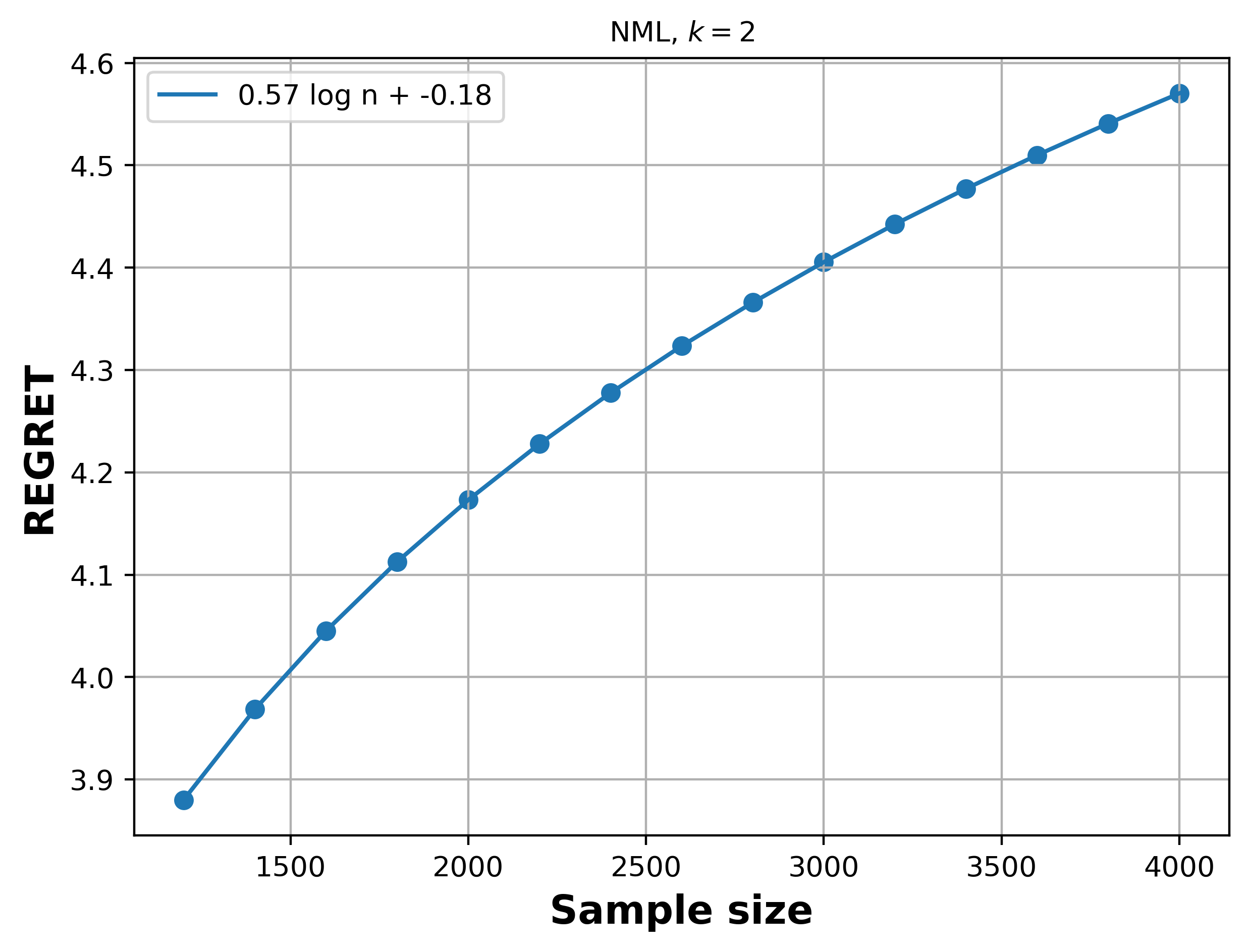}
    \end{minipage}
    \hfill
    \begin{minipage}{0.45\textwidth}
        \centering
        \includegraphics[width=0.9\textwidth]{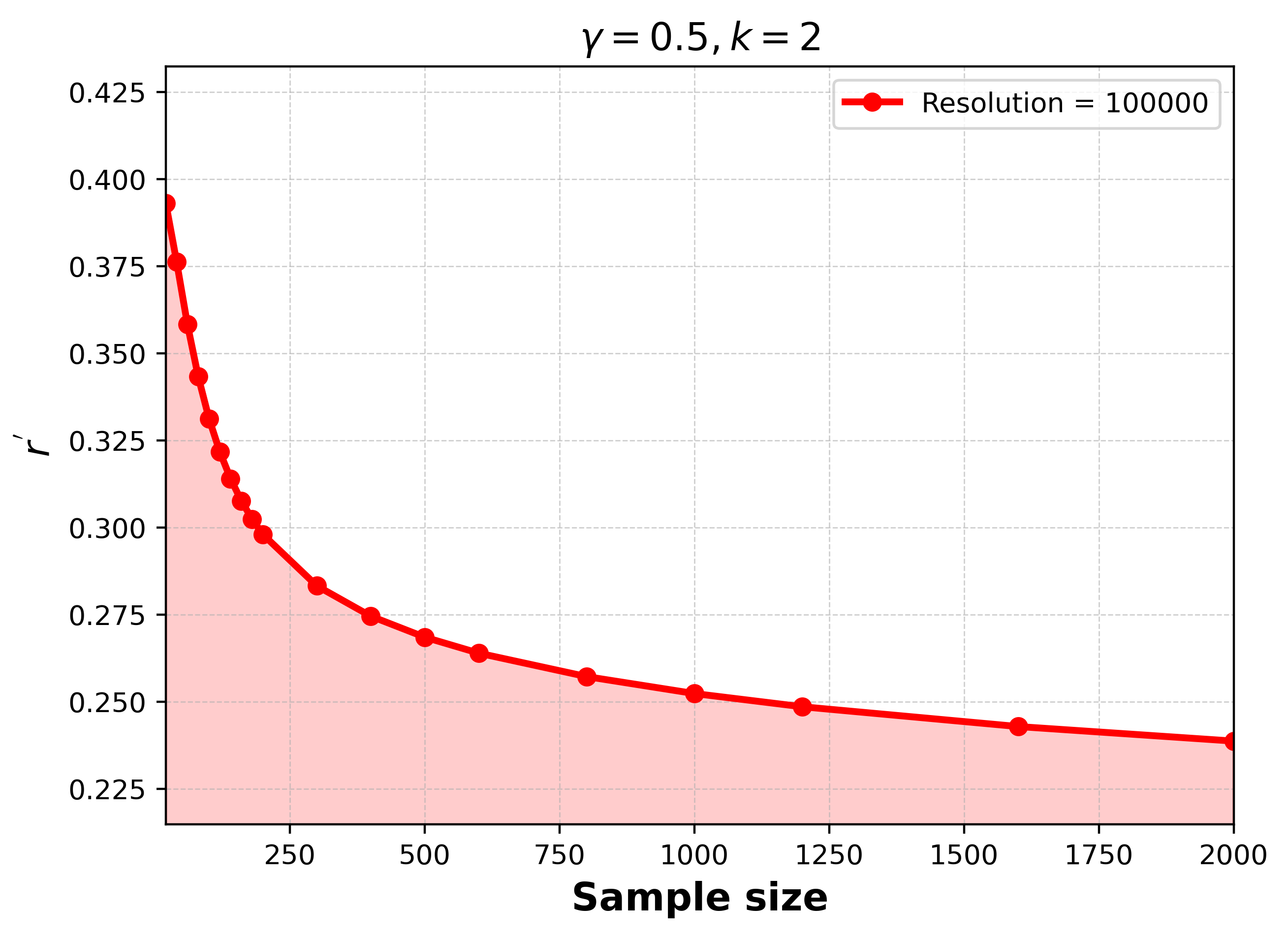} \\[2pt]
        \includegraphics[width=0.9\textwidth]{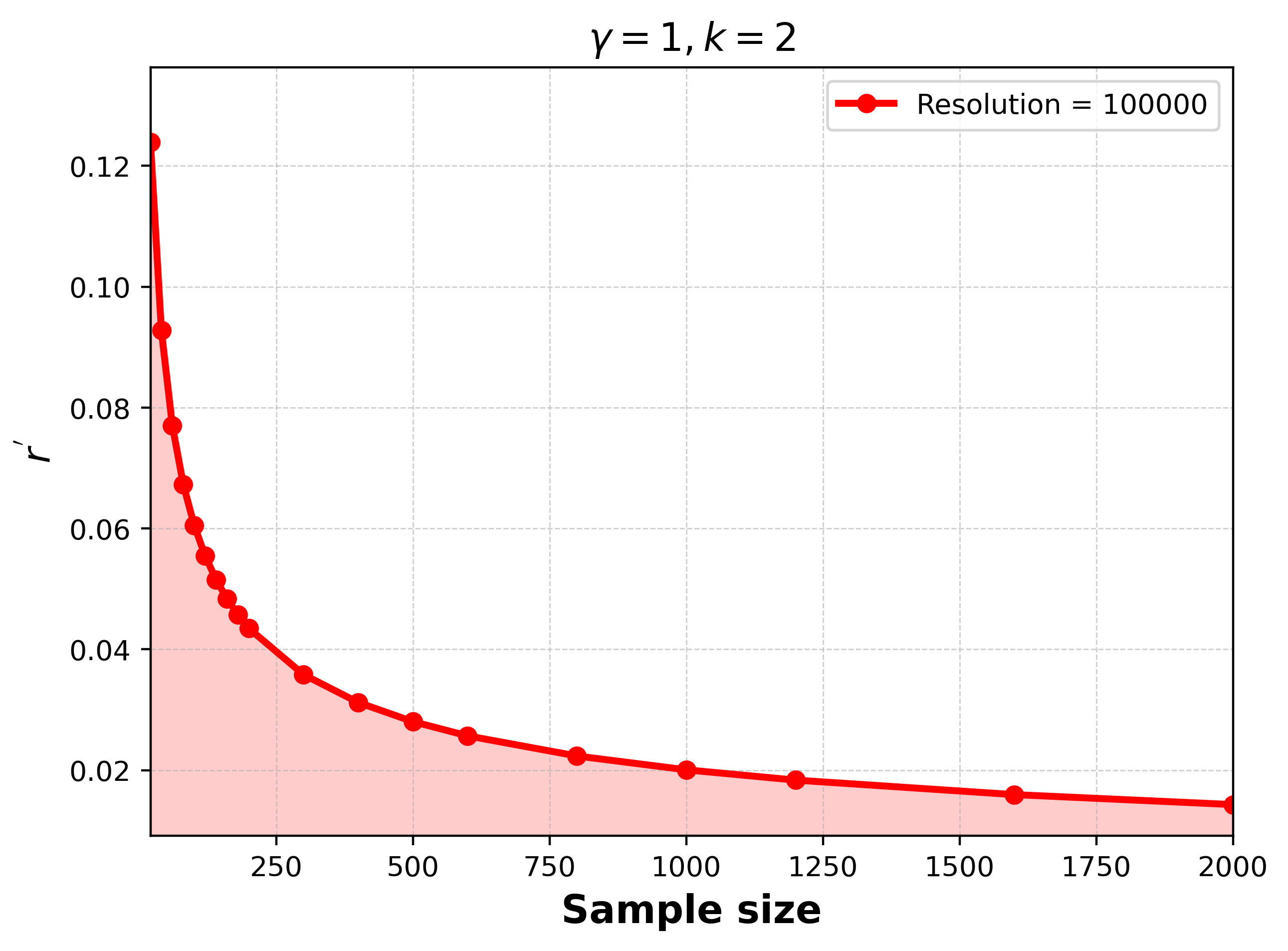} \\[2pt]
        \includegraphics[width=0.9\textwidth]{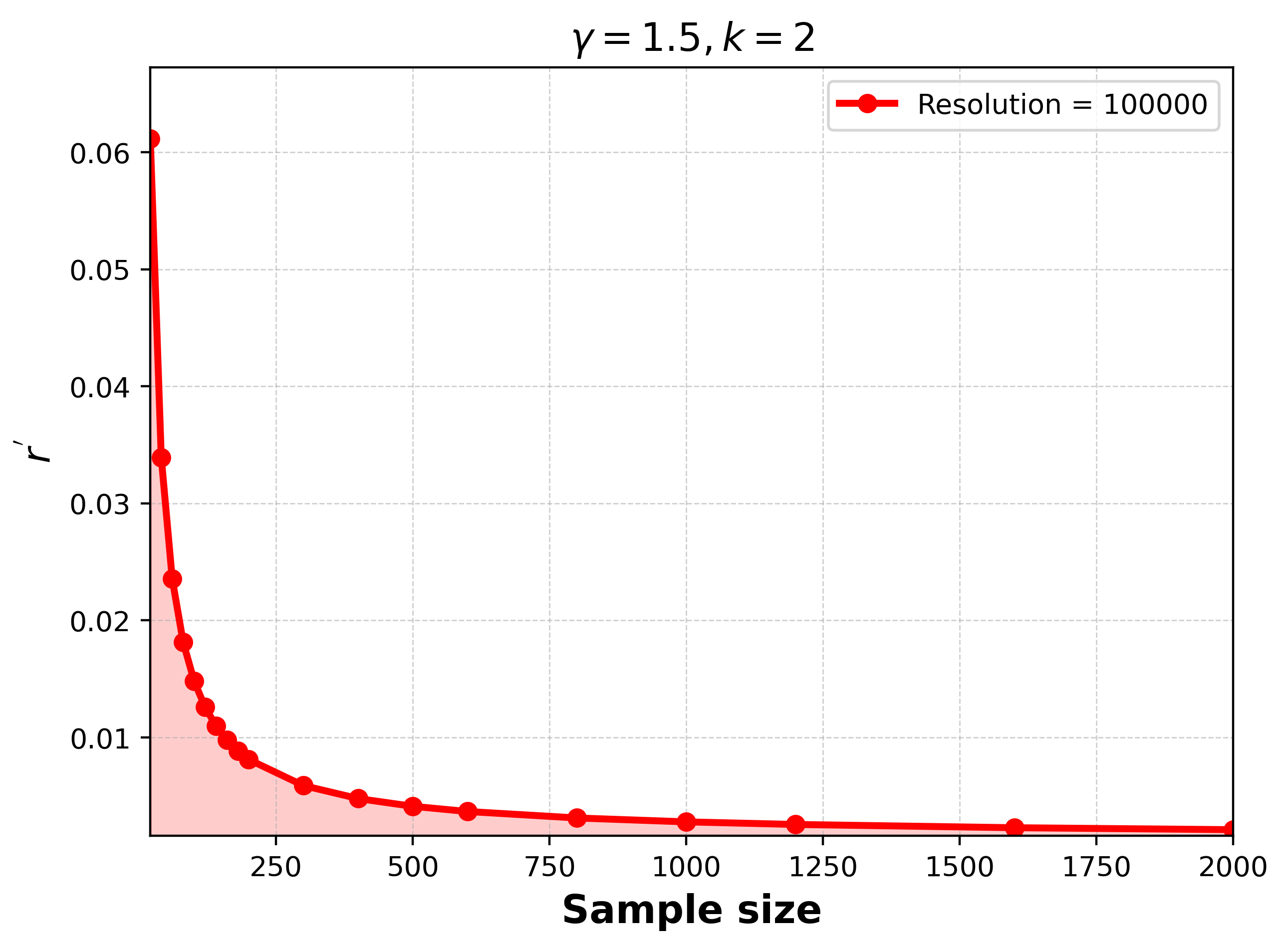} \\[2pt]
        \includegraphics[width=0.9\textwidth]{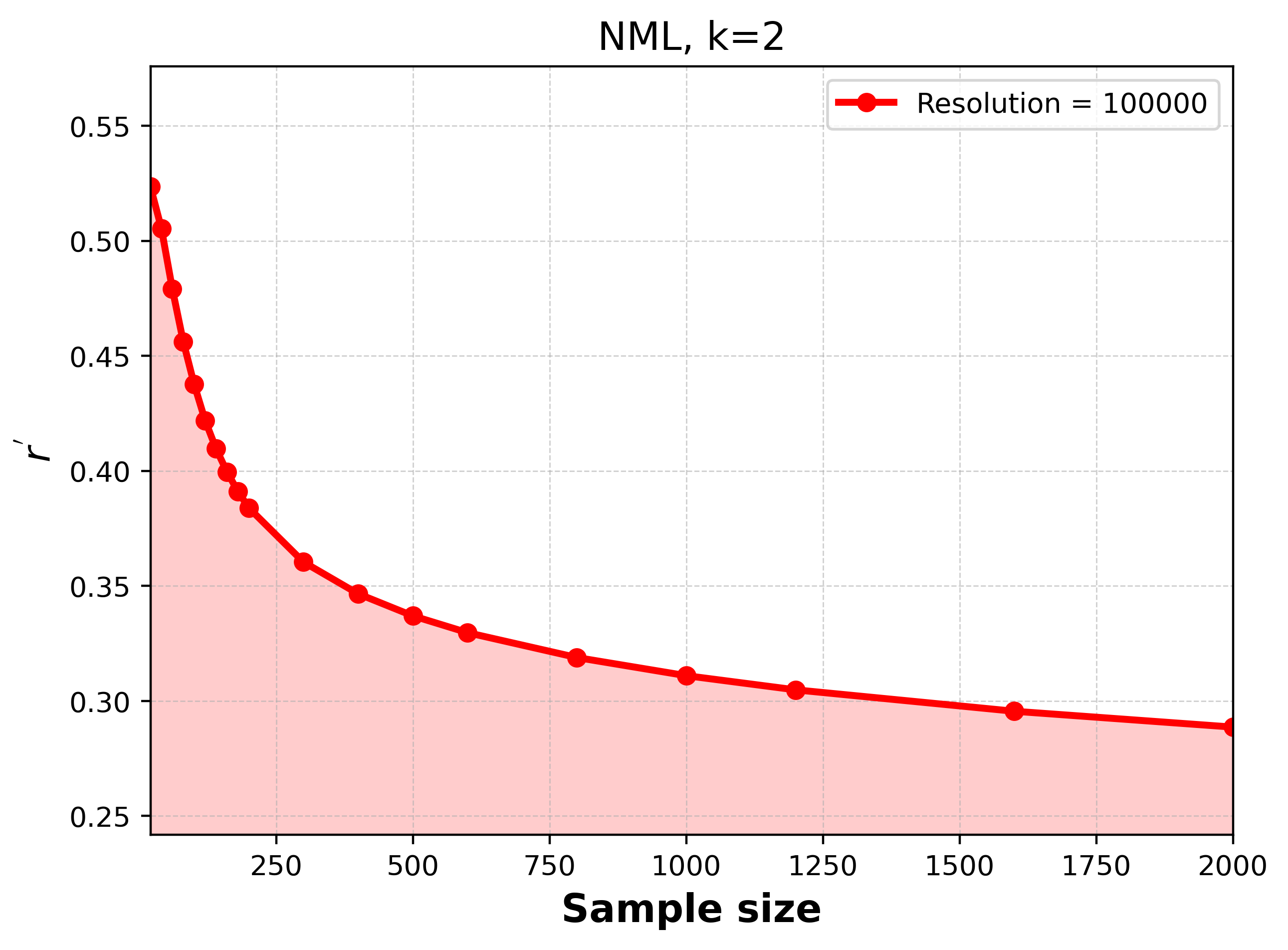}
    \end{minipage}
    \caption{Worst case regret (left) and convergence of $r'$ (right), for $2 \times 2$ tables with $n^a = n^b = m$ and sample size equal to $n = 2m$. Different choices of the alternative are considered: Bayesian with identical independent beta priors $B(\gamma, \gamma)$ with $\gamma = 0.5, \, 1, \, 1.5$, and NML. $r'$ is computed by considering $w^1_{\text{pseudo},0}$, obtained through a high resolution limit with resolution scale equal to $100000$.}
    \label{SM_fig: regret_r'}
\end{figure}

\begin{figure}[H]
\includegraphics[width=\textwidth
]{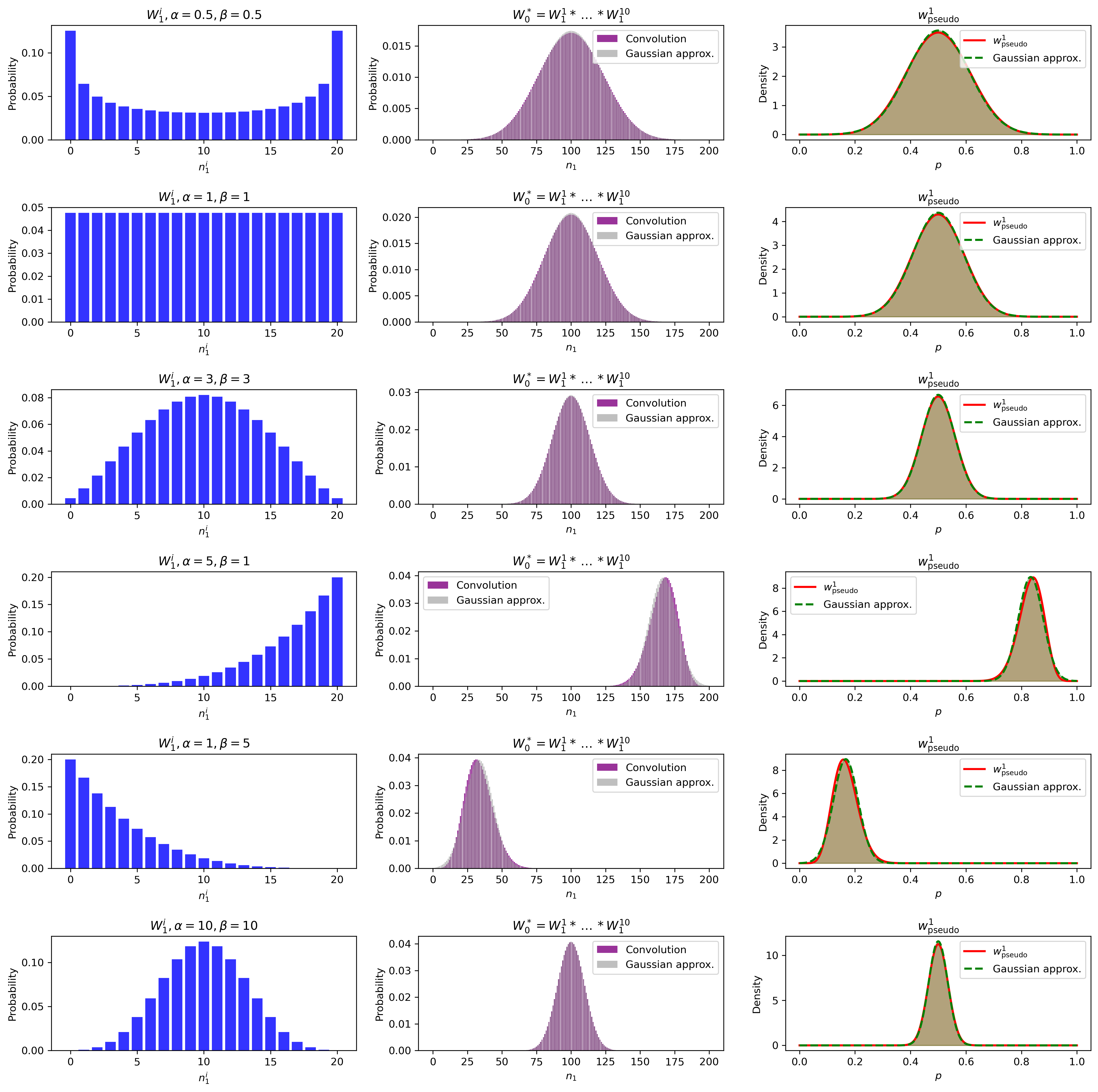}
\caption{The first column shows the discrete distribution of each component of the alternative sufficient statistics, here denoted simply by $W^i_{1}$, induced by independent identical beta priors on the alternative, for different prior parameters. The GRO-optimal microcanonical prior on the null $W_0^*$ (second column) for the $2\times k$ test is obtained by convolving these discrete distributions $k$ times. The pseudo prior density $w^1_{pseudo}$ (third column) is instead obtained by directly convolving $k$ times the corresponding beta priors. Both $W_0^*$ and $w^1_{pseudo}$ are shown together with their Gaussian approximations (discrete for $W_0^*$ and continuous for $w^1_{pseudo}$).  In this example, $k = 10$.}
\label{SM_fig_beta_convolutions_2x10}
\end{figure}

\begin{figure}[ht]
    \centering
    \begin{minipage}{0.45\textwidth}
        \centering
    \includegraphics[width=\textwidth]{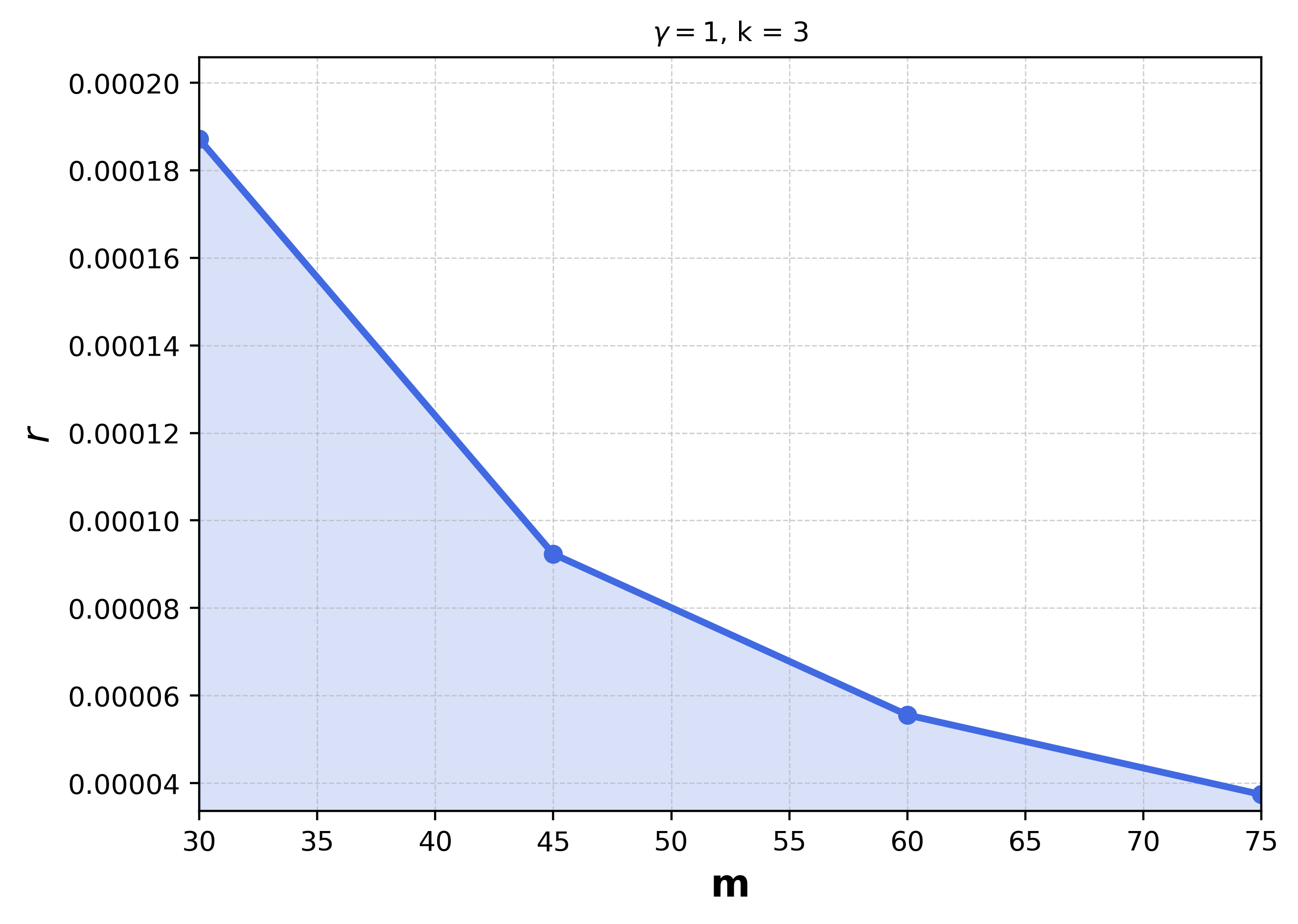} \\[2pt]
    \includegraphics[width=\textwidth]{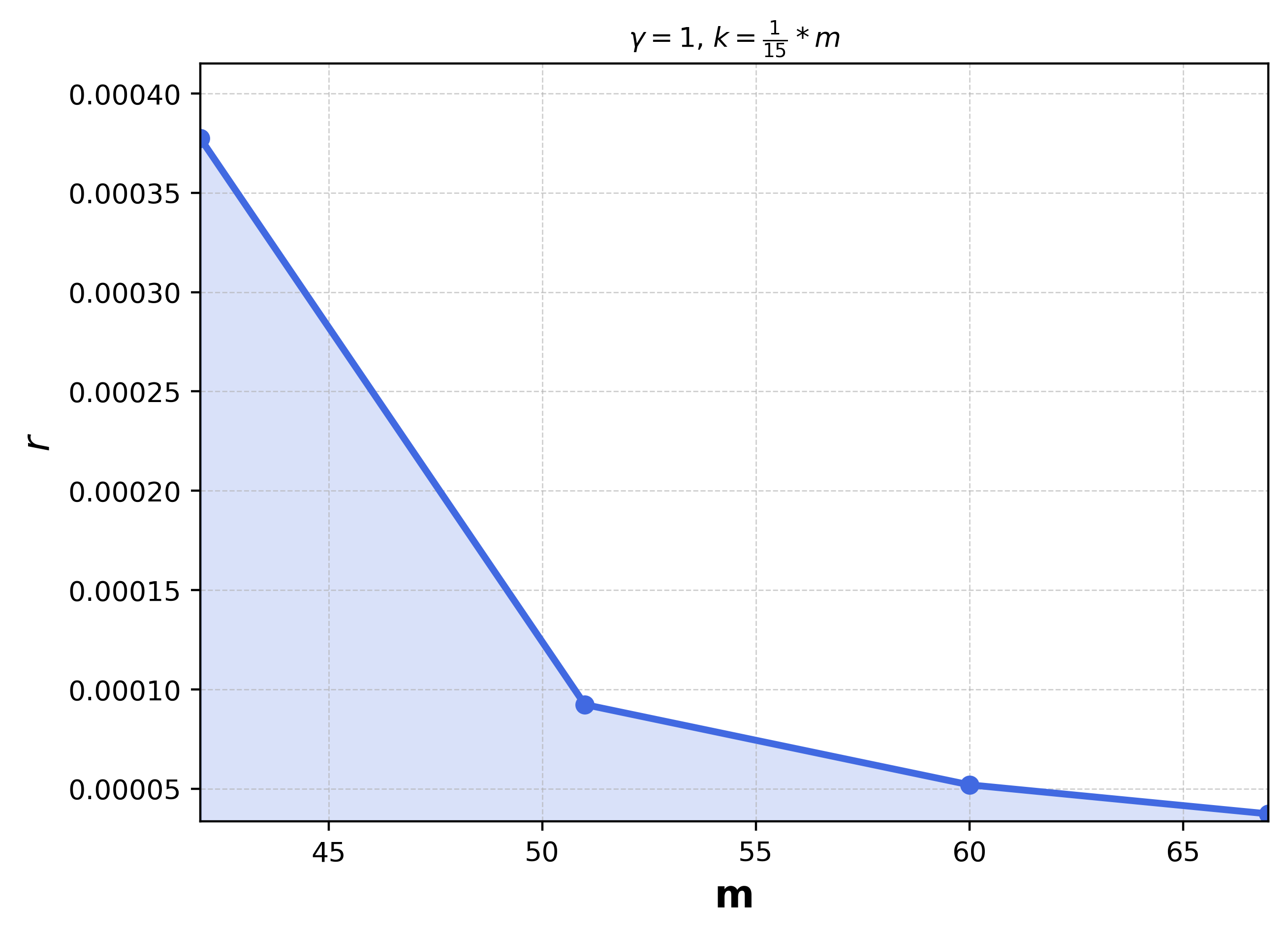} \\[2pt]
    \includegraphics[width=\textwidth]{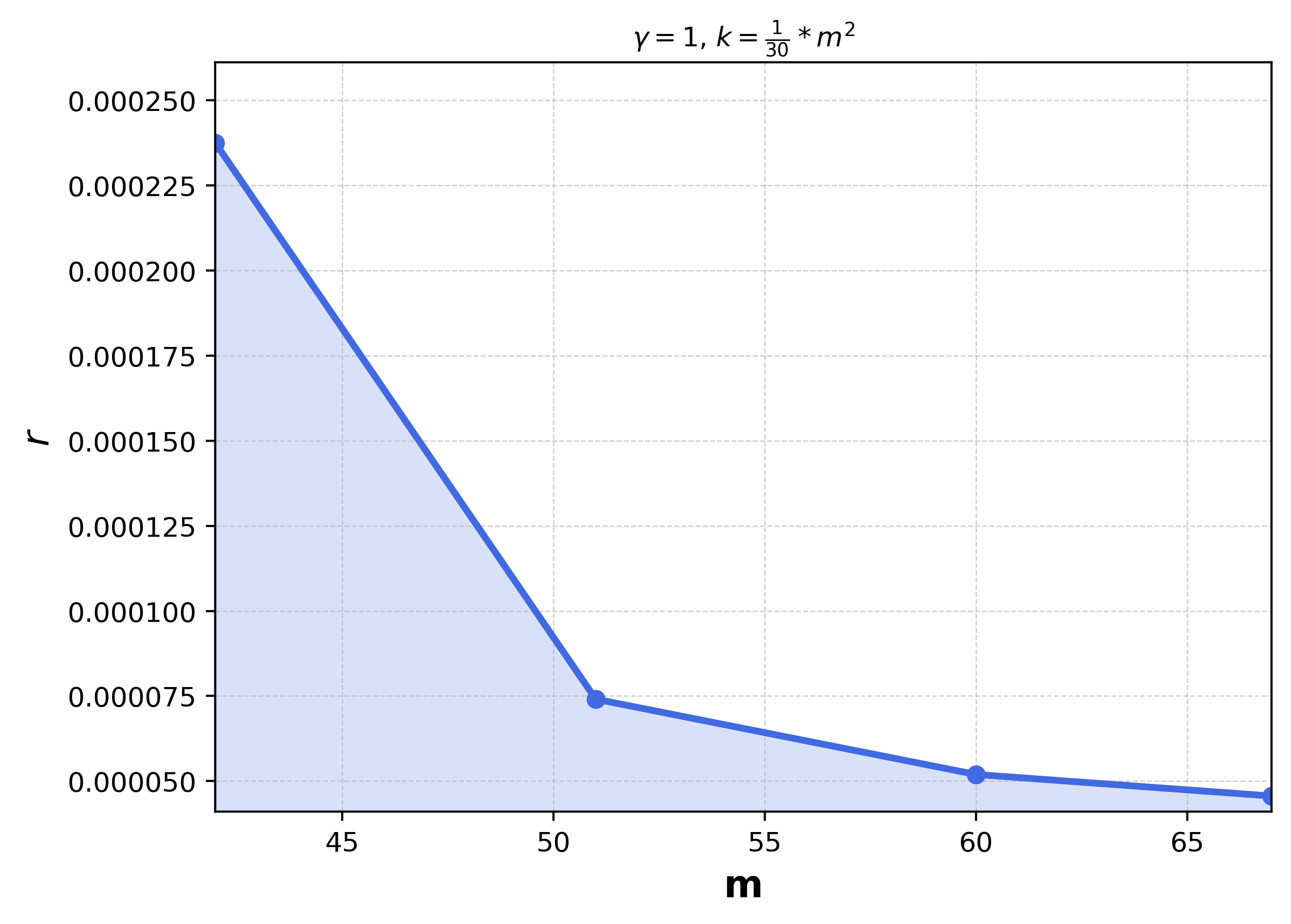} \\[2pt]
    \end{minipage}
    \hfill
    \begin{minipage}{0.45\textwidth}
        \centering
    \includegraphics[width=\textwidth]{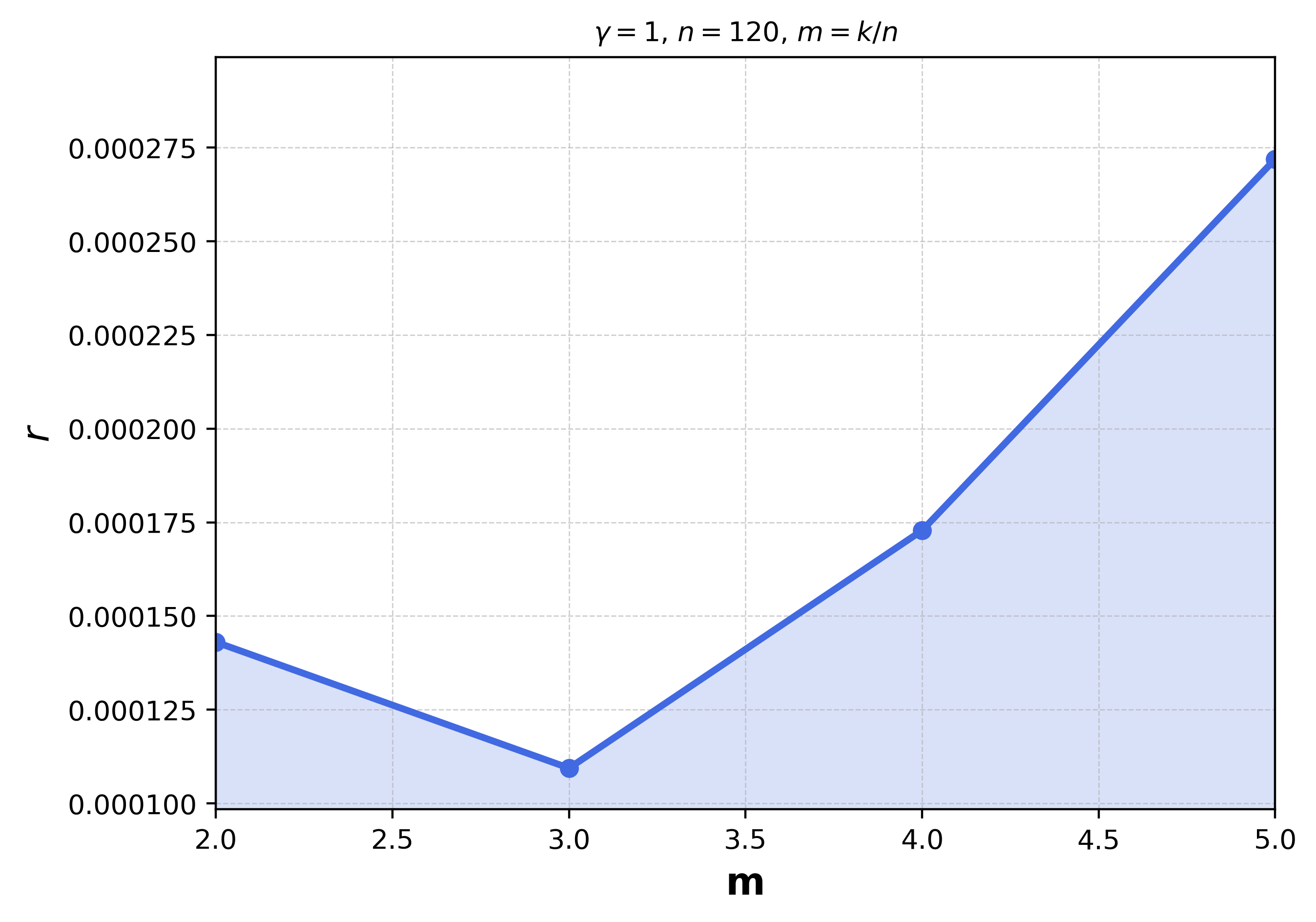} \\[2pt]
    \includegraphics[width=\textwidth]{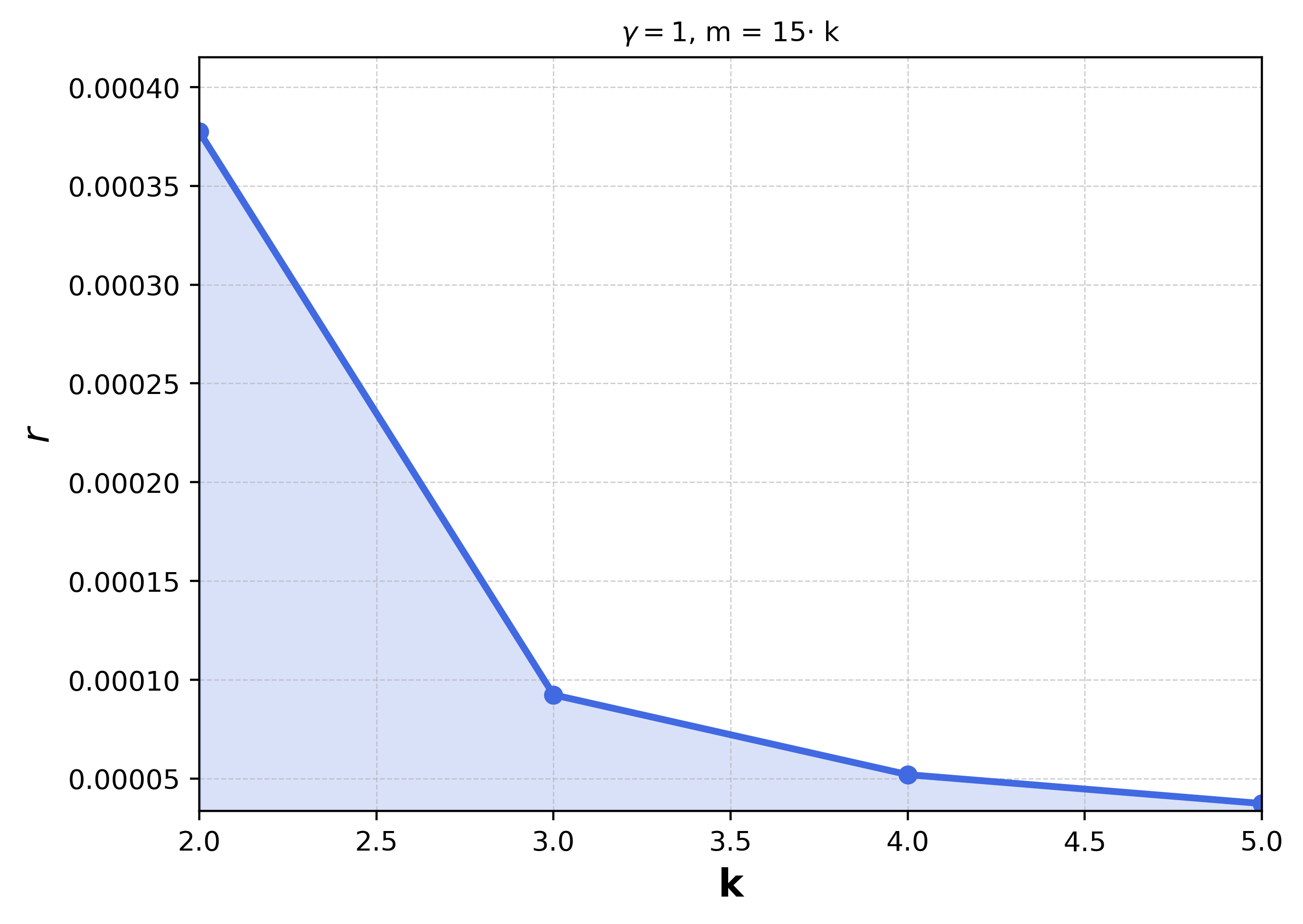} \\[2pt]
    \includegraphics[width=\textwidth]{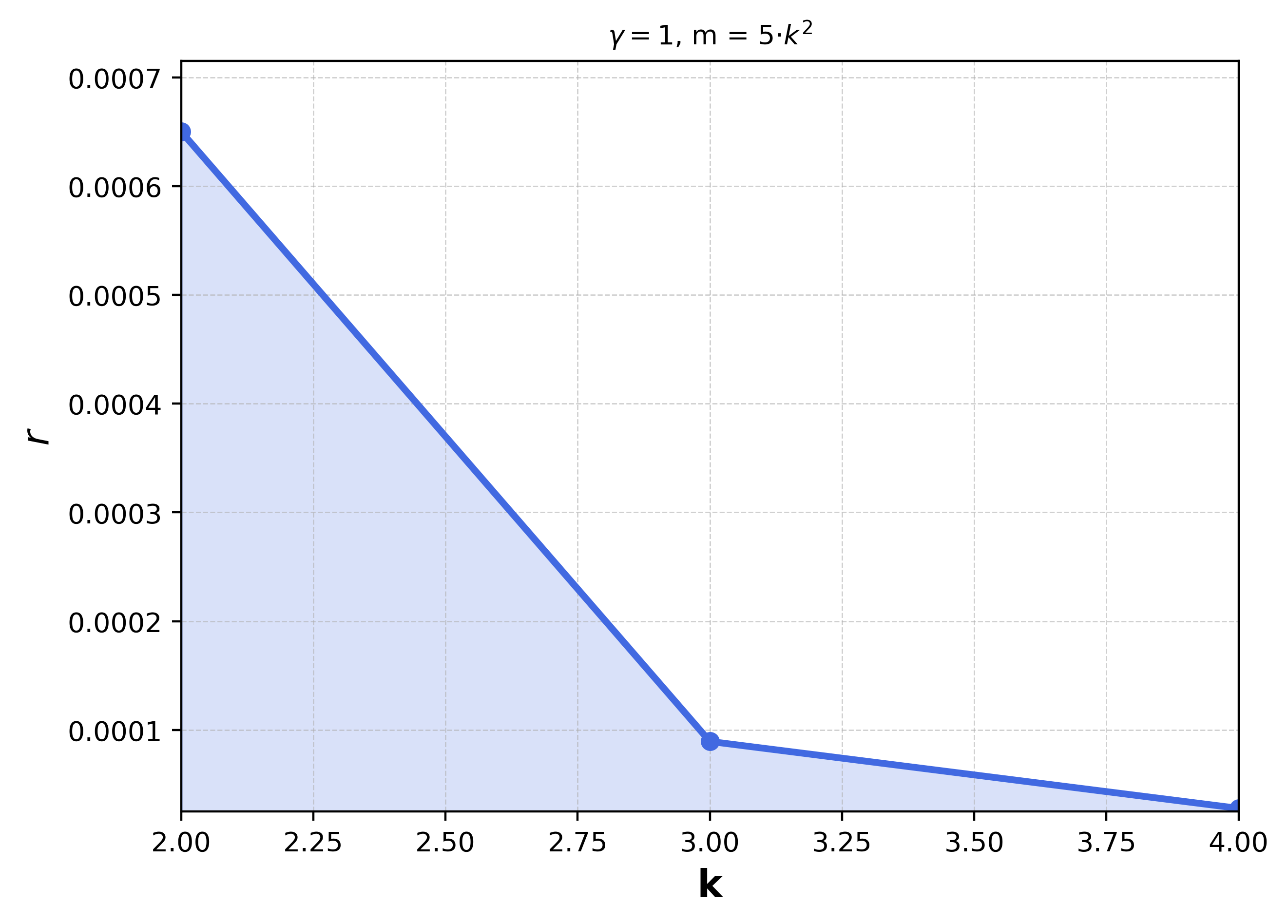} \\[2pt]
    \end{minipage}
\caption{Convergence to 0 of the interval width $r$ in the case of $2\times k$ contingency tables, where all groups have same size $m$, for different interplays between the number of groups $k$ and the size of each group $m$. Results are shown for identical independent beta priors on the alternative, with all parameters equal to $\gamma = 1$.}
\label{fig:r_2x2_k_and_m}
\end{figure}
\newpage

\end{document}